\documentclass[a4paper,fleqn,usenatbib]{mnras}

\usepackage[T1]{fontenc}
\usepackage{ae,aecompl}
\usepackage{float}
\usepackage{wrapfig}

\usepackage{graphicx}	
\usepackage{amsmath}	
\usepackage{amssymb}	

\title[Seismic emission in sunspots accompanying flares]{On the seismic emission in sunspots associated with Lorentz force changes accompanying major solar flares}

\author[H. Kumar and B. Kumar]{
Hirdesh Kumar,$^{1,2}$\thanks{E-mail: hirdesh@prl.res.in}
and Brajesh Kumar$^{1}$
\\
$^{1}$Udaipur
Solar Observatory, Physical Research Laboratory, Dewali, Badi Road, Udaipur 313004 Rajasthan, India\\
$^{2}$Department of physics, Indian Institute of Technology Gandhinagar, Gandhinagar 382 355 Gujarat, India\\
}

\date{Accepted XXX. Received YYY; in original form ZZZ}

\pubyear{2020}

\begin{document}
\label{firstpage}
\pagerange{\pageref{firstpage}--\pageref{lastpage}}
\maketitle


\begin{abstract}
Solar flares are known to generate seismic waves in the Sun. We present a detailed analysis of seismic emission in sunspots accompanying M- and X-class solar flares. For this purpose, we have used high-resolution Dopplergrams and line-of-sight magnetograms at a cadence of 45 s, along with vector magnetograms at a cadence of 135 s obtained from Helioseismic and Magnetic Imager (HMI) instrument aboard the {\em Solar Dynamic Observatory} ({\em SDO}) space mission. In order to identify the location of flare ribbons and hard X-ray foot-points, we have also used H$\alpha$ chromospheric intensity observations obtained from Global Oscillation Network Group (GONG) instruments and hard X-ray images in 12--25 KeV band from the {\em Reuvan Ramaty High Energy Solar Spectroscopic Imager} ({\em RHESSI}) spacecraft. The  Fast Fourier Transform (FFT) technique is applied to construct the acoustic velocity power map in 2.5--4 mHz band for pre-flare, spanning flare, and post flare epochs for the identification of seismic emission locations in the sunspots. In the power maps, we have selected only those locations which are away from the flare ribbons and hard X-rays foot-points. These regions are believed to be free from any flare related artefacts in the observational data. We have identified concentrated locations of acoustic power enhancements in sunspots accompanying major flares. Our investigation provides evidence that abrupt changes in the magnetic fields and associated impulsive changes in the Lorentz force could be the driving source for these  seismic emissions in the sunspots during solar flares.   
\end{abstract}




\section{Introduction}

Solar flares are the magnetized events in the solar atmosphere, in which previously-stored magnetic energy of the order of $10^{27}$--$10^{32}$ ergs are released into the solar atmosphere in the form of thermal radiation, mass motion and accelerated charged particles within few minutes to an hour. During  solar flares, the back bombarded charged particles gyrate along the magnetic fields lines and generate gyrosynchrotron emissions \citep{1972SoPh...23..155H}. The bremsstrahlung radiation generated by deacceleration of charged particles while striking the chromospheric plasma produces hard X-rays \citep{1971SoPh...18..489B}. It was suggested by \cite{1972ApJ...176..833W} that large solar flares could also stimulate free global oscillations in the Sun. Much later, \cite{1985ApJ...289..425F} proposed that during the course of solar flares, the bombardment of charged particles on the chromospheric layers heats the chromospheric plasma and, as result of that, chromospheric evaporation into the corona takes place. Further, the downward settlement of the condensed material can launch a shock on the photosphere thereby delivering large momentum to the photospheric layers \citep{1989ApJ...346.1019F} and this high amount of energy can excite the acoustic waves inside the Sun. \citet{1998Natur.393..317K} found the first instance of flare-induced seismic emission in the Sun during a moderate X-class solar flare, which occurred on 9 July 1996, using the data from Michelson Doppler Imager (MDI; \citet{1995SoPh..162..129S}) instrument aboard the {\em Solar and Heliospheric Observatory} ({\em SOHO}; \citet{1995SoPh..162....1D}) space mission. They identified this phenomenon as ``sunquake". Later on, there are several others cases reported by \cite{2005ApJ...630.1168D}, \cite{2006SoPh..239..113D}, \cite{2007MNRAS.374.1155M} and \cite{2006JApA...27..425K}, in which they have found acoustic emissions accompanying different classes of solar flares. \cite{2007ApJ...664..573Z} found acoustic emissions in a proton-rich solar flare and they explained these enhancements as due to the backreaction of the shock driven by high energetic charged particles impinging on the solar photosphere during the flare. Following this, \cite{2008MNRAS.387L..69V} observationally found that these high energetic charged particles can reach up to the photosphere from the chromosphere in a time span of about 1 minute and thus could be responsible for enhancing the velocity oscillations.
\cite{2008ASPC..383..221H} proposed an another mechanism, known as ``magnetic jerk", which could be responsible for the generation of seismic waves in sunspots, apart from the high energetic charged particles and back reaction of shock at the solar photosphere.\\

It is well established that during solar flares, magnetic fields of the corona change and the signatures of those appear in the form of the temporal evolution of the magnetic fields at the photosphere. The changes in the magnetic fields take place within short time scales before the flare, during the flare, and the post flare epochs. \cite{1981ApJ...243L..99P} was the first to report the rapid short term changes (transient changes) in the magnetic fields during solar flares. Further, \cite{1984ApJ...280..884P} illustrated that these observed transient changes in the magnetic fields could be due to emission in the core of the spectral line, which is used to derive the information of the magnetic fields. In addition, \cite{2003ApJ...599..615Q} discussed the possibility of change in line profiles during a transient polarity reversal observed in the photospheric magnetic field measurements obtained from the MDI instrument aboard {\em SOHO} space mission. \cite{1992SoPh..140...85W} and \cite{1994ApJ...424..436W,2002ApJ...576..497W} reported abrupt and permanent changes in the evolution of magnetic fields in the flaring region during the different class of solar flares using the data from Big Bear Solar Observatory. Later on, extensive analysis done by \cite{2005ApJ...635..647S} and \cite{2010ApJ...724.1218P} of X- and M-class solar flares using the line-of-sight (i.e., longitudinal) magnetic fields obtained from Global Oscillation Network Group (GONG; \cite{1996Sci...272.1284H}) instruments, showed abrupt and permanent changes in the magnetic fields in the flaring locations within 10-minute duration with a median magnitude of 100 Gauss. In addition, \cite{2016RAA....16..129K} also found abrupt and persistent changes in line-of-sight magnetic fields at different locations in active region during an M6.5 class solar flare using the data from Helioseismic and Magnetic Imager (HMI; \citet{2012SoPh..275..229S}) instrument aboard the {\em Solar Dynamic Observatory} ({\em SDO}; \citet{2012SoPh..275....3P}) space mission.\\

\cite{2008ASPC..383..221H} and \cite{2012SoPh..277...59F} proposed that abrupt changes in the magnetic fields can lead to an impulsive change in Lorentz force, which is also known as ``magnetic-jerk", and this can induce seismic emission in the sunspots. However, \cite{2014RAA....14..207R} observed distortions in the two circular polarization states of light, the Left Circular Polarization (LCP) and the Right Circular Polarization (RCP) observations in the flaring locations which showed transient changes in Doppler velocity and line-of-sight magnetic fields during an X-class flares, using the line profile observations from HMI instrument aboard the {\em SDO} space mission. They showed that observed transient changes in Doppler velocity and line-of-sight magnetic fields in flaring locations are prone to artefacts in the measurements.\\

Therefore, in this paper, we present a detailed analysis of seismic emission in the sunspots, which are away from the flare ribbons and hard X-ray foot-points accompanying large solar flares. Such acoustic enhancements are supposed to be free from any flare related problems in the observational data.\\

In the following Sections, we first describe the observational data and selection criteria for chosing the active regions aimed for our analysis. This is followed by methods of analysis and the results obtained. We finally conclude with discussions concerning our results.

\begin{table*}
	\centering
	\caption{Details of the active regions used in our analysis and information related to the flare evolution as seen in GOES soft X-ray     (1--8 {\AA} band).}
	\label{AR list}
	\begin{tabular}{lcccccr} 
		\hline
		Active Region & Location on the disc & Flare class & Date & Start time & Peak Time & Decay Time \\
		\hline
NOAA 11158 & S28W17 & X2.2 & 2011 February 15 & 01:44 UT & 01:56 UT & 02:06 UT \\
NOAA 11261 & N15W35 & M6.0 & 2011 August 03 & 13:17 UT & 13:48 UT & 14:10 UT \\
NOAA 11882 & S06E28 & M2.7 & 2013 October 28 & 14:46 UT & 15:01 UT & 15:04 UT\\
NOAA 11882	& S06E28&  M4.4 & 2013 October 28 & 15:07 UT & 15:15 UT & 15:21 UT\\
NOAA 12222 & S20W31 & M6.1 & 2014 December 04 & 18:05 UT & 18:25 UT & 18:56 UT\\
NOAA 12241 & S09W09 & M6.9 & 2014 December 18 & 21:41 UT & 21:58 UT & 22:25 UT \\	
NOAA 12242 & S21W24 & X1.8 & 2014 December 20 & 00:11 UT & 00:28 UT & 00:58 UT \\
NOAA 12297 & S16E13 & X2.0 & 2015 March 11 & 16:11 UT  & 16:22 UT &  16:29 UT \\
NOAA 12371 &  N13W06 & C3.9 & 2015 June 22 &  17:20 UT & 17:27 UT &17:33 UT \\
NOAA 12371 & N13W06 & M6.5 & 2015 June 22 & 17:39 UT & 18:23 UT & 18:23 UT\\
		\hline
	\end{tabular}
\end{table*}

\begin{figure*}
\centering
\includegraphics[width=0.55\textwidth]{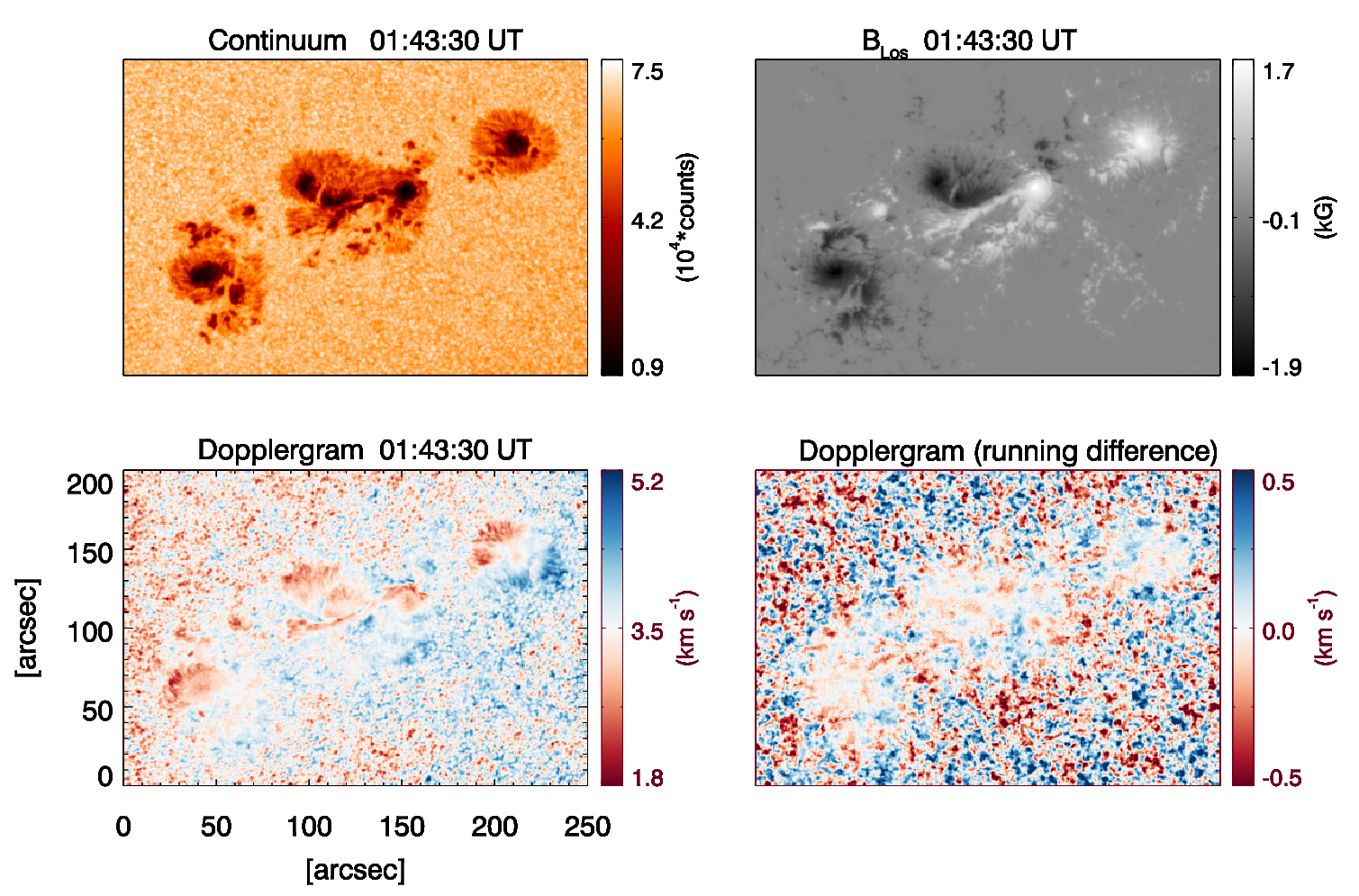}\hspace*{0.2cm}
\includegraphics[width=0.46\textwidth]{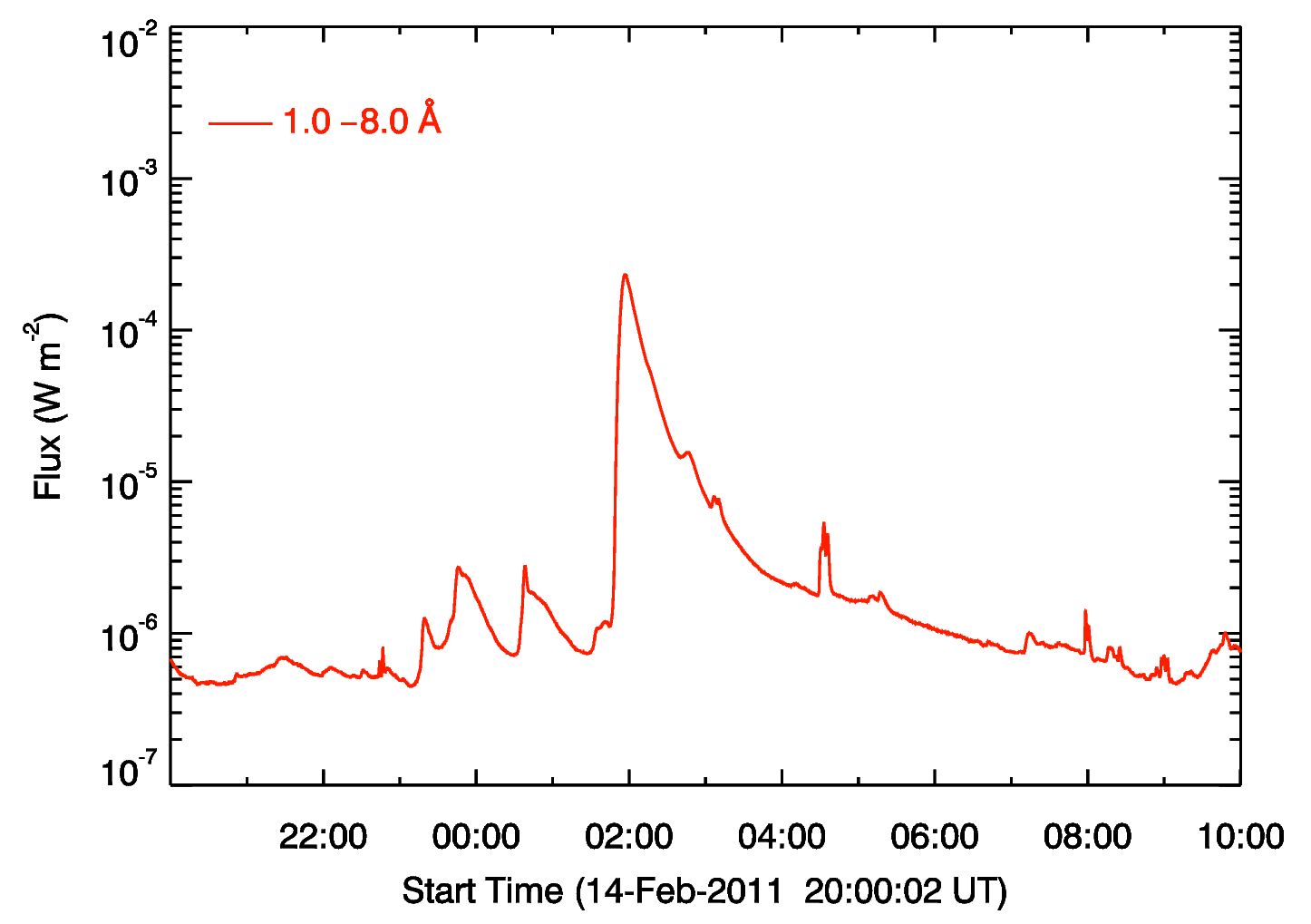}
\caption{\textit{Left panel}: Sample images of the active region NOAA 11158 showing continuum intensity image (top left panel), photospheric line-of-sight (i.e., longitudinal and scalar) magnetic fields (top right panel), Dopplergram (bottom left panel) and running difference of Doppler images (bottom right panel) acquired from HMI instrument aboard {\em SDO} spacecraft on 2011 February 15. \textit{Right panel}: Plot shows the temporal evolution of disk-integrated solar flux in 1--8 {\AA} band during an X2.2 class flare on 2011 February 15, obtained from {\em GOES} satellite.}
\label{fourimage11158GOES}
\end{figure*}

\begin{figure*}
\centering
\includegraphics[width=0.47\textwidth]{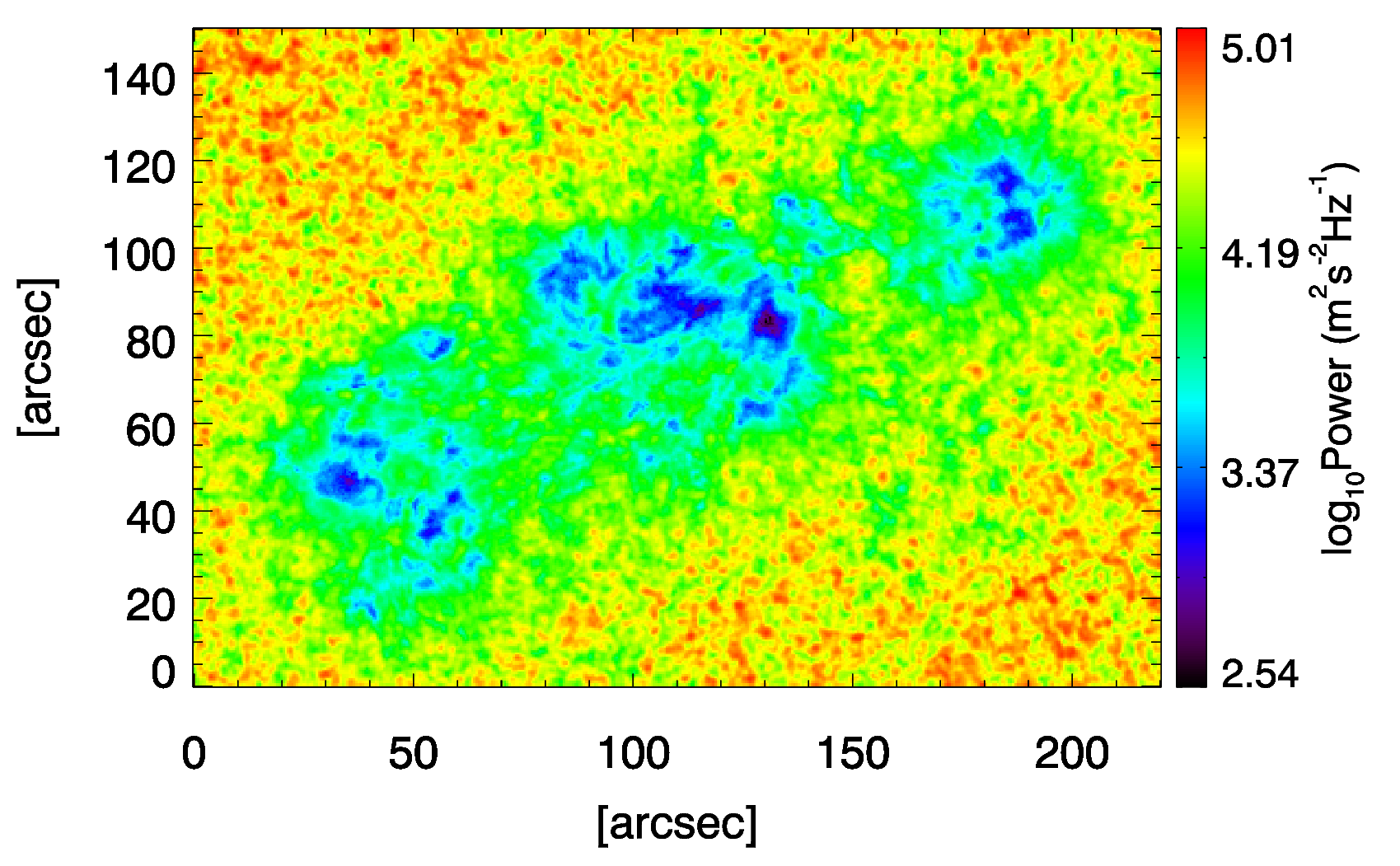}\hspace*{0.34cm}
\includegraphics[width=0.47\textwidth]{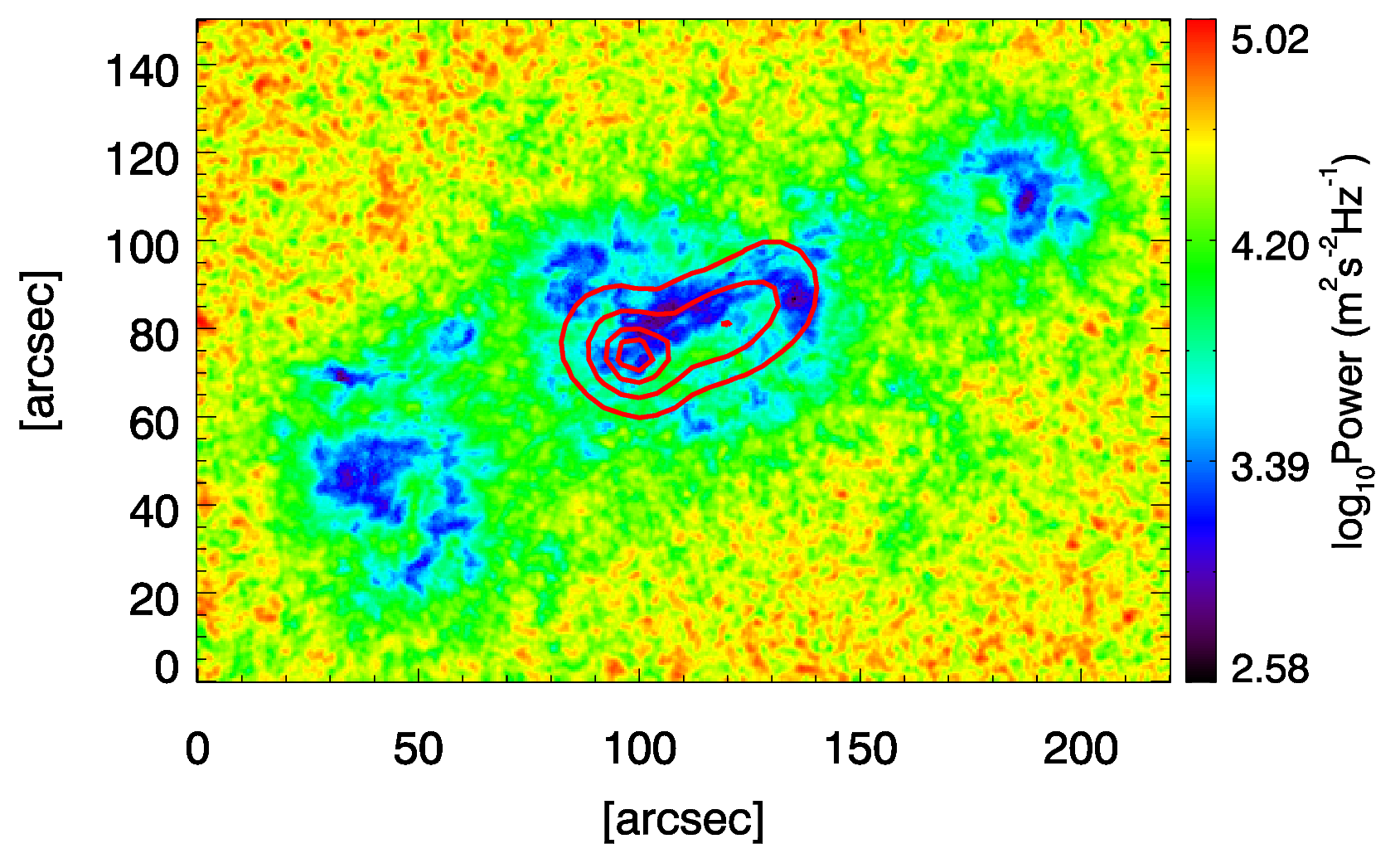}
\caption{\textit{Left panel}: Acoustic power map of active region NOAA 11158 in 2.5--4 mHz band estimated over 2 hour duration for pre-flare epoch. \textit{Right panel}: Same as shown in the left panel but for spanning the flare. Here, red contours represent hard X-rays foot-points at 25, 50, 75 and 90$\%$ of its maximuum as observed in 12--25 KeV band from {\em RHESSI} spacecraft.}
\label{fig: Prespanningflare}
\end{figure*}

\begin{figure*}
\centering
\includegraphics[width=0.52\textwidth]{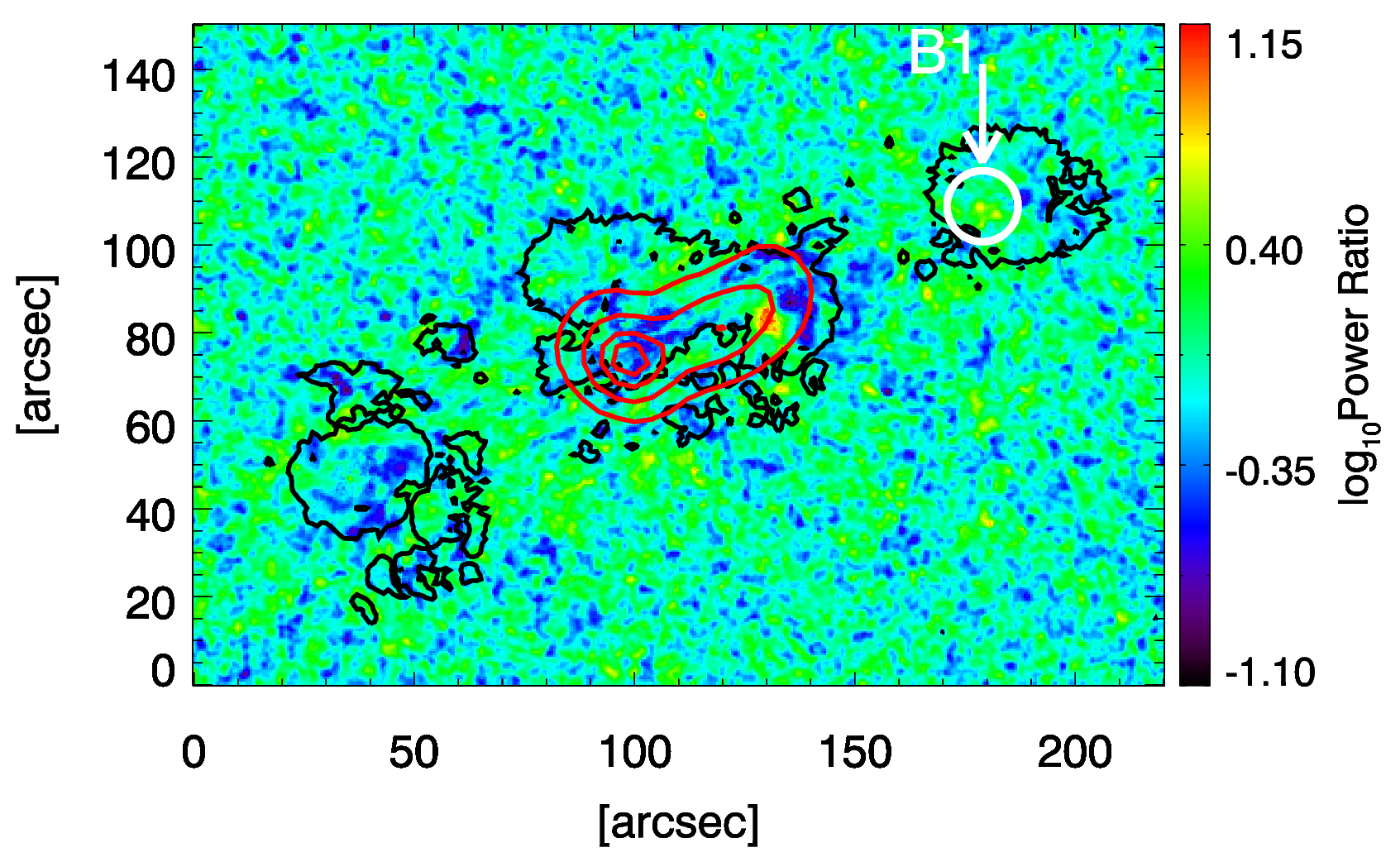}\hspace*{0.34cm}
\includegraphics[width=0.42\textwidth]{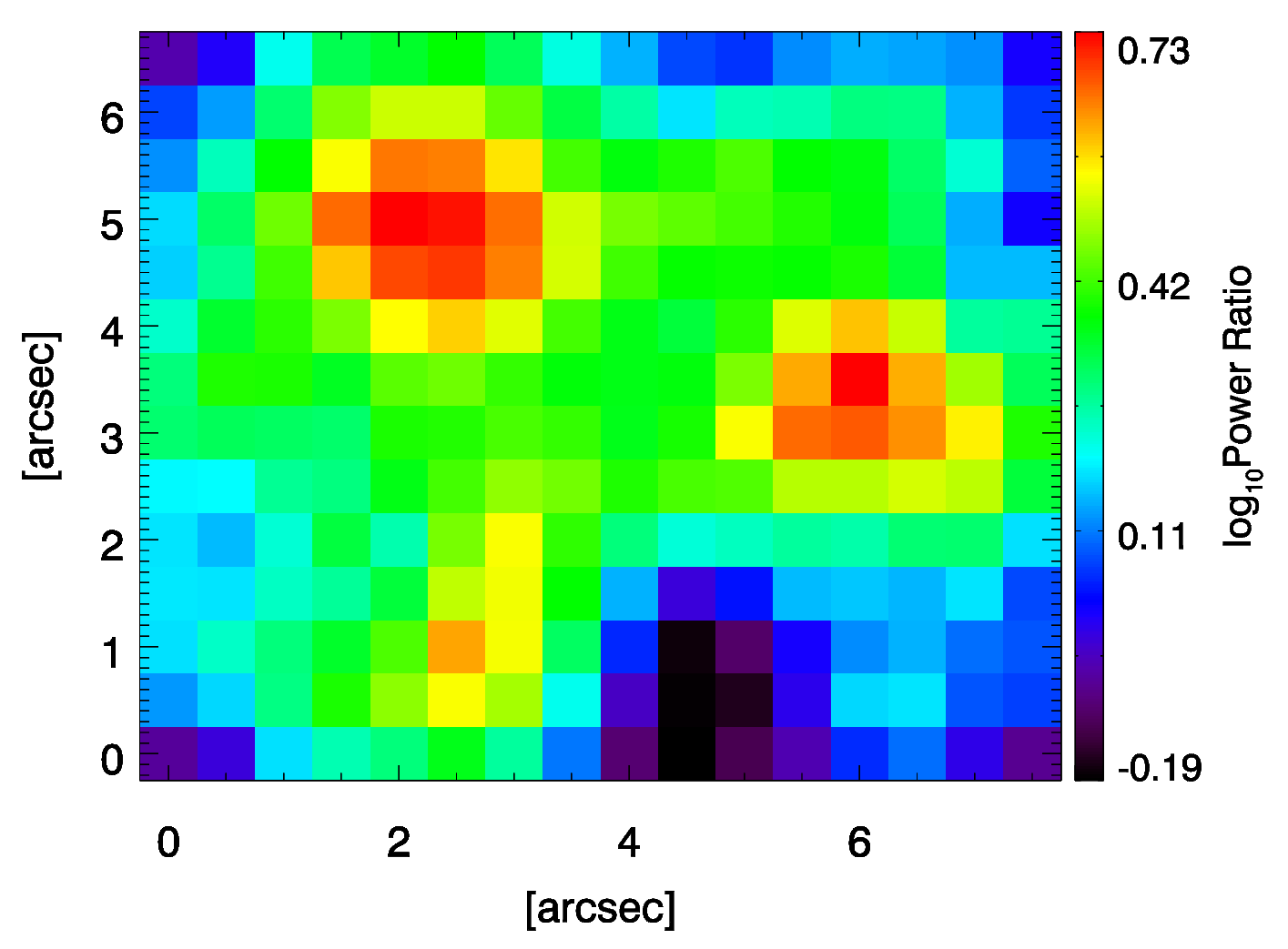}
\caption{\textit{Left panel}: This illustrates the ratio of acoustic power maps estimated for spanning flare and pre-flare epochs in the 2.5--4 mHz band. Here, black contours represent the outer boundary of the sunspot penumbra obtained from continuum intensity image whereas the red contours represent hard X-ray foot-points from {\em  RHESSI} spacecraft in 12--25 KeV band. \textit{Right panel}: Illustrates the blow-up region of `B1' enhanced location in the sunspot as indicated in the power map ratio.}
\label{fig: ratiopowermapB1}
\end{figure*}

\begin{figure*}
\centering
\includegraphics[width=0.45\textwidth]{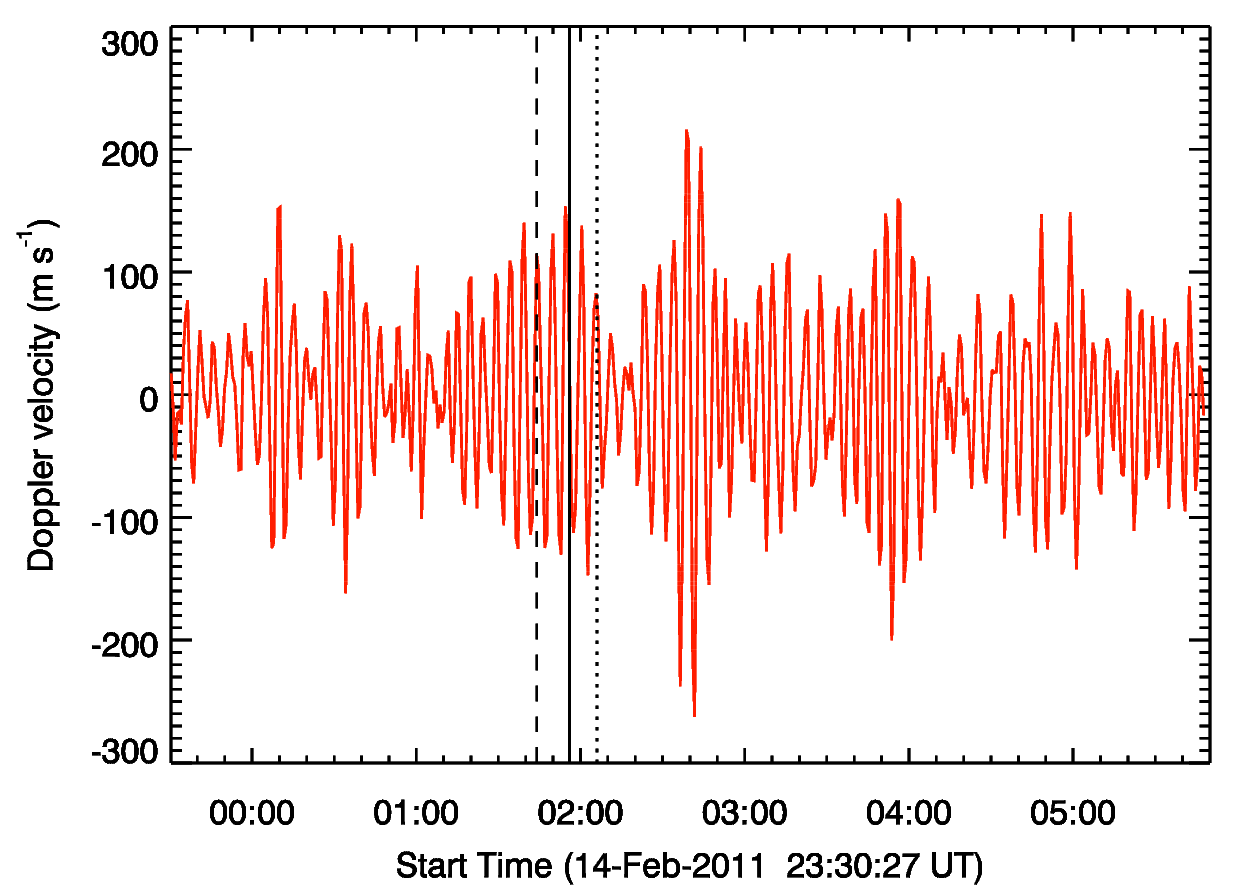}\hspace*{0.34cm}
\includegraphics[width=0.45\textwidth]{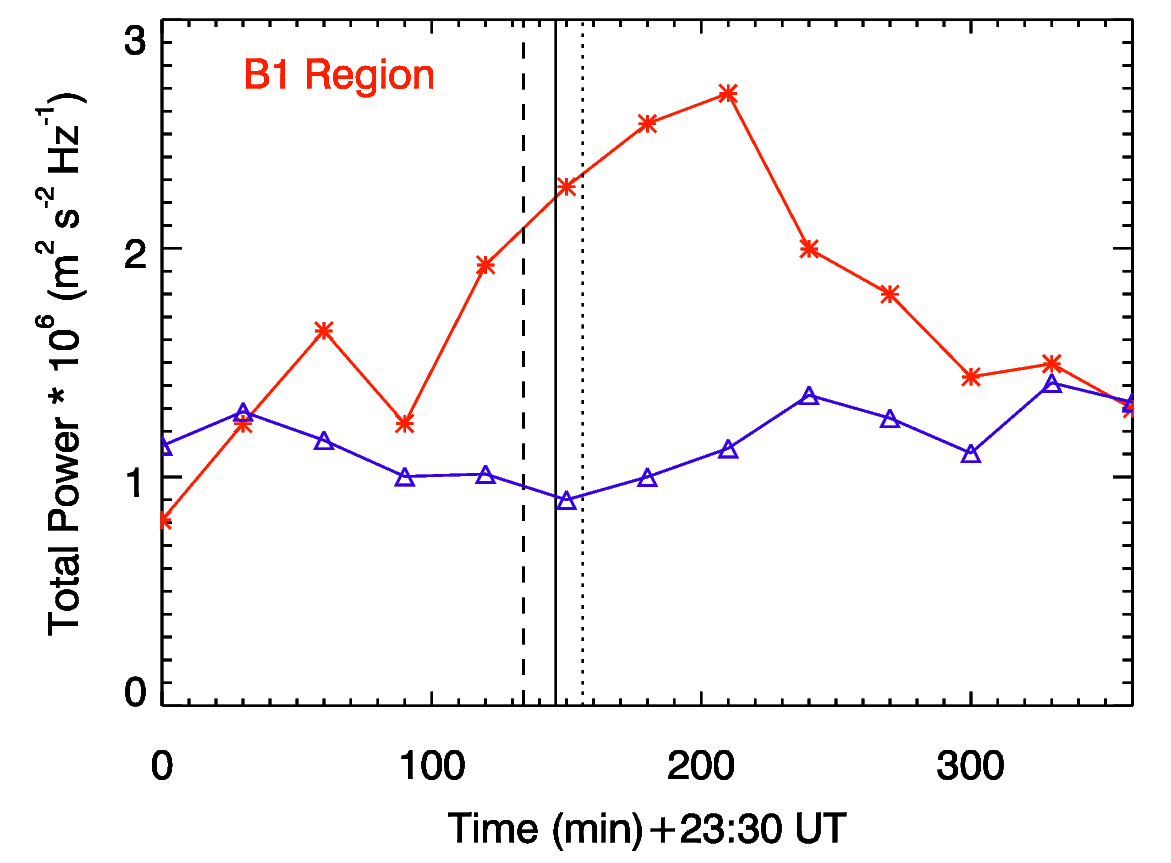}
\caption{\textit{Left panel}: Plot in red colour shows the temporal evolution of Doppler velocity at a cadence of 45 s in the `B1' location of active region NOAA 11158. The dashed, solid and dotted vertical lines represent onset, peak and decay time of the flare. \textit{Right panel}: Plot showing the temporal evolution of integrated acoustic power over the `B1' location (red colour with asterisks) whereas that shown in blue colour with triangles represents evolution of total power in an unaffected region in the same sunspot. It is to be noted that there is a time offset of about $\pm$ 30-minutes between the acoustic power variation and the {\em GOES} flare-time.}
\label{fig: DVpowerB1}
\end{figure*}

\begin{figure*}
\centering
\includegraphics[width=0.45\textwidth]{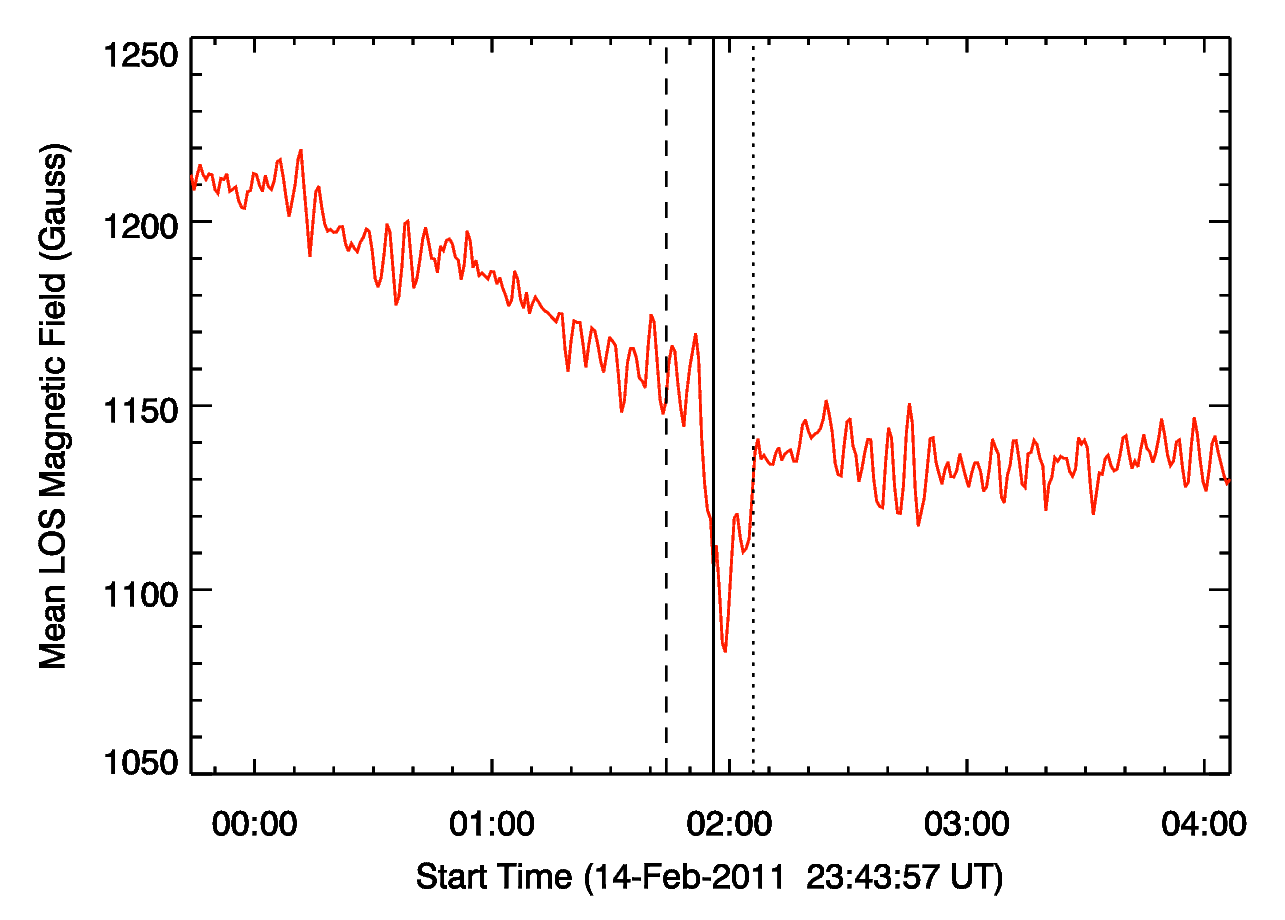}\hspace*{0.34cm}
\includegraphics[width=0.45\textwidth]{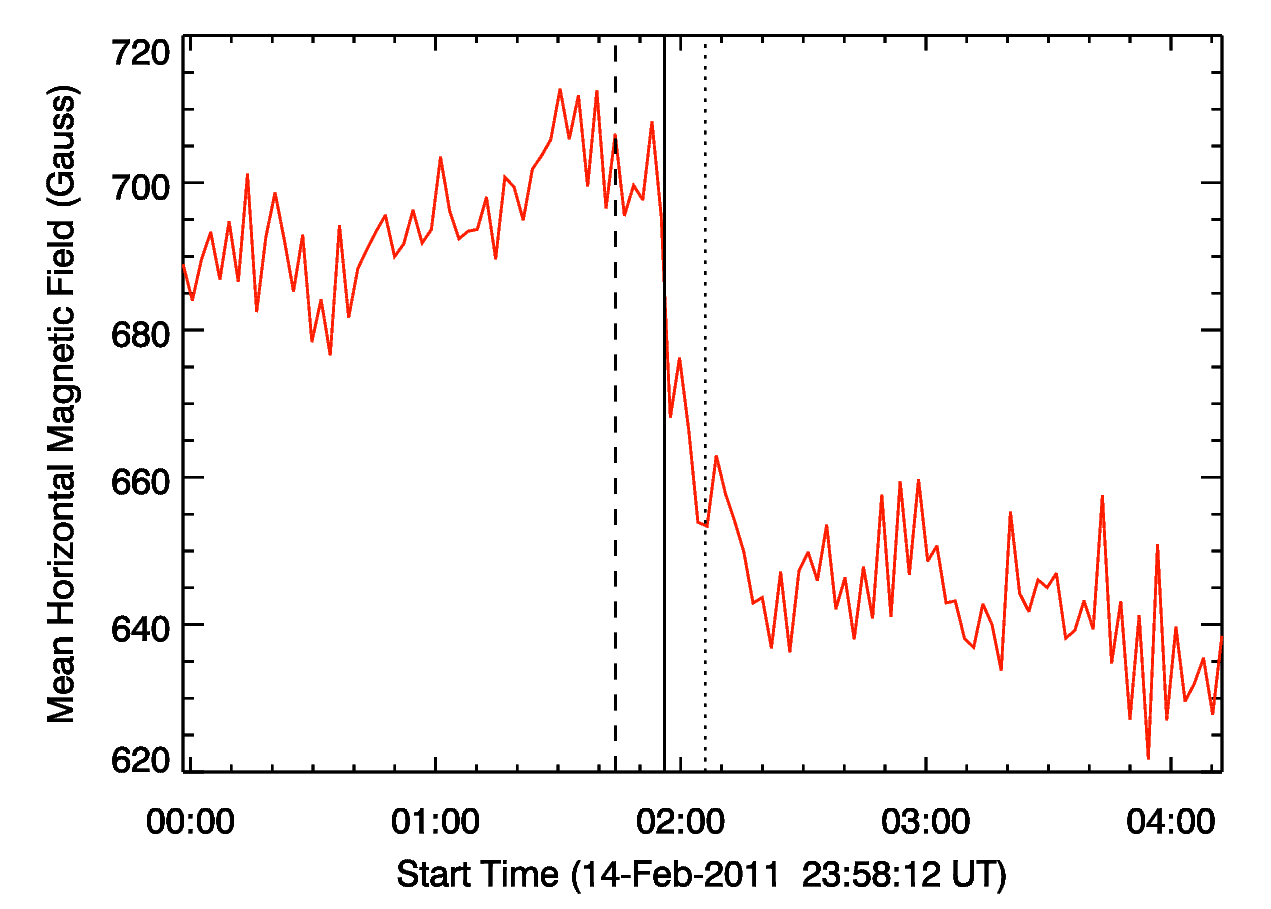}
\caption{\textit{Left panel}: Plot in red colour shows the temporal evolution of line-of-sight (i.e., longitudinal and scalar) magnetic fields at a cadence of 45 s in the `B1' location of active region NOAA 11158. {\textit{Right panel}}: Plot in red colour shows the temporal evolution of horizontal (i.e., transverse) component of vector magnetic fields at a cadence of 135 s in the `B1' location. The dashed, solid and dotted vertical lines represent the onset, peak and decay time of the flare.}
\label{fig: magfield}
\end{figure*}

\begin{figure*}
\centering
\includegraphics[width=0.45\textwidth]{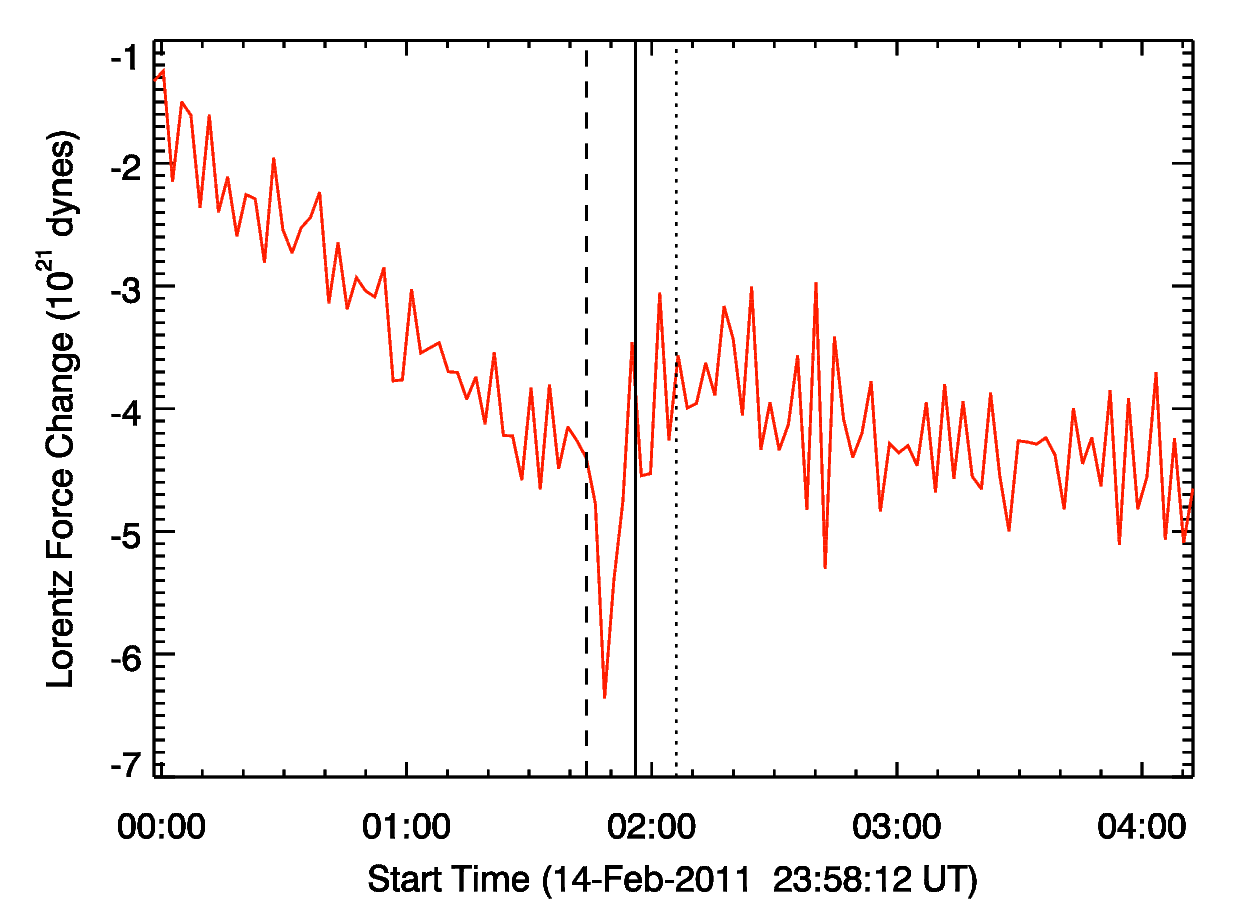}\hspace*{0.34cm}
\includegraphics[width=0.45\textwidth]{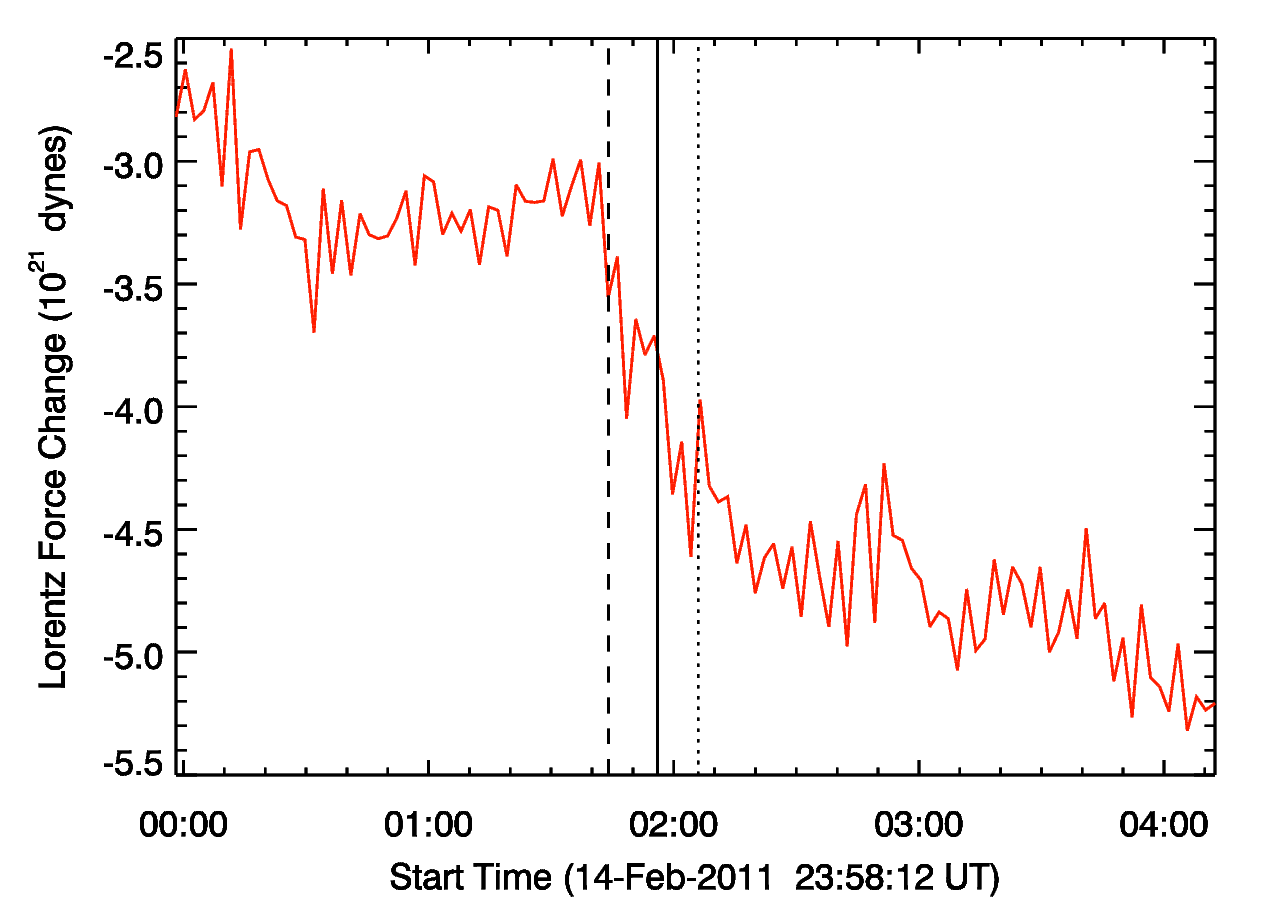}
\caption{\textit{Left panel}: Plot in red colour shows the temporal evolution of change in radial (i.e., upward) component of Lorentz force in the `B1' location of the active region NOAA 11158. \textit{Right panel}: Same as shown in the left panel but for change in horizontal (i.e., transverse) component of Lorentz force in the aforementioned location. The dashed, solid and dotted vertical lines represent the onset, peak and decay time of the  flare.}
\label{fig: lorentzforcechange}
\end{figure*}


\begin{figure*}
\centering
\includegraphics[width=0.52\textwidth]{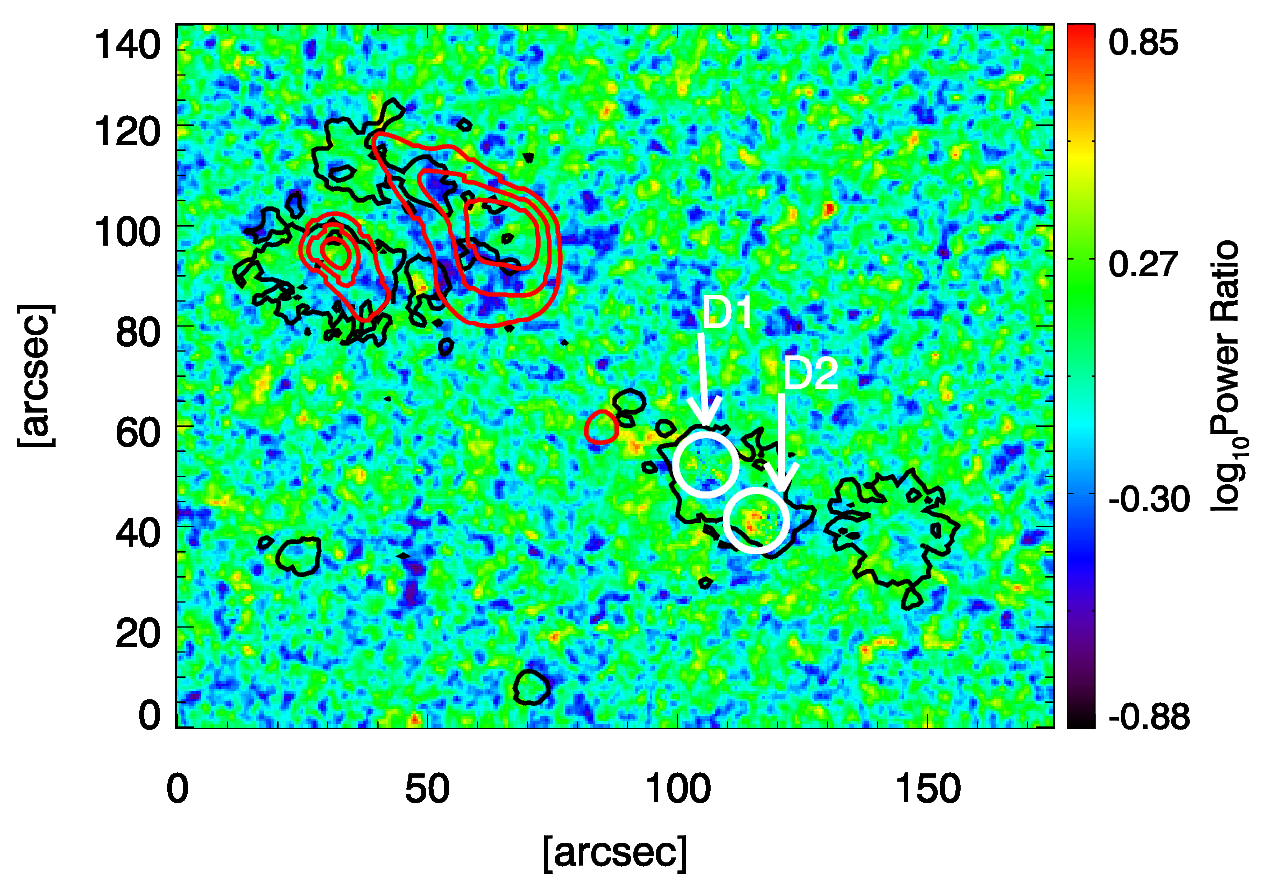}\hspace*{0.34cm}
\includegraphics[width=0.37\textwidth]{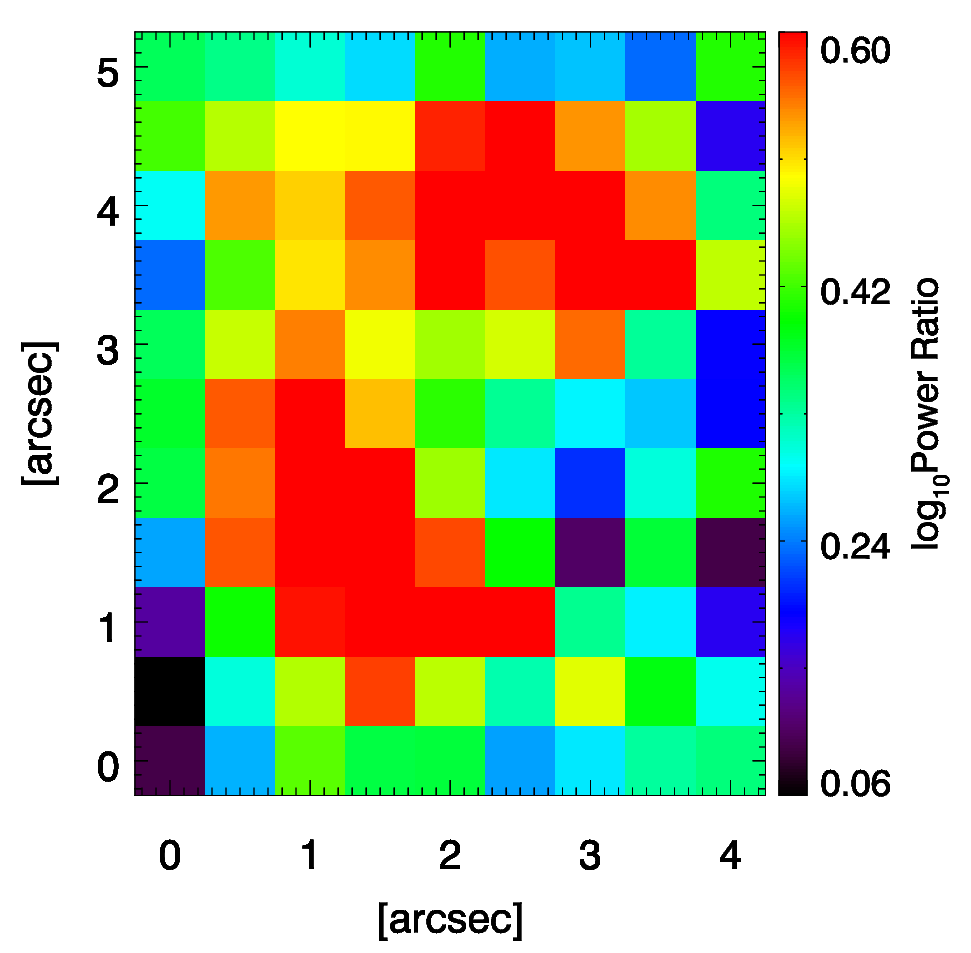}\\
\vspace*{0.5cm}
\includegraphics[width=0.45\textwidth]{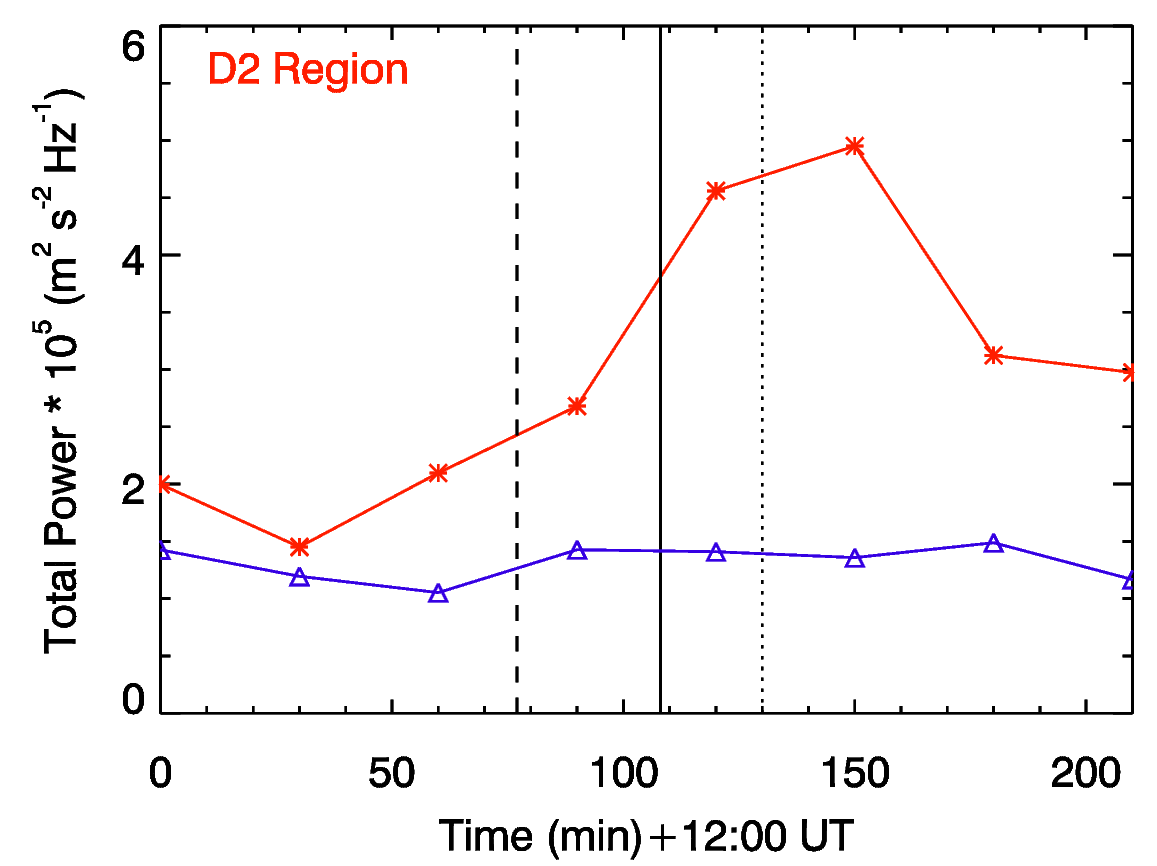}\hspace*{0.34cm}
\includegraphics[width=0.45\textwidth]{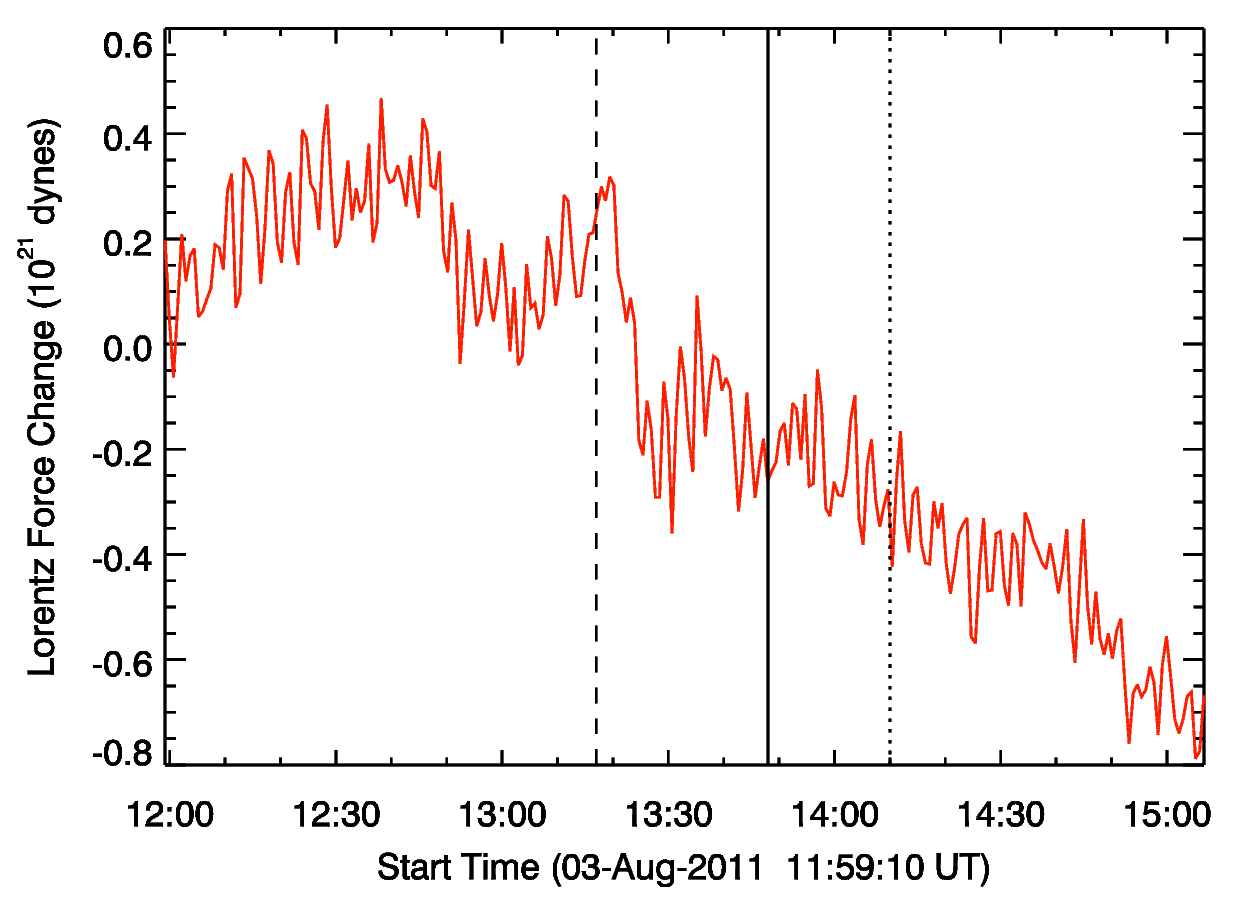}
\caption{A brief illustration of important results obtained for the  active region NOAA 11261. \textit{Top left panel}: This illustrates the ratio of acoustic power maps estimated for spanning flare and pre-flare epochs in the 2.5--4 mHz band. Here, black contours represent the outer boundary of the sunspot penumbra obtained from continuum intensity image whereas the red contours represent flare-ribbon locations from H$\alpha$ chromospheric intensity observations at 70, 80 and 90 $\%$ of its maximum value as observed by the GONG instrument. \textit{Top right panel}: Illustrates the blow-up region of `D2' enhanced location in the sunspot as indicated in the power map ratio. \textit{Bottom left panel}: Plots showing the temporal evolution of integrated acoustic power over the `D2' location (red colour with asterisks) whereas that shown in blue colour with triangles represents evolution of total power in an unaffected region in the same sunspot. It is to be noted that there is a time offset of about $\pm$ 30-minutes between the acoustic power variation and the {\em GOES} flare-time. \textit{Bottom right panel}: Plot in red colour shows the temporal evolution of change in radial (i.e., upward) component of Lorentz force in the `D2' location. The dashed, solid and dotted vertical lines represent the onset, peak and decay time of the flare. The remaining maps and plots have been provided in the online supplementary material.}
\label{AR11261}
\end{figure*}

\begin{figure*}
\centering
\includegraphics[width=0.5\textwidth]{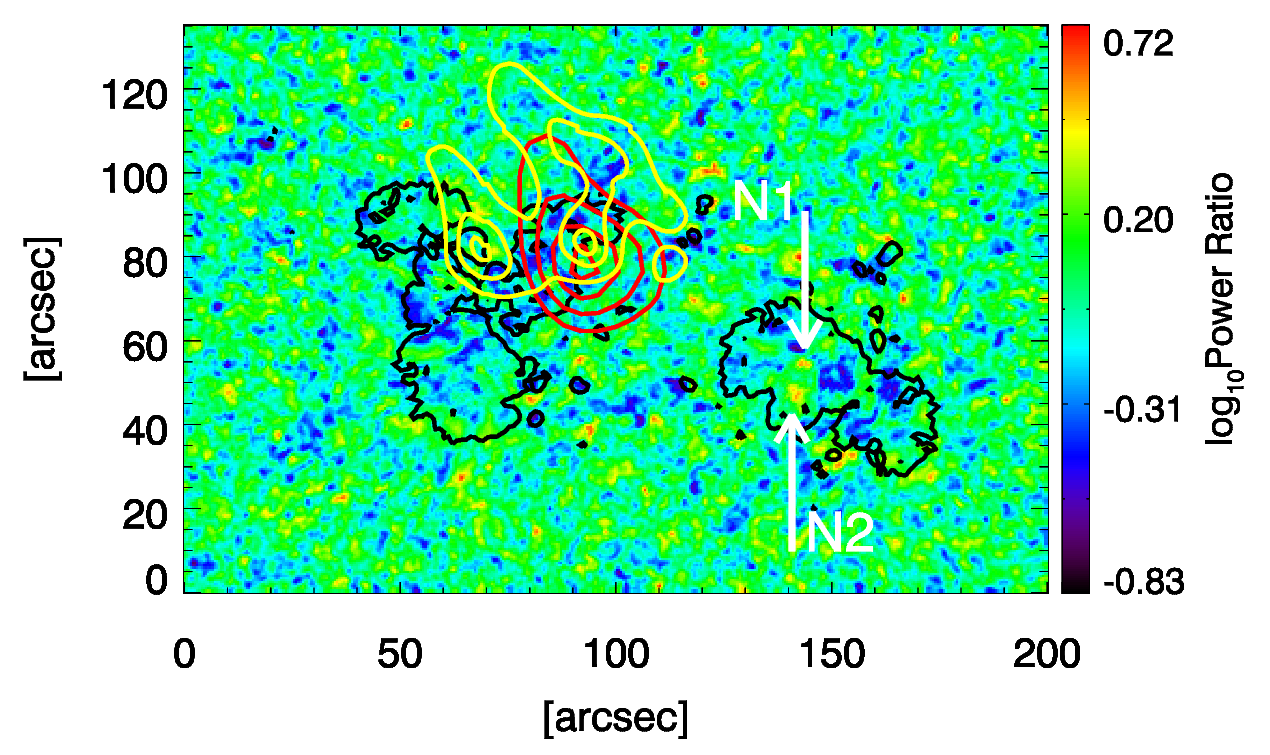}\hspace*{0.34cm}
\includegraphics[width=0.45\textwidth]{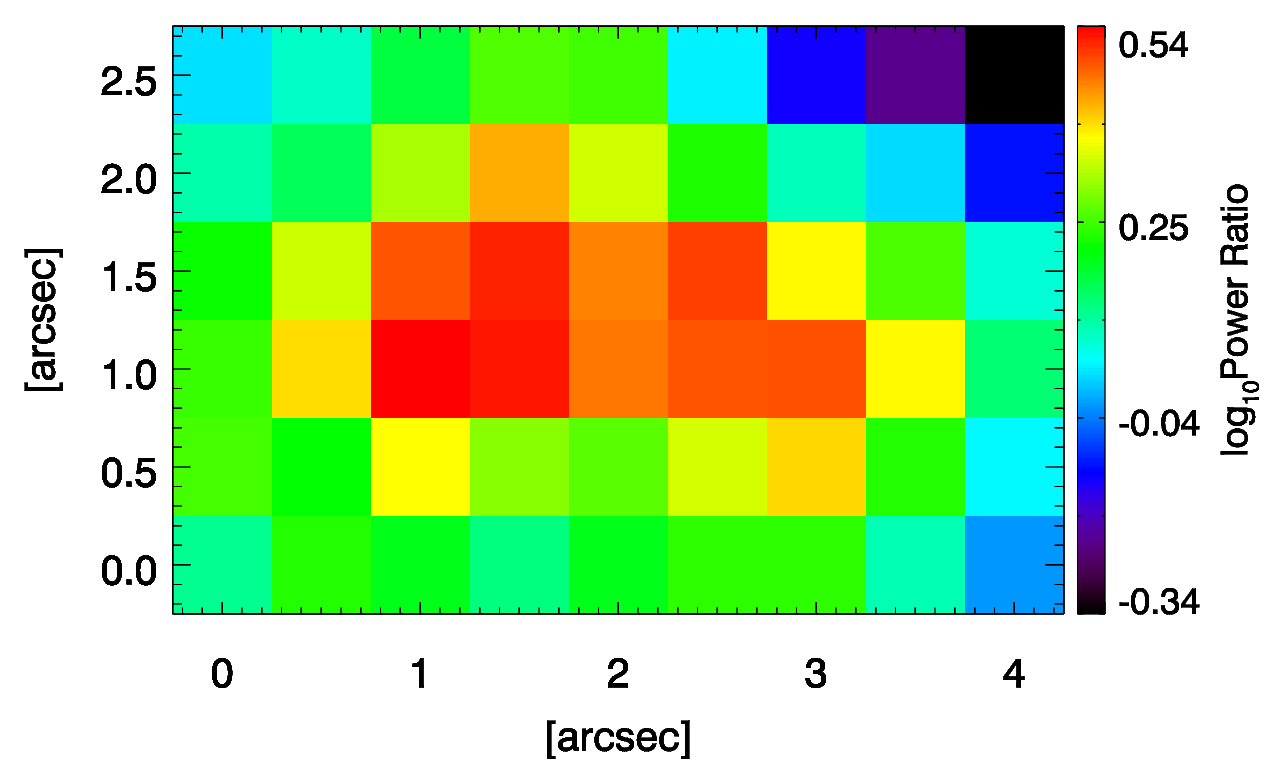}\\
\vspace*{0.5cm}
\includegraphics[width=0.45\textwidth]{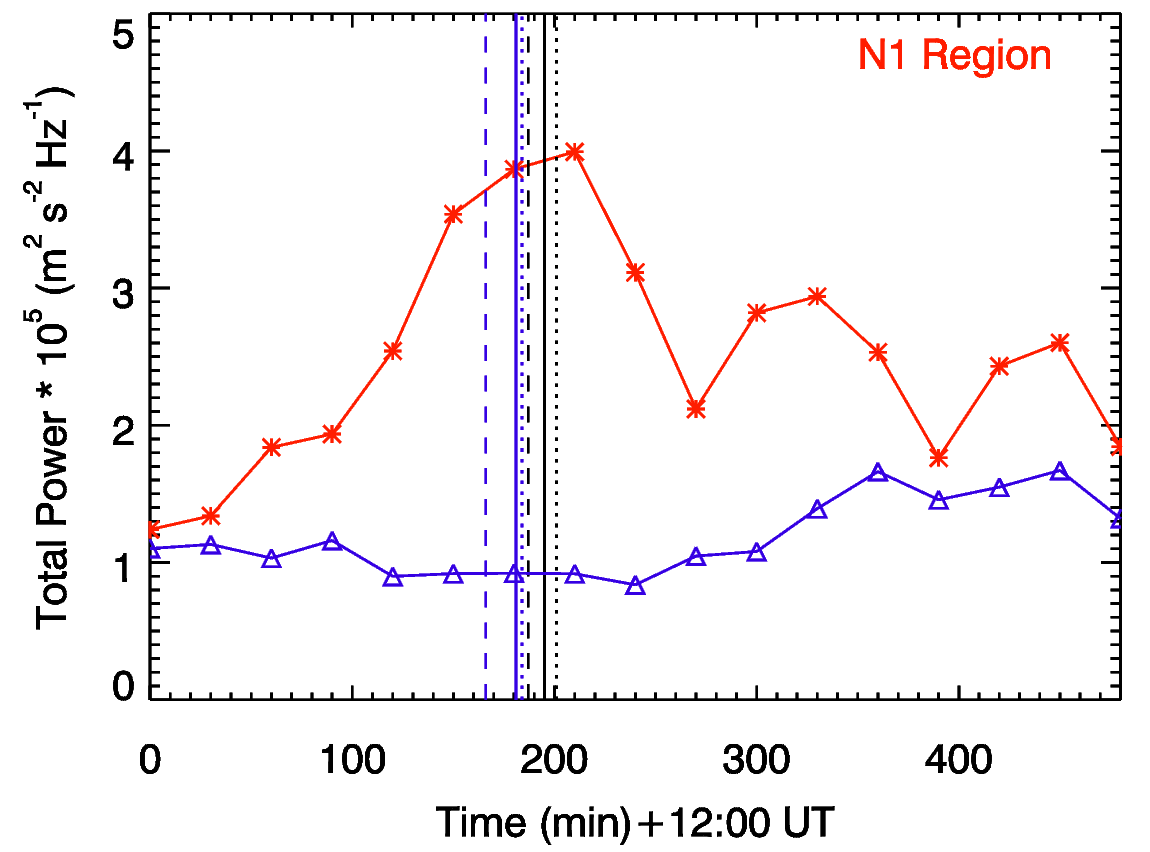}\hspace*{0.34cm}
\includegraphics[width=0.45\textwidth]{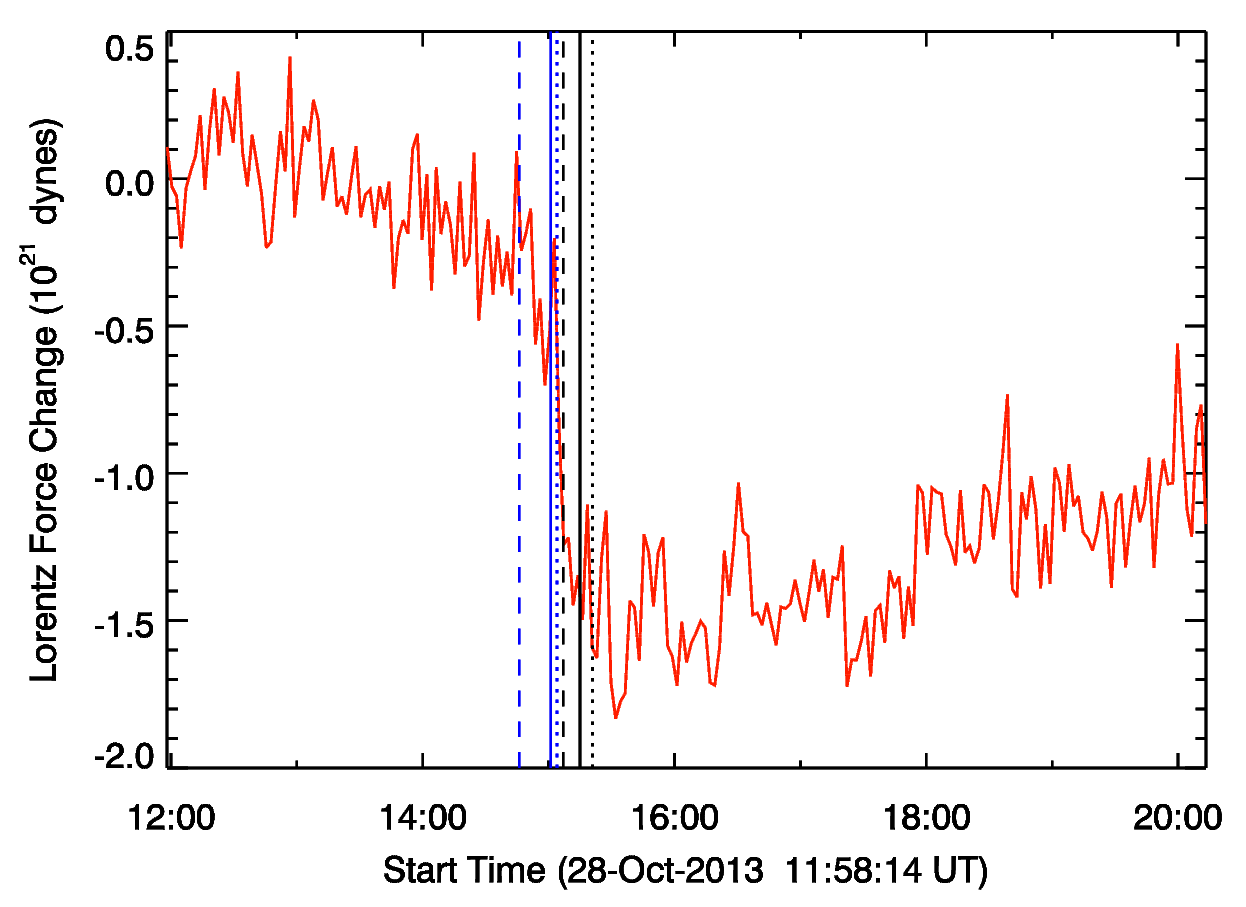}
\caption{Same as Figure \ref{AR11261}, but for `N1' location of active region NOAA 11882. It is to be noted that there is a time offset of about $\pm$ 30-minutes between the acoustic power variation and the {\em GOES} flare-time in the bottom left panel. The bottom right panel illustrates the change in horizontal (i.e., transverse) component of Lorentz force in the aforementioned location. The dashed, solid and dotted blue and black vertical lines represent the onset, peak and decay time of M2.7 and M4.4 class flares, respectively. The remaining maps and plots have been provided in the online supplementary material.}
\label{AR11882}
\end{figure*}

\begin{figure*}
\centering
\includegraphics[width=0.54\textwidth]{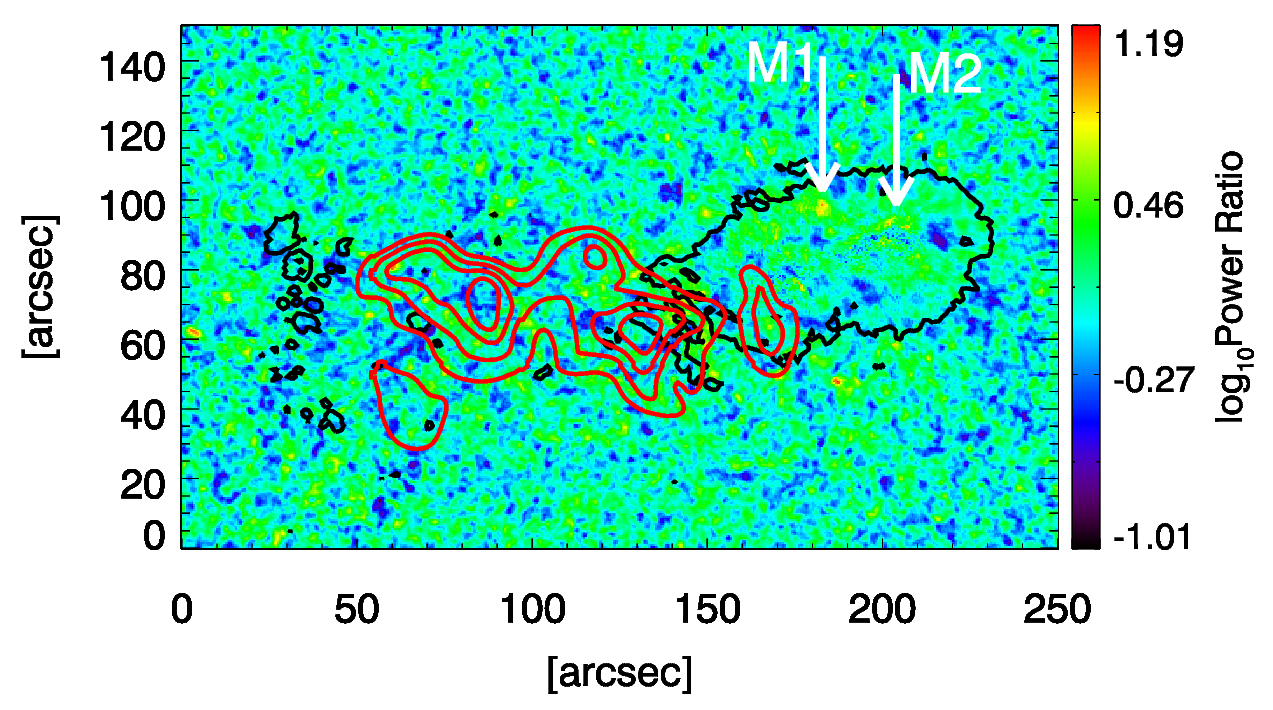}\hspace*{0.34cm}
\includegraphics[width=0.32\textwidth]{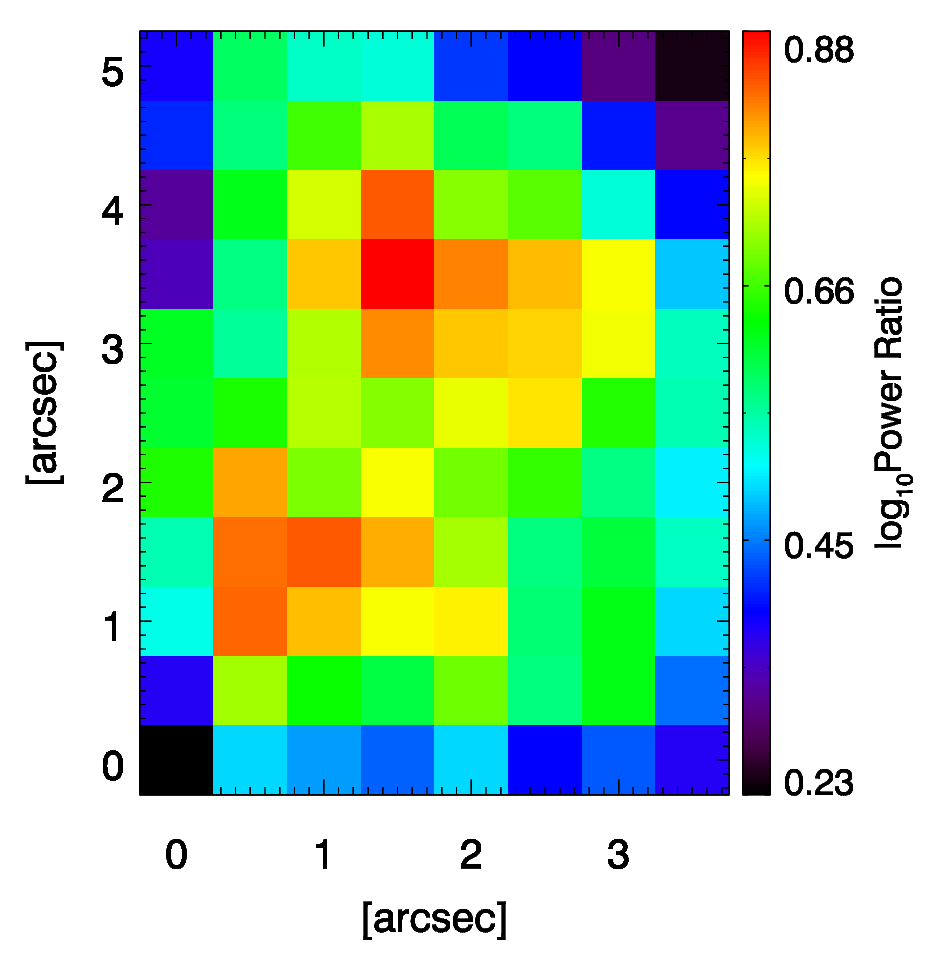}\\
\vspace*{0.5cm}
\includegraphics[width=0.45\textwidth]{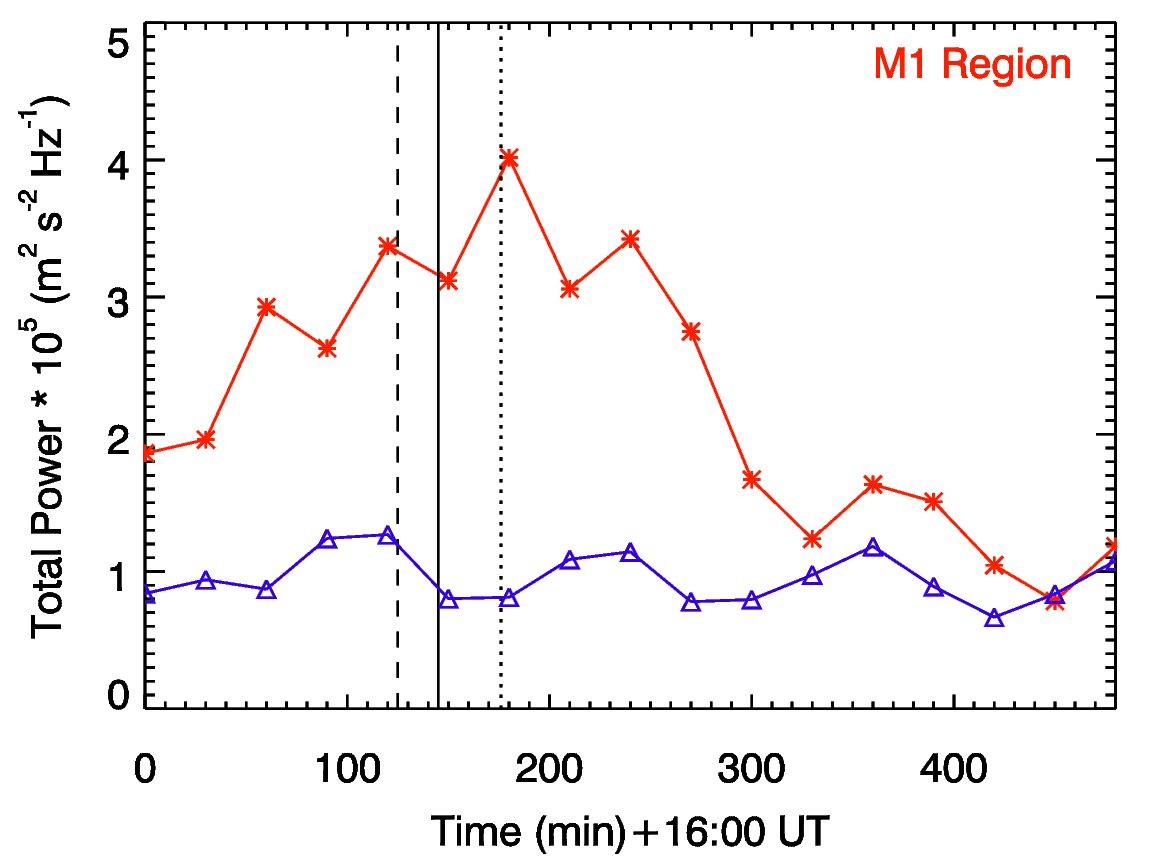}\hspace*{0.34cm}
\includegraphics[width=0.45\textwidth]{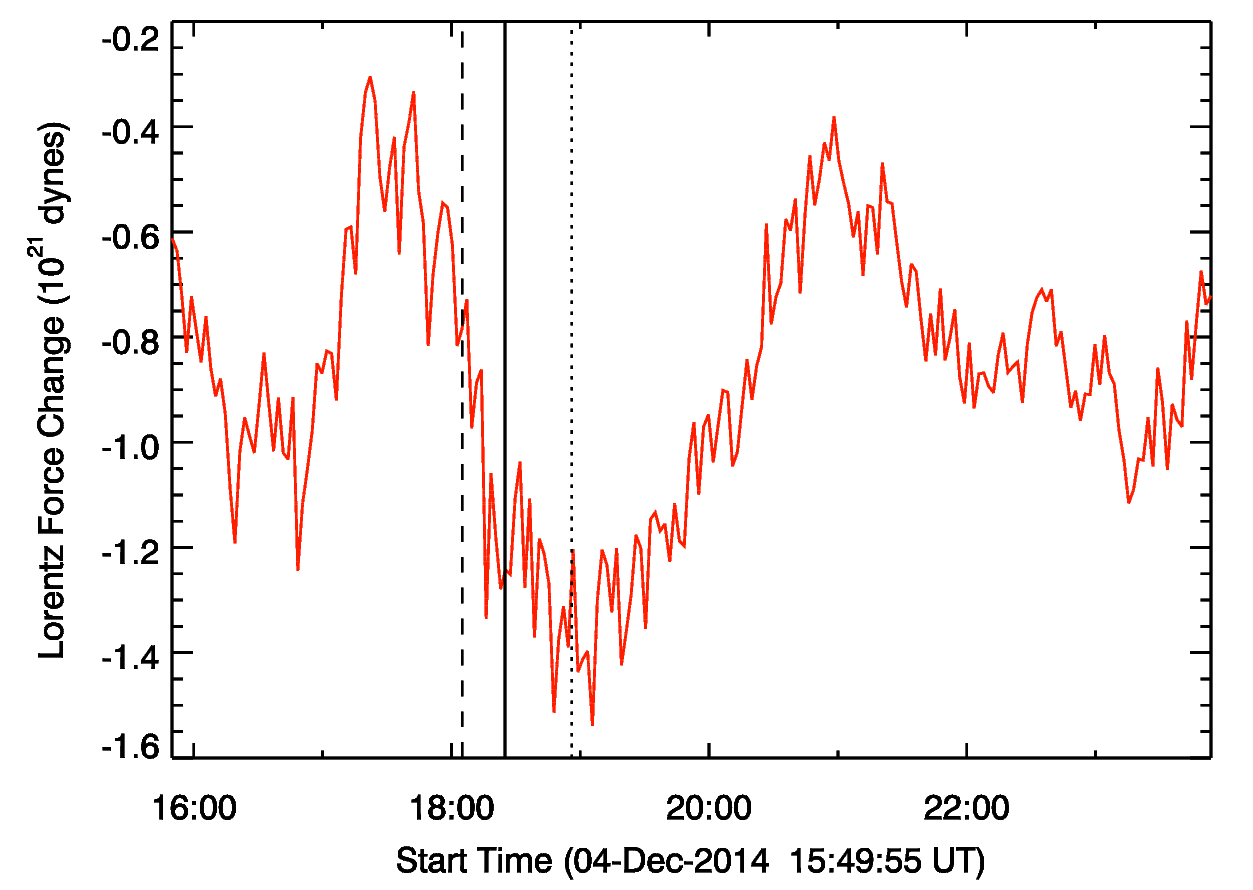}\hspace*{0.34cm}
\caption{Same as Figure \ref{AR11261}, but for `M1' location of active region NOAA 12222. It is to be noted that there is a time offset of about $\pm$ 30-minutes between the acoustic power variation and the {\em GOES} flare-time in the bottom left panel. The bottom right panel illustrates the change in radial (i.e., upward) component of Lorentz force in the aforementioned location. The remaining maps and plots have been provided in the online supplementary material.}
\label{AR12222}
\end{figure*}

\begin{figure*}
\centering
\includegraphics[width=0.52\textwidth]{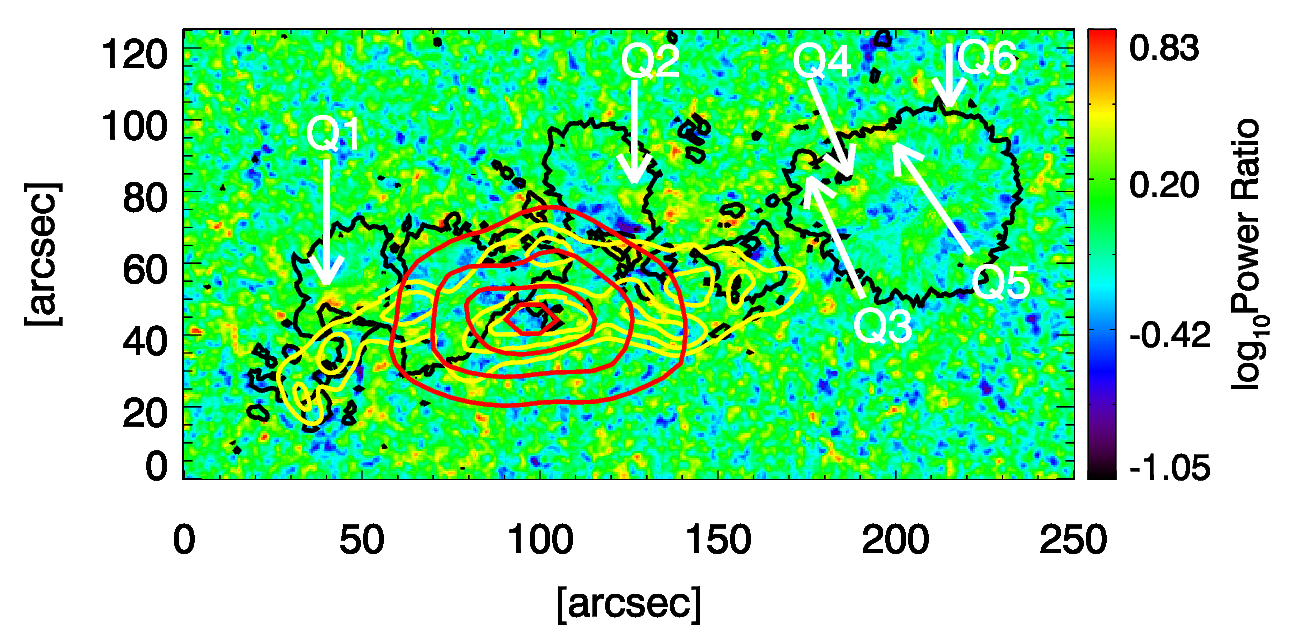}\hspace*{0.34cm}
\includegraphics[width=0.45\textwidth]{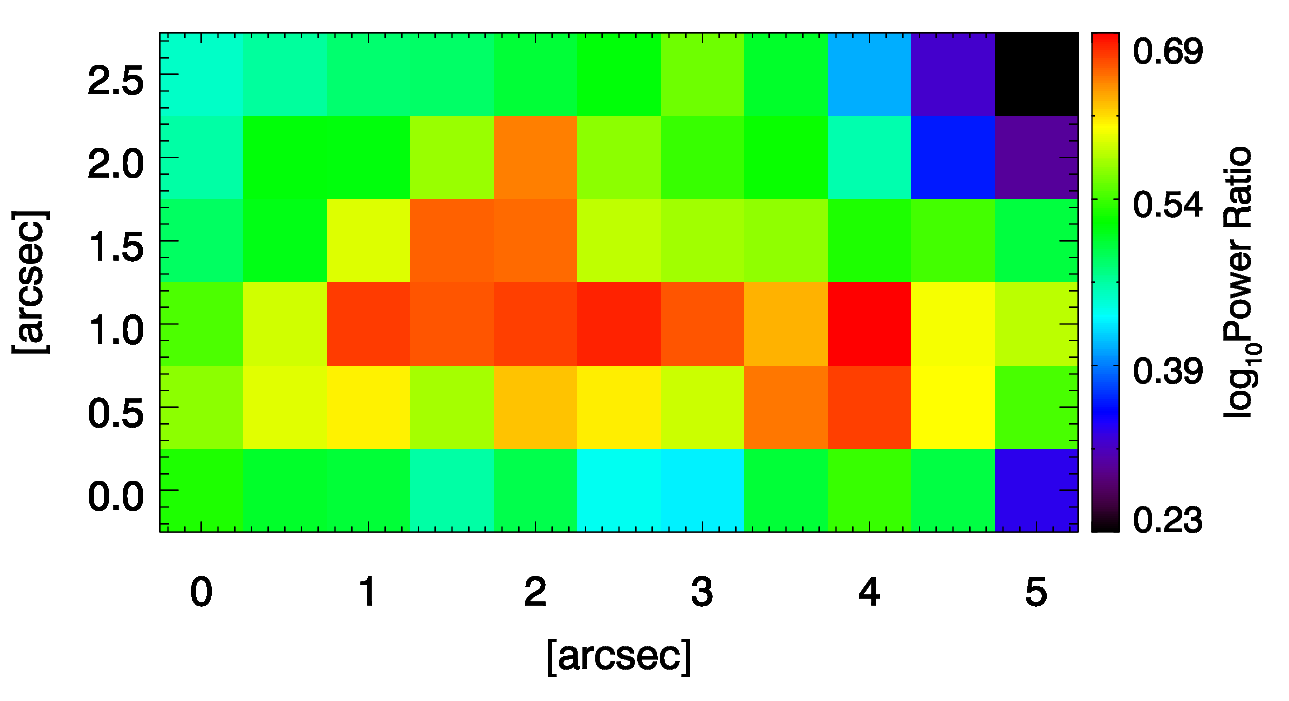}\\
\vspace*{0.5cm}
\includegraphics[width=0.47\textwidth]{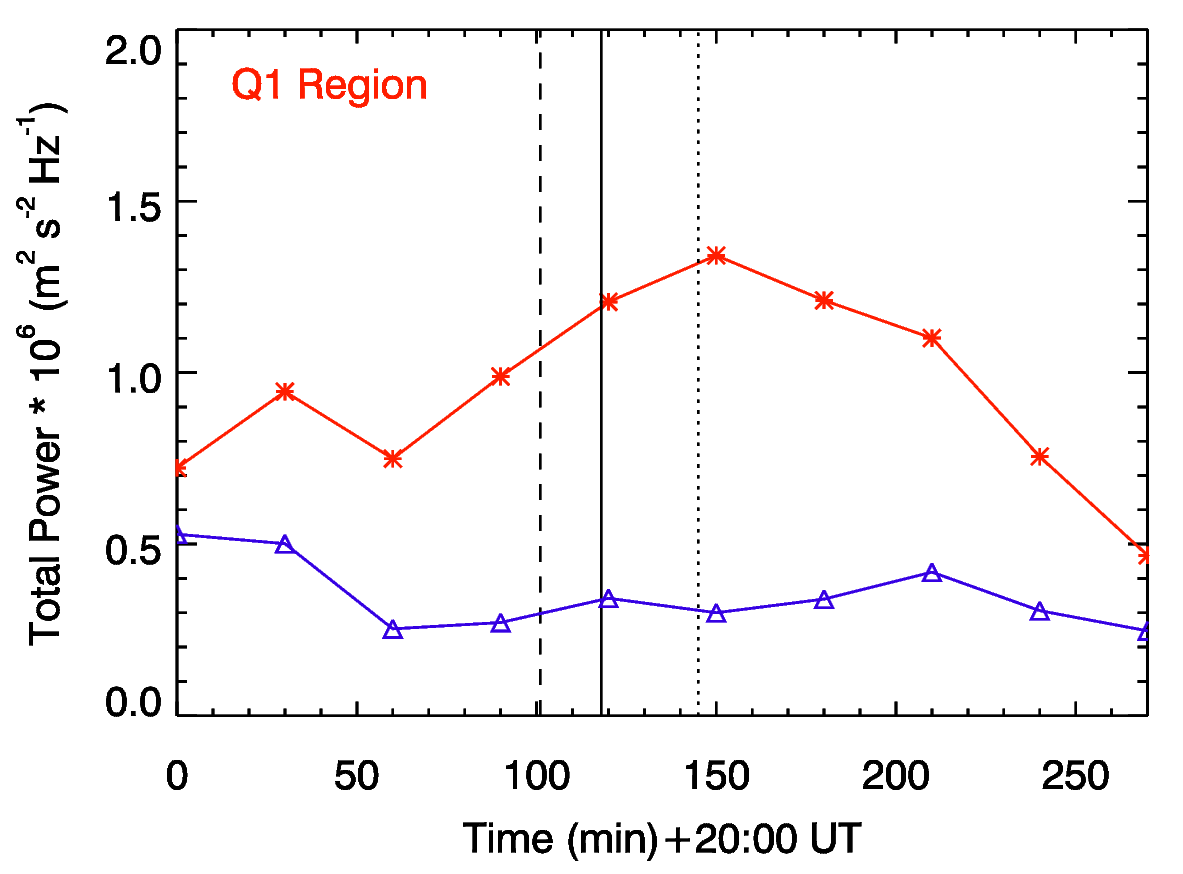}\hspace*{0.34cm}
\includegraphics[width=0.47\textwidth]{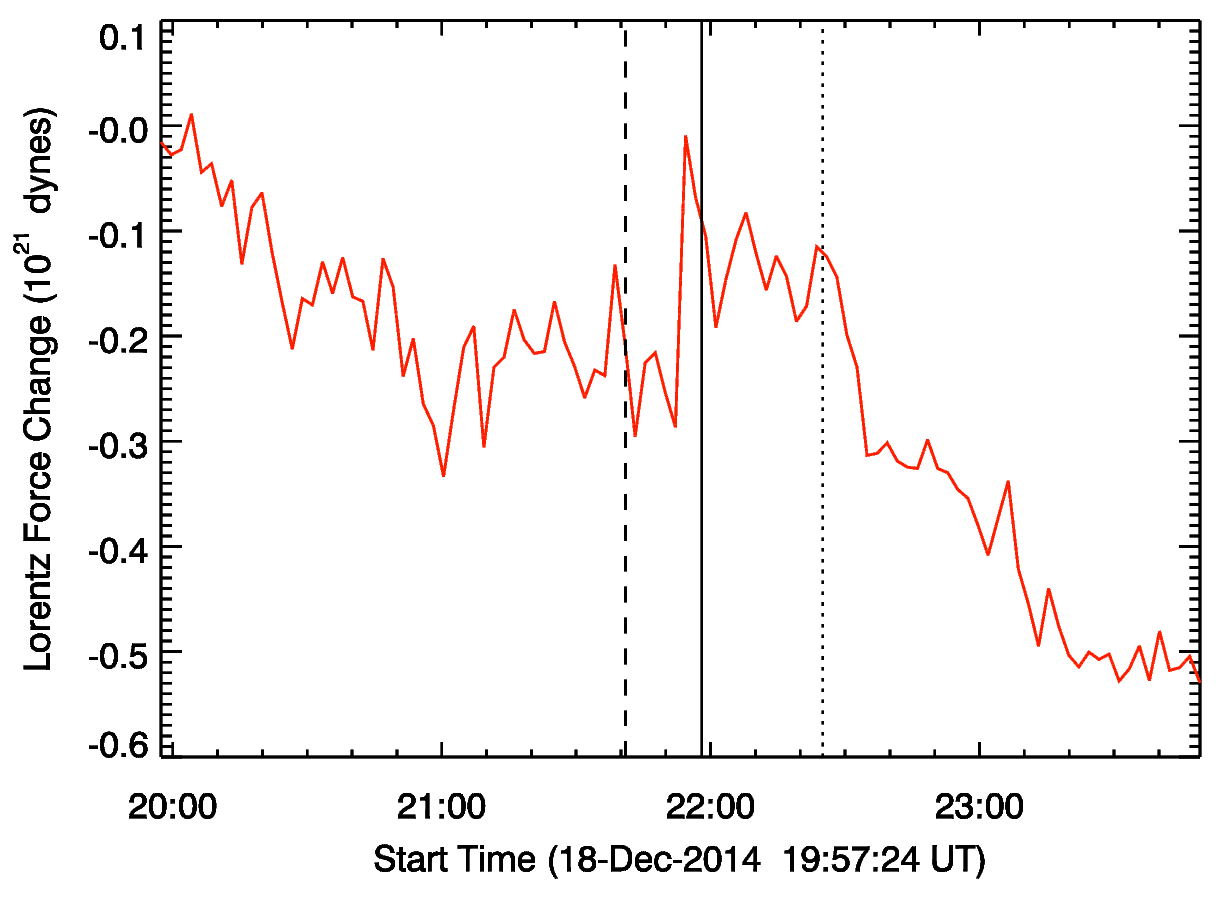}\hspace*{0.34cm}
\caption{Same as Figure \ref{AR11261}, but for `Q1' location of active region NOAA 12241. It is to be noted that there is a time offset of about $\pm$ 30-minutes between the acoustic power variation and the {\em GOES} flare-time in the bottom left panel. The bottom right panel illustrates the change in horizontal (i.e., transverse) component of Lorentz force in the aforementioned location. The remaining maps and plots have been provided in the online supplementary material.}
\label{AR12241}
\end{figure*}

\begin{figure*}
\centering
\includegraphics[width=0.5\textwidth]{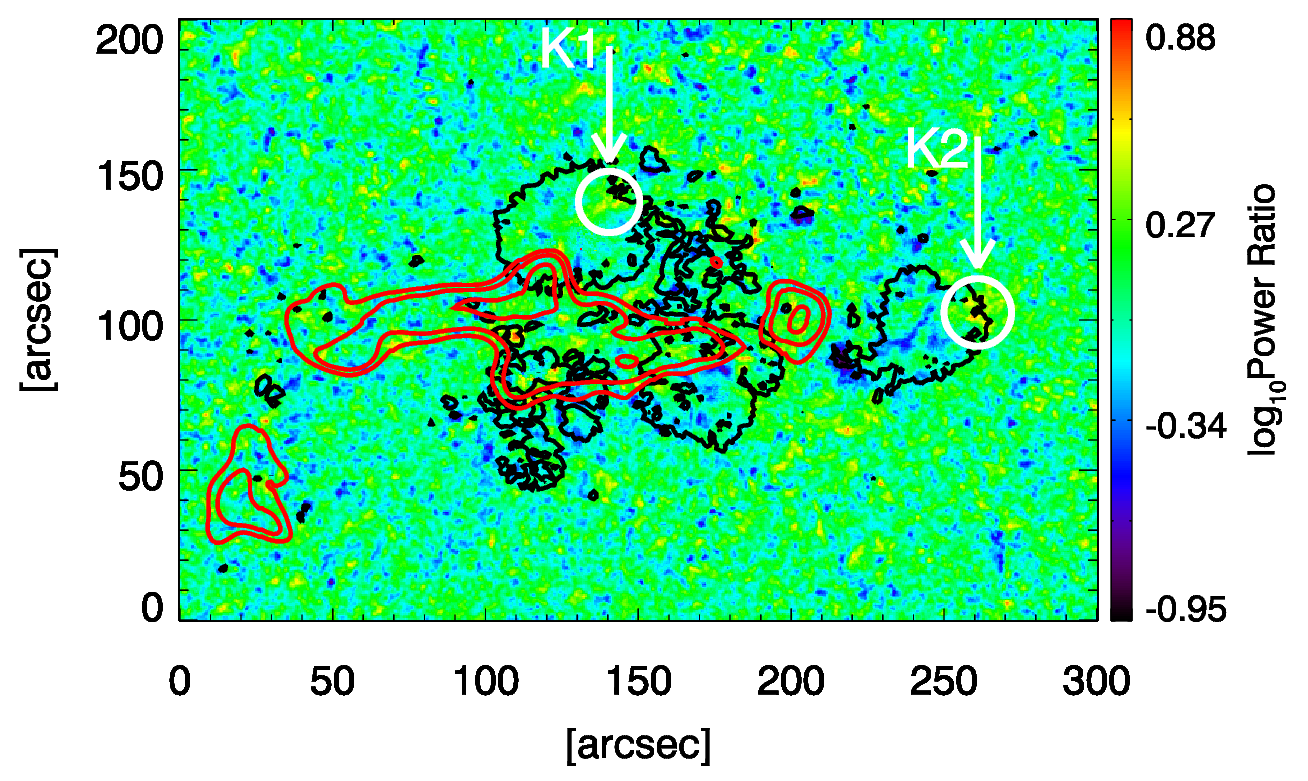}\hspace*{0.34cm}
\includegraphics[width=0.47\textwidth]{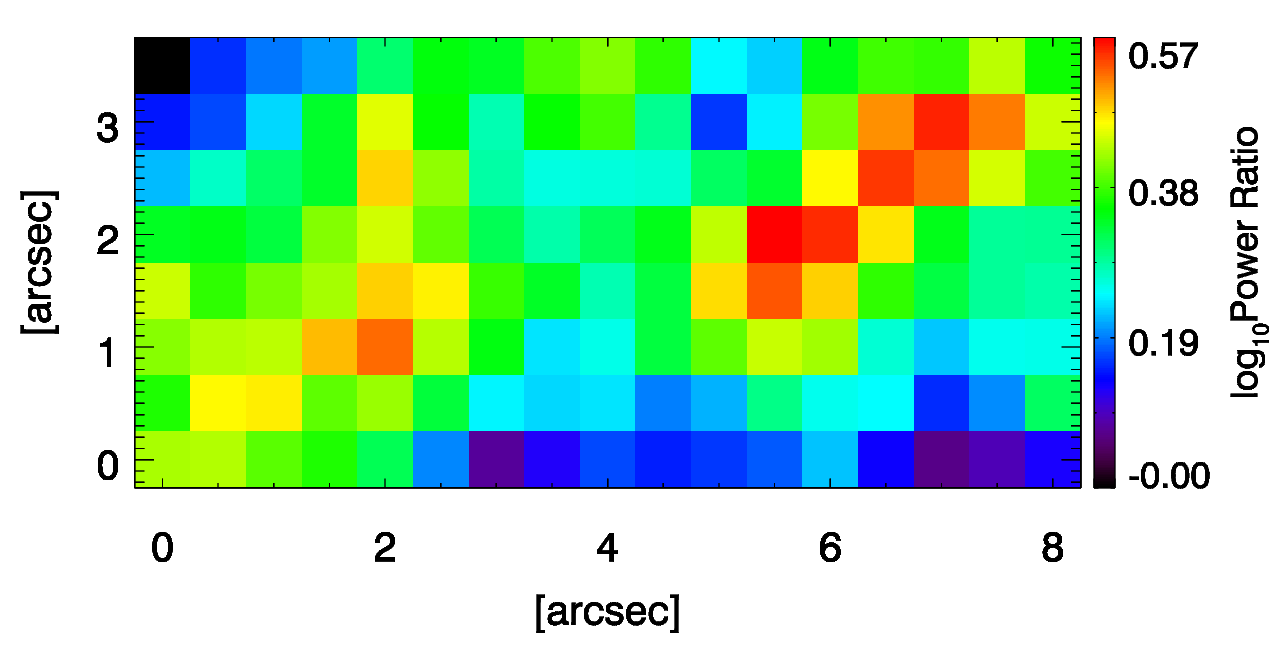}\\
\vspace*{0.5cm}
\includegraphics[width=0.45\textwidth]{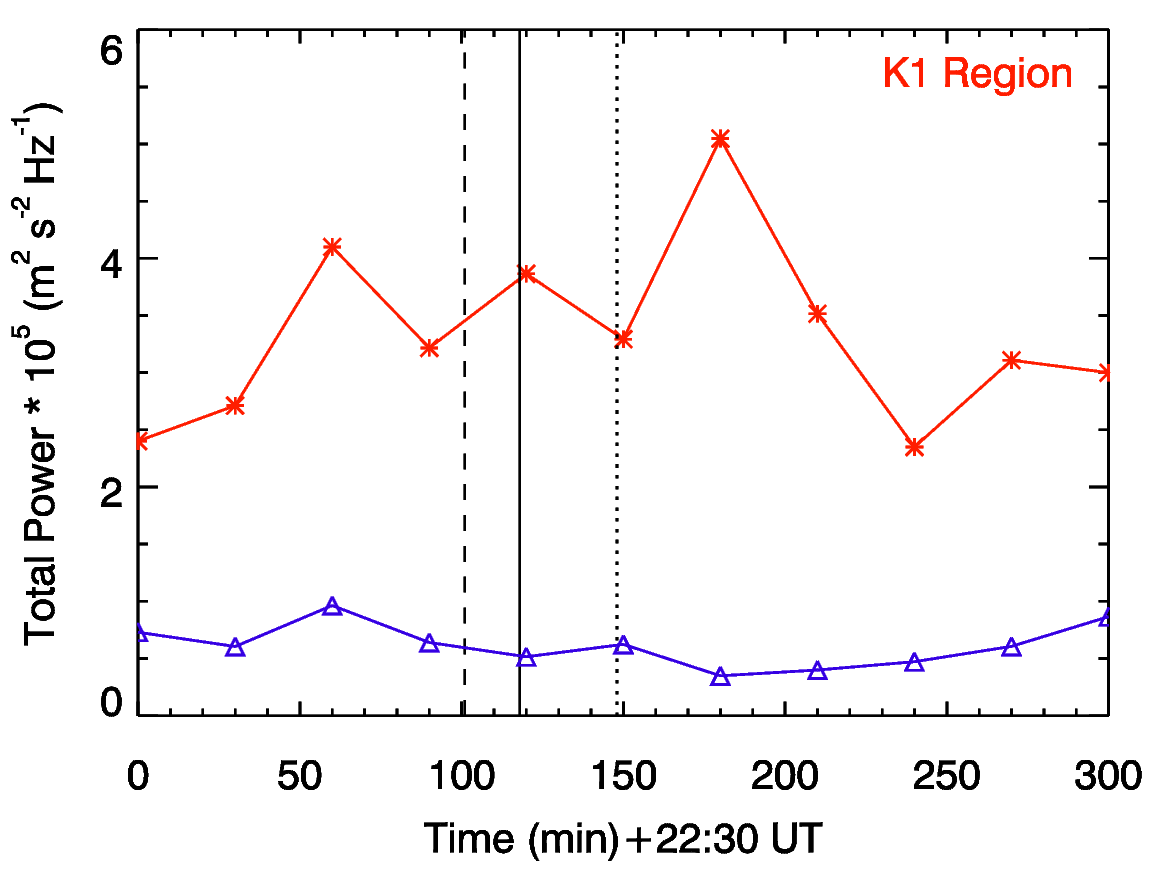}\hspace*{0.34cm}
\includegraphics[width=0.45\textwidth]{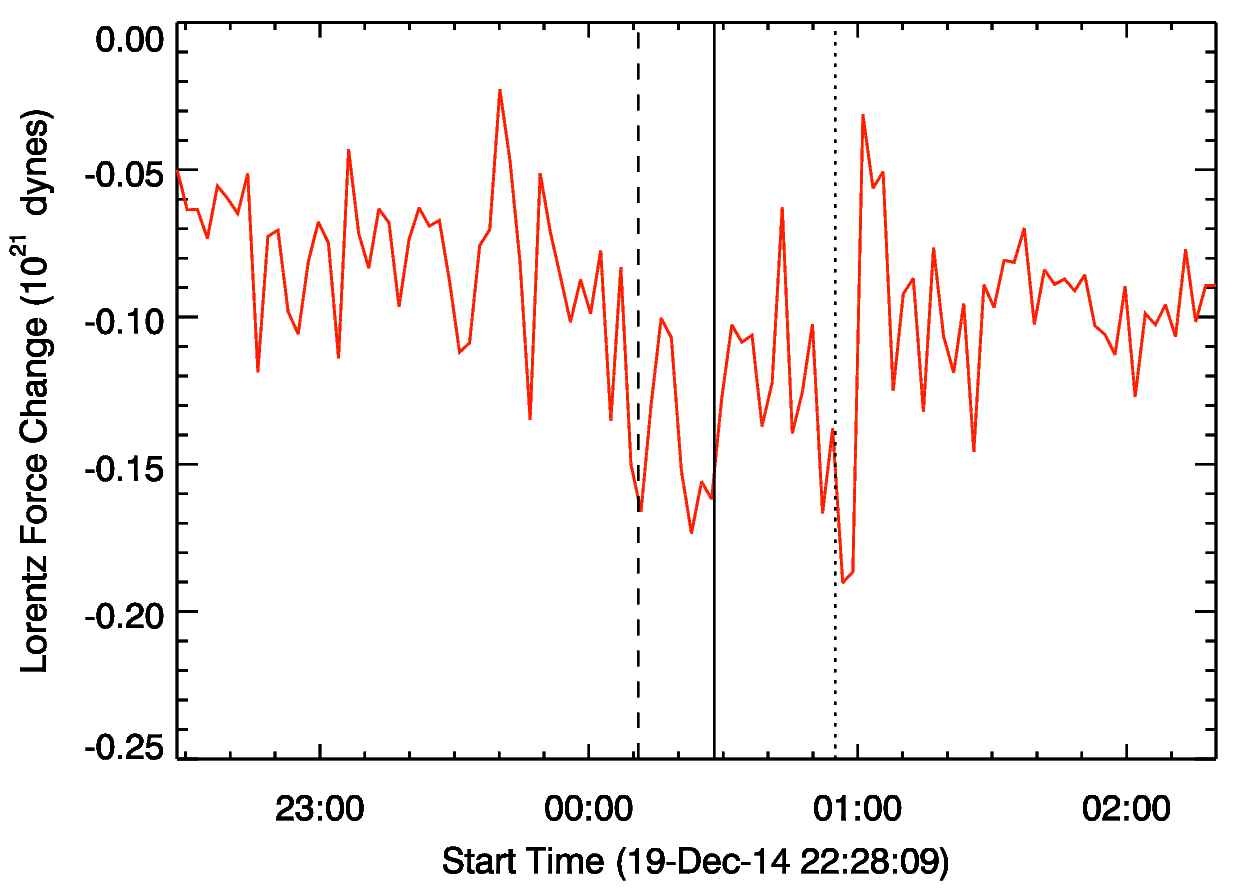}\hspace*{0.34cm}
\caption{Same as Figure \ref{AR11261}, but for `K1' location of active region NOAA 12242. It is to be noted that there is a time offset of about $\pm$ 30-minutes between the acoustic power variation and the {\em GOES} flare-time in the bottom left panel. The bottom right panel illustrates the change in horizontal (i.e., transverse) component of Lorentz force in the aforementioned location. The remaining maps and plots have been provided in the online supplementary material.}
\label{AR12242}
\end{figure*}

\begin{figure*}
\centering
\includegraphics[width=0.55\textwidth]{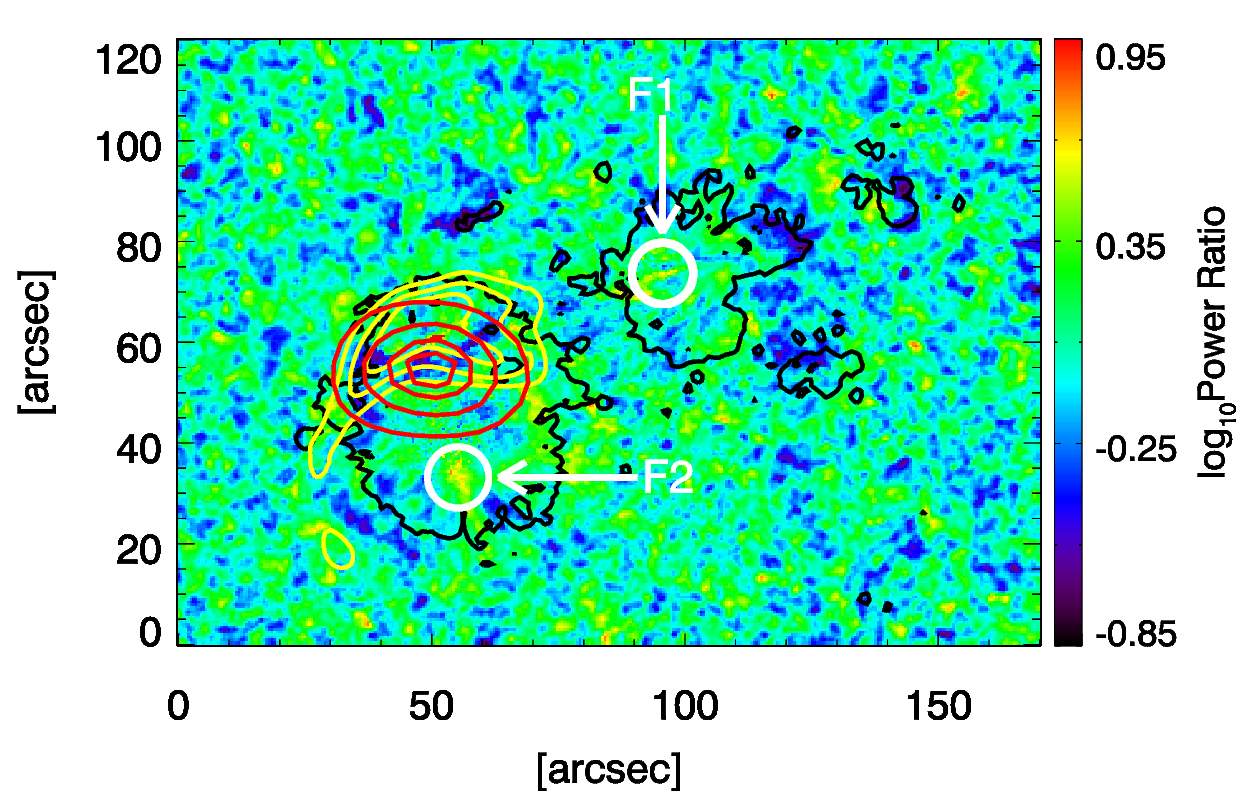}\hspace*{0.34cm}
\includegraphics[width=0.3\textwidth]{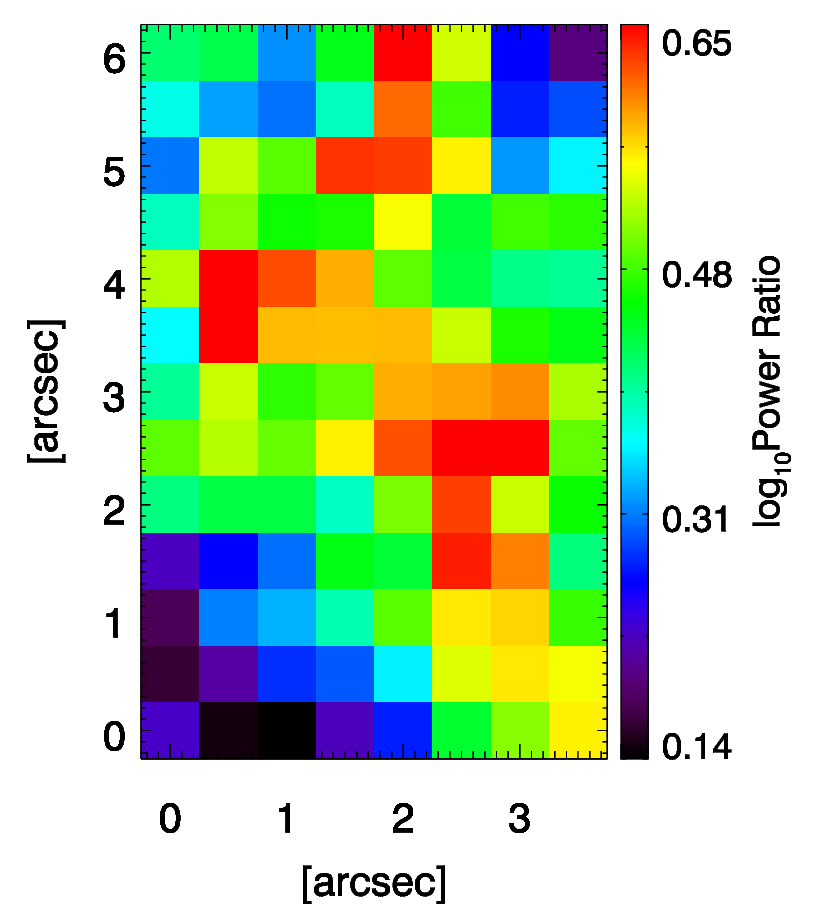}\\
\vspace*{0.3cm}
\includegraphics[width=0.45\textwidth]{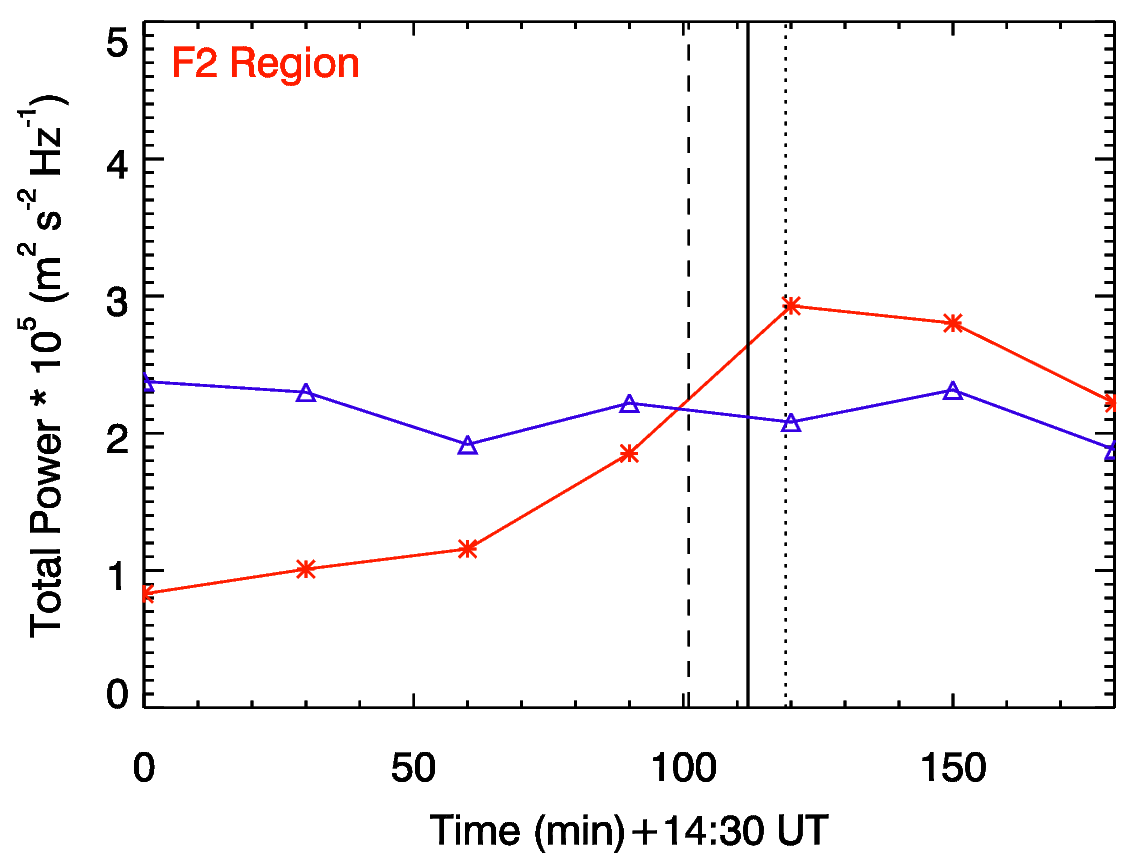}\hspace*{0.34cm}
\includegraphics[width=0.45\textwidth]{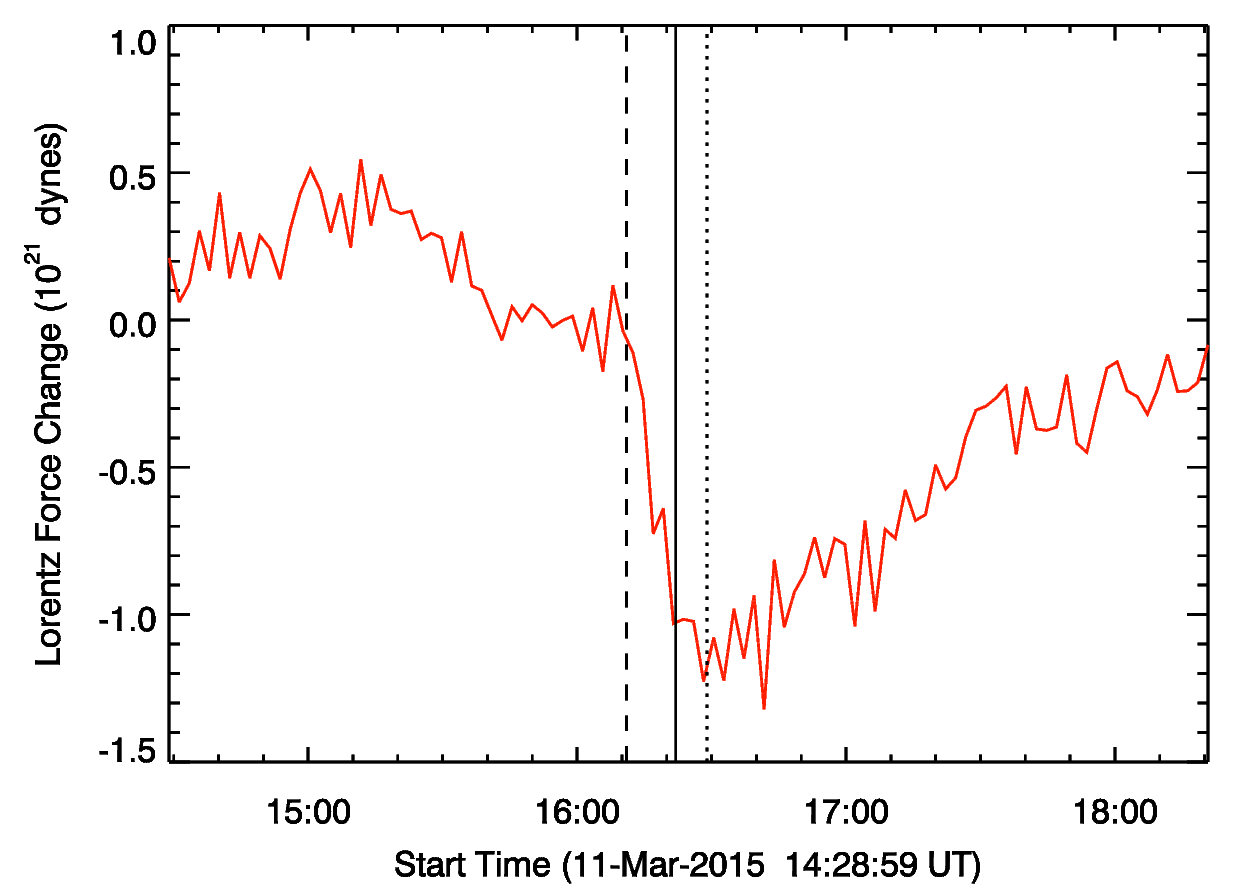}
\caption{Same as Figure \ref{AR11261}, but for `F2' location of active region NOAA 12297. It is to be noted that there is a time offset of about $\pm$ 30-minutes between the acoustic power variation and the {\em GOES} flare-time in the bottom left panel. The bottom right panel illustrates the change in horizontal (i.e., transverse) component of Lorentz force in the aforementioned location. The remaining maps and plots have been provided in the online supplementary material.}
\label{AR12297}
\end{figure*}

\begin{figure*}
\centering
\includegraphics[width=0.53\textwidth]{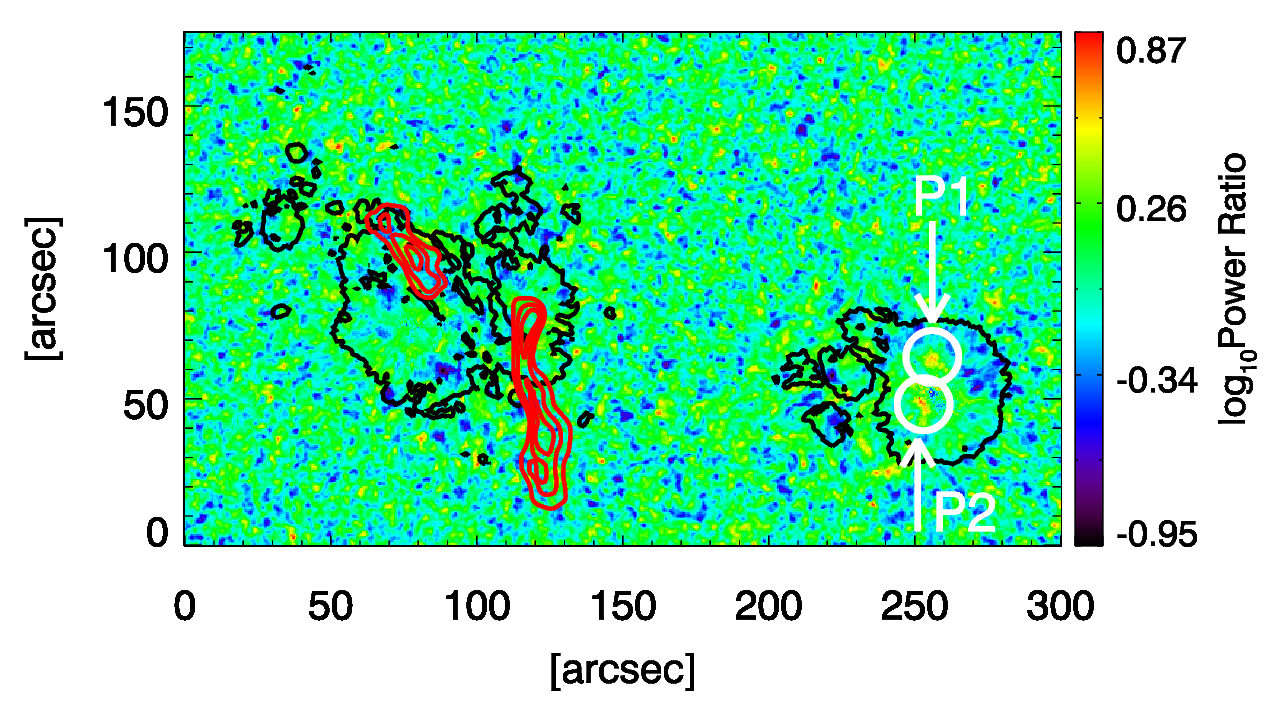}\hspace*{0.34cm}
\includegraphics[width=0.38\textwidth]{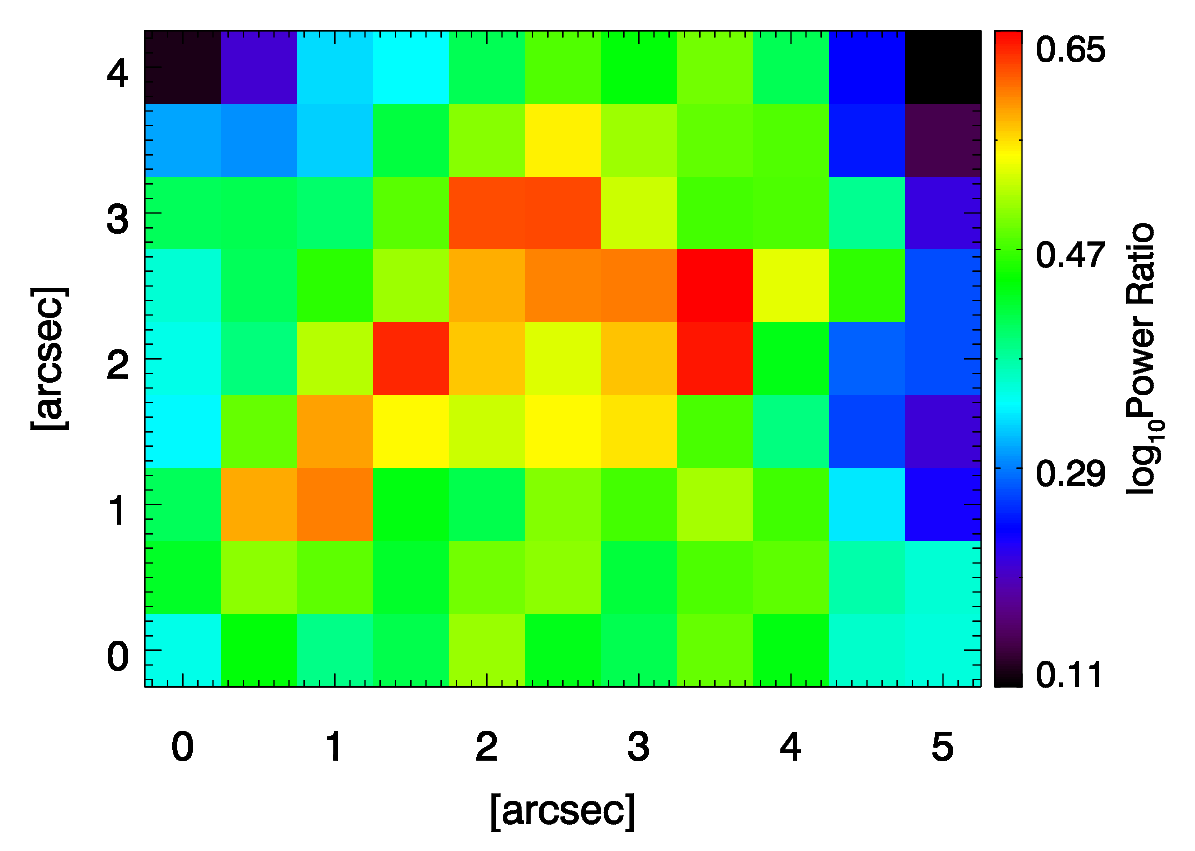}\\
\vspace*{0.3cm}
\includegraphics[width=0.45\textwidth]{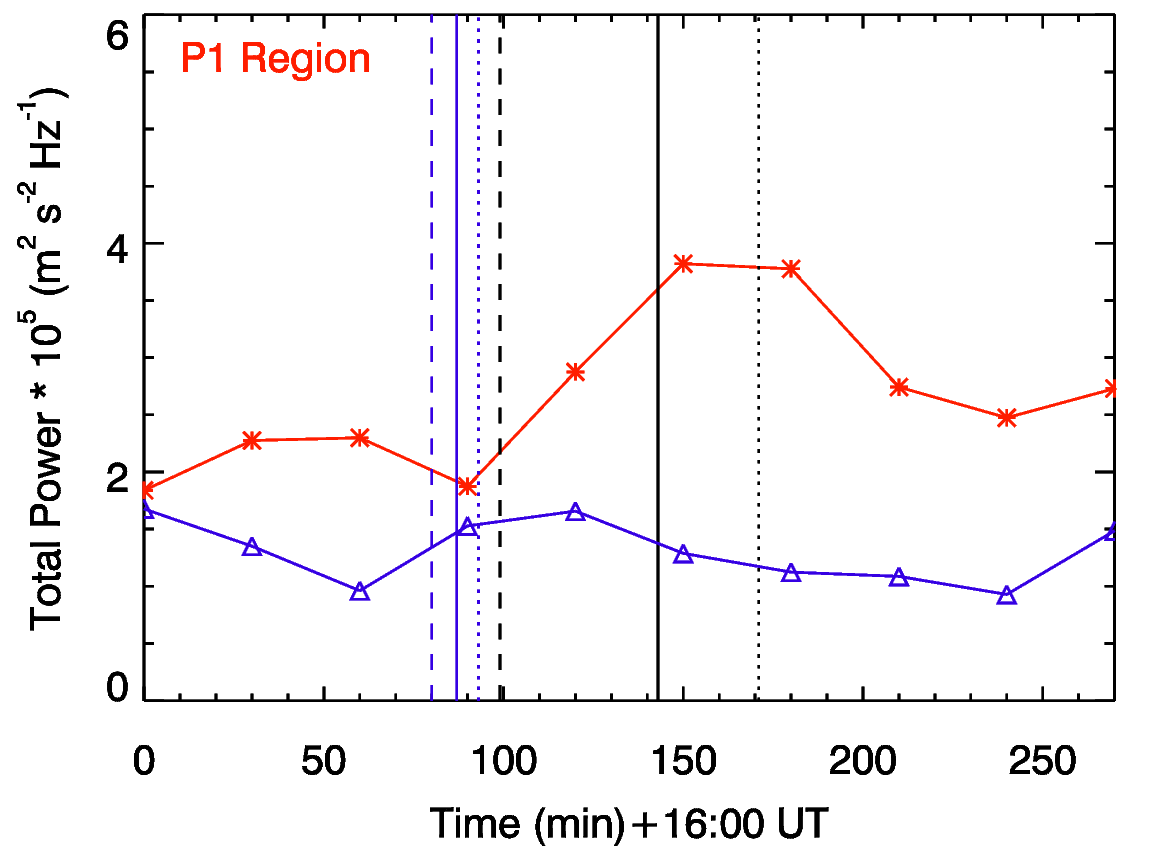}\hspace*{0.34cm}
\includegraphics[width=0.45\textwidth]{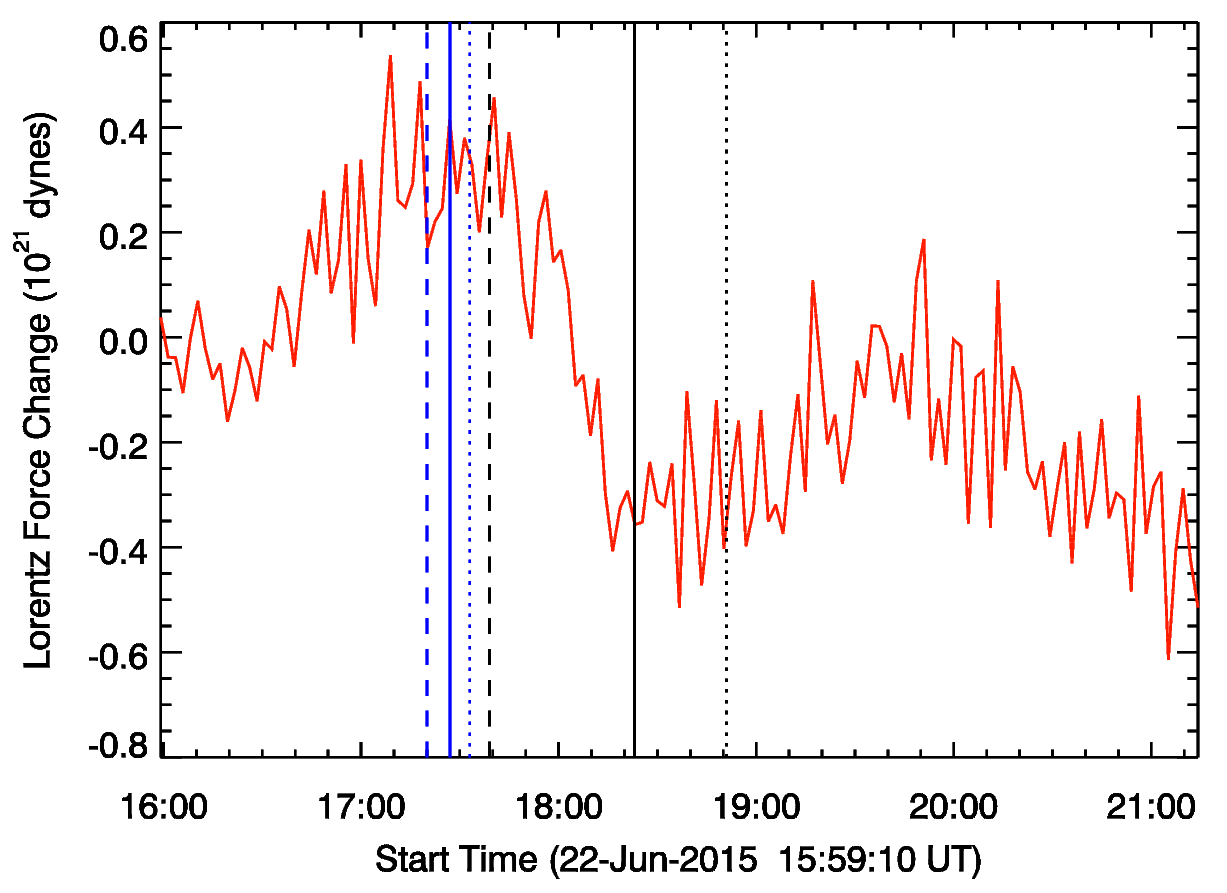}
\caption{Same as Figure \ref{AR11261}, but for `P1' location of active region NOAA 12371. It is to be noted that there is a time offset of about $\pm$ 30-minutes between the acoustic power variation and the {\em GOES} flare-time in the bottom left panel. The bottom right panel illustrates the change in radial (i.e., upward) component of Lorentz force in the aforementioned location. The dashed, solid and dotted blue and black vertical lines denote onset, peak and decay time of C3.9 and M6.5 class flares, respectively. The remaining maps and plots have been provided in the online supplementary material.}
\label{AR12371}
\end{figure*}

\section{The Observational data}

We have extensively used observations from the HMI instrument aboard the {\em SDO} spacecraft, which is a solar space mission of the National Aeronautics and Space Administration (NASA). The HMI instrument observes the photosphere of the Sun in Fe I 6173 {\AA} absorption spectral line at six different wavelengths positions in $\pm$ 172 m{\AA} wavelength window in two different cameras. One of the cameras is mainly dedicated to the 45 s cadence observations, which provides full-disc Dopplergrams, line-of-sight (i.e., longitudinal and scalar) magnetograms and continuum intensity images at spatial sampling of 0.5$\arcsec$ per pixel. The second camera provides vector magnetic field observations of the solar photosphere with the same spatial sampling. The Stokes parameters (I, Q, U, V) are observed at six different wavelength positions from the centre of Fe I 6173 {\AA} absorption line, which requires 135 s to complete the line profiles \citep{2017ApJ...839...67S}. Then, to extract the vector magnetic field components from these  Stokes parameters, Very Fast Inversion of the Stokes algorithm code (\citet{2011SoPh..273..267B}) is used to derive vector magnetic field components at the photosphere. The remaining $180^{\circ}$ disambiguity in the azimuthal field component is resolved using the minimum-energy method (\citet{1994SoPh..155..235M}; \citet{2009SoPh..260...83L}). A coordinate transformation for remapping the vector fields onto the Lambert cylindrical equal-area projection is carried out, and finally, the vector fields are transformed into heliocentric spherical coordinates (B$_r$, B$_\theta$, B$_\phi$). These co-ordinates are approximated to (B$_z$, -B$_y$, B$_x$) in heliographic cartesian coordinates for ready use in various parameter studies \citep{1990SoPh..126...21G}. Thus, we have used the tracked photospheric continuum intensity (hmi.Ic$\_45$s), Dopplergrams (hmi.V$\_45$s), and line-of-sight magnetograms (hmi.M$\_45$s) at a cadence of 45 s and high cadence  vector magnetograms (hmi.B$\_135$s) at a cadence of 135 s, acquired from the HMI instrument aboard the {\em SDO} spacecraft for our motivated analysis. The left panel of Figure \ref{fourimage11158GOES} shows the sample images of the National Oceanic and Atmospheric Administration (NOAA) solar active region 11158 from the HMI instrument, which produced an X2.2 class solar flare on 2011 February 15. In the continuum intensity image, we can see the complex structure of this active region while the morphology of the magnetic fields is seen in the line-of-sight magnetogram. The Dopplergram shows the Doppler velocity flows in the active region while the running difference image of Dopplergrams shows the suppression of velocity oscillations in the sunspots in contrast with the dominant velocity oscillations in the quiet regions. For flare-ribbon information at the chromospheric layer of the Sun, we have used H$\alpha$ chromospheric intensity observations obtained from GONG instruments at 60 s cadence having a spatial sampling of 1.05$\arcsec$ per pixel (in full-disc mode). In order to identify hard X-ray foot-point locations during the flare evolution, we have used hard X-rays images from the {\em Reuven Ramaty High Energy Solar Spectroscopic Imager} ({\em RHESSI}; \citet{ 2002SoPh..210...61H}) spacecraft in 12--25 KeV energy band. We have also used disk-integrated soft X-ray solar flux in 1--8 {\AA} range from the {\em Geostationary Orbital Environmental Satellite} ({\em GOES}; \cite{1994SoPh..154..275G}) for having information on the temporal evolution of the flare and flare-class. The right panel of Figure \ref{fourimage11158GOES} demonstrates the temporal evolution of soft X-rays (1--8 \AA) from GOES satellite, which shows enhancement in flux around 01:44 UT and further provides information on the peak and decay time of an X2.2 class flare on 2011 February 15.\\

We have adopted selection criteria to choose NOAA active regions for our analysis, which are as follows:\\

(1) We have selected only those active regions which are within $40^{\circ}$ latitude and longitude of the solar disc since \textit{p}-mode oscillations are dominant near the disk centre and as we go away from the disc centre, these signals suffer from projection effects. Therefore, we restricted ourselves to within the $\pm$ $40^{\circ}$ of the disc centre. \\

(2) We have analyzed only those active regions which have produced M- and X-class flares.\\

(3) We have restricted our analysis to only those active regions in which magnetic field strength does not exceed 3000 Gauss and the Doppler velocity is less then $\pm$ 6 km $s^{-1}$, including the orbital velocity of the spacecraft. It is to be noted that in the active regions where the magnetic field strength is more than $\pm$ 3800 Gauss or the total Doppler velocity exceeds $\pm$ 8.5 km $s^{-1}$, the Fe I 6173 {\AA} spectral line can go outside the observing window in HMI instrument thereby causing saturation problem in the observational data \citep{2012SoPh..278..217C}. \\

On applying the aforementioned selections criteria, we have selected eight active regions for our analysis, the details of which are as shown in Table \ref{AR list}.

\section{Analysis and Results}
The main objective of this work is to identify the seismic emissions in the sunspots, accompanying major solar flares and to further investigate the changes in magnetic fields in those affected locations. We also aim to estimate the corresponding changes in Lorentz force in the aforementioned locations.

In the following Sections, we describe the analysis procedure of the Dopplergrams, line-of-sight magnetograms and the vector magnetograms obtained from HMI instrument aboard the {\em SDO} spacecraft.

\subsection{Analysis of Dopplergram data}

We have analyzed the Dopplergrams to identify any seismic emissions in the sunspots. For this purpose, we have analyzed several active regions as listed in Table \ref{AR list}. Here we present in detail the analysis procedure and results of the active region NOAA 11158, however a brief overview of important results concerning other active regions are also presented. The remaining results are being provided in the online supplementary material. \\ 

The active region NOAA 11158 produced several flares during its passage on the solar disc. This was the first active region of the Solar Cycle 24, which produced an X2.2 class flare on 2011 February 15 around 01:44 UT (c.f. right panel of Figure \ref{fourimage11158GOES}). We have used the tracked  Dopplergrams (hmi.V$\_45$s) at a cadence of 45 s (c.f. left panel of  Figure \ref{fourimage11158GOES}) obtained from HMI instrument aboard the {\em SDO} spacecraft. In order to study the \textit{p}-mode oscillations, we have applied a two-point backward difference filter which removes the slowly varying features from the Dopplergrams. Those filtered Dopplergrams are then subjected to Fast Fourier Transform (FFT) for estimating power spectrum at each pixel. Following this, we construct power maps in the frequency range 2.5--4 mHz band (\textit{p}-mode oscillations) for pre-flare and spanning flare epochs (c.f. Figure \ref{fig: Prespanningflare}). Further, we have overplotted the hard X-ray contours from {\em RHESSI} instrument in 12--25 KeV energy band on the spanning flare power map (c.f. right panel of Figure \ref{fig: Prespanningflare}) in order to know the hard X-ray foot-point locations. These acoustic power maps (c.f. Figure \ref{fig: Prespanningflare}), demonstrate that there is a suppression of \textit{p}-mode power in the sunspot region, while it is dominant in the quiet region. In the right panel of Figure \ref{fig: Prespanningflare} (spanning flare power map), red contours represent hard X-rays from {\em RHESSI} instrument in 12--25 KeV band at 25, 50, 75, and 90$\%$ of its maximum level. It is believed that corresponding to the hard X-ray foot-points, high energetic charged particles can reach up to the photosphere and they can induce photospheric \textit{p}-mode oscillations. However, in the flaring region/hard X-ray foot-point locations, it has been demonstrated that there could be distortions in the line profiles during the flares due to the change in the local thermodynamic conditions \citep{2014RAA....14..207R}. Therefore, we have looked for only those acoustically enhanced locations in the sunspots which are away from HXR footpoints and flare ribbons.\\

Although, by visualization of these two power maps (c.f. Figure \ref{fig: Prespanningflare}), it is difficult to infer any adequate changes in the acoustic power in the sunspot regions during the flare. Hence, we have taken the ratio of these two power maps (spanning flare to pre-flare) as shown in the left panel of Figure \ref{fig: ratiopowermapB1}. Again, we have over-plotted hard X-rays foot-point contours (red colour) from {\em RHESSI} instruments in 12--25 KeV band on this power map ratio. In addition, we have also overplotted contours of the outer boundary of the sunspot penumbra (black colour) from HMI continuum intensity image on the power map ratio. Thus, we selected the acoustically enhanced `B1' location in the sunspot, which is away from the flaring regions. The blow-up region of the `B1' location shows enhancement in power ratio in patches (right panel of Figure \ref{fig: ratiopowermapB1}). The temporal evolution of Doppler velocity in `B1' location shows enhancement in \textit{p}-mode velocity oscillations after the solar flare (c.f. left panel of Figure \ref{fig: DVpowerB1}). We have also examined the temporal evolution of total power integrated over `B1' region as shown in red colour with asterisks in the left panel of Figure \ref{fig: DVpowerB1}. This is done by estimating the power spectrum for 1-hour duration and then shifting it for every 30-minutes during the observation period. Thus, it will have a time offset of about $\pm$ 30-minutes with respect to onset, peak and decay time of the flare. The temporal evolution of total power over `B1' location demonstrates significant enhancement spanning the flare. On the other hand, the plot shown in blue colour with triangles representing the evolution of total power for an unaffected region in the same sunspot shows normal evolution over the whole duration. We have also calculated the percentage change in total power of the regions showing acoustic enhancement in sunspots in power map ratios of the other active regions considered in our analysis. All the seismic emission regions and corresponding percentage changes in total power spanning the flare are listed in Table \ref{percentagepower}, which demonstrate significant values. We further examined the cause of these seismic emissions in the sunspot regions, which are away from flaring sites. For this purpose, we have analysed the scalar and vector magnetograms acquired from HMI instrument aboard the {\em SDO} spacecraft. In Section 3.2, we describe the analysis procedure of the magnetogram data.\\

\subsection{Analysis of magnetogram data} 

We have analysed sequences of tracked line-of-sight (i.e., longitudinal and scalar) magnetograms (hmi.M$\_45$s) at a cadence of 45 s and vector magnetograms (hmi.B$\_135$s) at a cadence of 135 s obtained from HMI instrument aboard the {\em SDO} spacecraft, in order to examine the evolution of magnetic fields corresponding to the seismic emission regions in the sunspots accompanying the flares. The left panel of Figure \ref{fig: magfield} represents the temporal evolution of line-of-sight magnetic fields over the `B1' location. Here, we find that before the flare the magnetic field is evolving normally, while there is an abrupt change in the magnetic field of the order of 80--90 Gauss within the time scale of 10--20 minutes spanning the flare and again there is a normal evolution after the flare. Further, in order to have the complete information on the evolution of magnetic fields in the seismic emission locations, we also examined the vector magnetogram data. The right panel of Figure \ref{fig: magfield} shows the temporal evolution of horizontal (i.e., transverse) component of vector magnetic fields over `B1' location. Here, we observe that there is a step-like change in the horizontal magnetic fields of the order of 60--70 Gauss within 10--20 minutes during the flare. We have done similar analysis of magnetogram data for the seismic emission locations for others active regions, which are mentioned in Table \ref{AR list}. The temporal evolution of magnetic fields corresponding to seismic emission locations in all active regions shows similar changes in the magnetic fields. These results have been provided in the online supplementary material.\\

\cite{2008ASPC..383..221H} and \cite{2012SoPh..277...59F} proposed that abrupt changes in the magnetic fields can lead to an impulsive change in the Lorentz force (known as ``magnetic-jerk") of the order of 10$^{22}$ dynes, which can excite seismic emission in the sunspots accompanying flares. Hence, we have estimated the change in Lorentz force from the available photospheric   vector magnetograms at a cadence of 135 s corresponding to the identified seismic emission locations as shown in Table \ref{percentagepower}. \\

To calculate the change in Lorentz force acting on the solar photosphere, we have used the Equations (17) and (18) introduced by \cite{2012SoPh..277...59F}, which are as follows:\\

\begin{equation}
\delta F_{r},_{interior} =  \frac{1}{8 \pi} \int_{A_{ph}} dA(\delta B^{2}_{r} - \delta B^{2}_h)
\end{equation}   

\begin{equation}
\delta F_{h},_{interior} = \frac{1}{4 \pi} \int_{A_{ph}} dA \delta(B_{r}B_{h})
\end{equation}

Where, B$_{r}$ and B$_{h}$ are the radial (i.e., upward) and horizontal (i.e., transverse) components of vector magnetic fields, respectively, $\delta F_{r}$ $\&$ $\delta F_{h}$ are the changes in radial and horizontal components of Lorentz force acting on the solar interior, dA is the elementary surface area over the photosphere and $A_{ph}$ represents the area of the photospheric domain containing the active region. \\

It is to be noted that \citet{2014SoPh..289.3663P} further discussed the application of formulation introduced by \cite{2012SoPh..277...59F} within a subdomain of active region. \cite{2014SoPh..289.3663P} concluded that these expressions could be used in a subdomain of active region at photosphere if the horizontal length scale of the structure is much more than 300 km and magnetic field strength is greater than 630 G at the height of the observations. In our analysis, the horizontal length scale and magnetic field  strength are more than the aforementioned thresholds in all the seismic emission locations.

By using Equations (1) and (2), we have calculated changes in both the components of Lorentz force. Figure \ref{fig: lorentzforcechange} shows the temporal evolution of changes in radial and horizontal components of Lorentz force over `B1' location. These plots demonstrate that there is an abrupt change in radial component of Lorentz force and step-like change in horizontal component of Lorentz force of the order of 10$^{21}$ dynes within 10--20 minutes duration spanning the flare. We have found similar changes in the Lorentz force corresponding to the analysis of other active regions (c.f. Table \ref{percentagepower}), which are in close approximation to the estimates of \cite{2008ASPC..383..221H} and \cite{2012SoPh..277...59F}. The plots of change in Lorentz force for the other active regions are shown in the Figures \ref{AR11261} to \ref{AR12371} and also have been provided in the online supplementary material.

\begin{table*}
	\centering
	\caption{Details of the active regions and flare produced, identified seismic emission locations in the sunspots, observed changes in the total power and estimated changes in Lorentz force corresponding to the  seismic emission locations accompanying the flares.}

	\label{percentagepower}
	\begin{tabular}{lcccccr} 
		\hline
Active Region & Flare class &Seismic emission locations & Change in power & Change in Lorentz Force \\
           &       &         &   (percent) & (dynes)          \\
		\hline
NOAA 11158 & X2.2 & B1 & $\approx$ 113 & $\approx$ 2.0 $\times$ $10^{21}$  \\
NOAA 11261 & M6.0 & D1& $\approx$ 30 & $\approx$ 1.0 $\times$ $10^{21}$\\
 & M6.0 & D2& $\approx$ 162 & $\approx$ 0.5 $\times$ $10^{21}$\\
 
NOAA 11882 & M2.7, M4.4 & N1 & $\approx$ 56 & $\approx$ 1.5 $\times$ $10^{21}$\\
      & M2.7, M4.4 & N2 & $\approx$ 123 & $\approx$ 0.2 $\times$ $10^{21}$ \\
  
NOAA 12222 & M6.1 & M1& $\approx$ 284 & $\approx$ 1.0 $\times$ $10^{21}$\\

  & M6.1 & M2& $\approx$ 151 & $\approx$ 1.0 $\times$ $10^{21}$\\

NOAA 12241 & M6.9 & Q1 & $\approx$ 192 & $\approx$ 0.3 $\times$ $10^{21}$ \\
      & M6.9 & Q2 & $\approx$ 100 & $\approx$ 0.6 $\times$ $10^{21}$ \\
      & M6.9 & Q3 & $\approx$ 130 & $\approx$ 0.3 $\times$ $10^{21}$ \\
      & M6.9 & Q4 & $\approx$ 166  & $\approx$ 0.2 $\times$ $10^{21}$\\
      & M6.9 & Q5 & $\approx$ 213 & $\approx$ 0.3 $\times$ $10^{21}$\\
      & M6.9 & Q6 & $\approx$ 101 & $\approx$ 0.5 $\times$ $10^{21}$\\  
  
NOAA 12242 & X1.8 & K1 & $\approx$ 114 & $\approx$ 0.1 $\times$ $10^{21}$ \\	
      & X1.8 & K2 & $\approx$ 156 & $\approx$ 0.4 $\times$ $10^{21}$ \\
     
NOAA 12297 & X2.0 & F1 & $\approx$ 76 & $\approx$ 3.0 $\times$ $10^{21}$ \\
& X2.0 & F2& $\approx$ 190 & $\approx$ 1.0 $\times$ $10^{21}$ \\

NOAA 12371 & C3.9, M6.5 & P1 & $\approx$ 169 & $\approx$ 0.7 $\times$ $10^{21}$\\
      & C3.9, M6.5 & P2 & $\approx$ 123 & $\approx$ 0.3 $\times$ $10^{21}$\\

		\hline
	\end{tabular}
\end{table*}

\section{Discussion and Conclusions}
We have studied seismic emission in the sunspots accompanying large solar flares for several active regions as referred in Table \ref{AR list} using the high resolution observations from HMI instrument aboard the {\em SDO} space mission. The summary of our analysis, chief findings and interpretation of results are as follows:\\

(1) We have analyzed the Dopplergrams of the active regions at a cadence of 45 s to examine acoustic enhancements in sunspots using power maps in 2.5--4 mHz band for the pre-flare and spanning flare epochs. We have selected only those locations for our study (c.f. Table \ref{percentagepower}), which are within the sunspots and away from the flare ribbons and hard X-ray foot-points, since these are supposed to be free from any flare related artefacts in the observational data. The temporal evolution of total power corresponding to these seismic emission locations shows  enhancement in power accompanying the flare and it tends to return to normal value after the flare. We also note that the temporal evolution of Doppler velocity in the seismic emission location shows enhancement during the flare and post-flare epochs. Additionally, the identified seismic emission locations in the sunspots show significant quantity of percentage change in total acoustic power (c.f. Table \ref{percentagepower}) during the flares. Thus, we believe that these seismic emissions in the sunspots during the flare are solar in nature.\\

(2) In order to understand the cause of these seismic emissions, we have analysed line-of-sight (i.e., longitudinal) magnetic fields at a cadence of 45 s as well as high cadence (135 s) vector magnetograms in these identified locations. In most of the cases, the temporal evolution of these magnetic fields corresponding to seismic emission locations shows abrupt and persistent changes of the order of 50--100 Gauss within a duration of 10--20 minutes during the flare and post flare epochs. The plots of the evolution of magnetic fields in `B1' location are shown here (c.f. Figure \ref{fig: magfield}) whereas those for the other identified locations are provided in the online supplementary material. These results are consistent with the earlier flare related magnetic field changes reported by \cite{2005ApJ...635..647S} and \cite{2010ApJ...724.1218P}. \\

(3) We have estimated the changes in Lorentz force corresponding to the seismic emission locations. Our analysis shows changes in the Lorentz force of the order of 10$^{21}$ dynes in the seismic emission locations in the sunspots (c.f. Table \ref{percentagepower}). The magnitude of change in Lorentz force in the identified seismic locations as obtained in our analysis is an order lower as compared to that estimated by \cite{2008ASPC..383..221H} and other previous studies. Apparently, this is because our locations are away from the centres of the active regions where the observed magnetic field changes are relatively smaller. On the other hand, those reported in earlier studies (\cite{2012ApJ...759...50P}, \cite{2014SoPh..289.3663P} $\&$ \cite{2018SoPh..293...16S}) are mostly along the polarity inversion lines (PILs) where the magnetic field vectors become stronger and more horizontal during the flares and hence the magnetic field changes are more pronounced. 
The plots of changes in the Lorentz force in `B1' location of active region NOAA 11158 and brief overview of results of the analysis of the NOAA active regions 11261, 11882, 12222, 12241, 12242, 12297 and 12371 are shown in Figures \ref{fig: lorentzforcechange} to \ref{AR12371}. The remaining maps and plots of the analysis of these active regions are provided in the online supplementary material. We observe a good correspondence between the enhancement in acoustic power in the seismic emission locations and the impulsive and other episodic changes of similar size in the Lorentz force in these identified locations. Therefore, our investigation indicates that ``magnetic-jerk" is the driving force for seismic emissions in the sunspots away from flare ribbons and hard X-ray foot-point locations observed during the flares. \\   

(4) We have also estimated the work done by change in Lorentz force in the seismic emission location `B1' in the active region NOAA 11158 and compared this available energy budget with the acoustic emission computed in one of the kernels (5$\times$5 pixels) in the aforementioned seismic location. It is to be noted that \cite{2008ASPC..383..221H} have computed the work done by the change in Lorentz force ($\approx 10^{22}$ dyne) as estimated from the results of \cite{2005ApJ...635..647S} and considering the displacement of the photosphere to be $\approx$ 3 km as observed in the amplitude of seismic wave produced during an X2.6 class flare in active region NOAA 7978 \citep{1998Natur.393..317K}. Thus, they arrived at an energy budget of $\approx$ 3$\times$10$^{27}$ erg, which was found to be comparable to the energy of acoustic emissions reported in previous studies \citep{2006SoPh..239..113D,{2007MNRAS.374.1155M}}. Following \cite{2008ASPC..383..221H}, we have estimated the work done (W = $\delta$F.$r$) by the change in Lorentz force ($\delta$F $\approx$ 2$\times$10$^{21}$ dyne) as obtained for the seismic emission location `B1' in displacing the photoshperic plasma by, say, $r$ $\approx$ 3 km, which yields W $\approx$ 6$\times$10$^{26}$ dyne cm. We further compute the excess acoustic energy in the aforementioned kernel, which could be represented by, $\delta$E = $\rho.\delta p.A.d$, where `$\rho$' is the mean density of the solar photosphere ($\approx$ 2$\times$10$^{-7}$ g cm$^{-3}$), `$\delta p$' is the change in acoustic power in the identified kernel in the frequency band 2.5-4.0 mHz band ($\approx$ 5$\times$10$^6$ cm$^2$ s$^{-2}$), `A' is the area of the kernel ($\approx$ 3.2$\times$10$^{16}$ cm$^2$), and `d' is the depth of the kernel which could be considered approximately equal to its linear size ($\approx$ 1.8$\times$10$^8$ cm), assuming that as much acoustic energy travels vertically from the source as in each horizontal direction. Thus, we obtain $\delta$E $\approx$ 5.7$\times$10$^{24}$ erg. This is the approximate energy budget of the acoustic emission observed in the identified kernel, which is lower than the magnitude of the work done by the change in Lorentz force in this seismic  location. Thus, our results suggest that the observed seismic emission has been induced by the impulsive changes in Lorentz force in the sunspot region during the flare.\\

(5) We have also tried to investigate a relationship between the change in Lorentz force and percentage change in seismic power, however we could not find any explicit relation between these two parameters. This could be due to the reason that in our photospheric Doppler observations, we are able to observe only the trapped acoustic modes, which is some fraction of the induced seismic emission by ``magnetic-jerk". The remaining fraction of these acoustic waves would propagate higher into the solar atmosphere along the magnetic field lines in the form of magnetoacoustic waves depending on their inclination and interaction with these acoustic waves. Hence, it would be possible to conduct a detailed study of the relation between these two, only if we have simulatneous Dopplergrams available for the layers above the photosphere.\\

(6) These seismic emissions in the sunspots are essential to study because these enhanced locations can give better information about the deep dynamics in the active regions during the flares. In addition, since these ``magnetic-jerk" driven seismic waves can also propagate from the photosphere to higher solar atmospheric layers along the magnetic field lines in the form of magnetoacoustic waves, hence it can contribute to the heating of the solar atmosphere. Thus, considering its aforementioned importance this study will be further carried out by using the upcoming facility of simultaneous velocity observations of the photosphere and chromosphere from the Multi Application Solar Telescope \citep{2009ASPC..405..461M, 2017SoPh..292..106M, Venkat17} operational at the Udaipur Solar Observatory, India.  

\section*{Acknowledgements}

We acknowledge the use of data from HMI instrument aboard the {\em SDO} spacecraft. Our sincere thanks goes to HMI team for providing high cadence (135 s) vector magnetogram data for our analysis. We acknowledge the use of H$\alpha$ data from Global Oscillations Network Group. We also acknowledge the use of soft X-ray data from {\em GOES-15} satellite and hard X-ray data from {\em RHESSI} spacecraft. We are thankful for the support being provided by Udaipur Solar Observatory/Physical Research Laboratory. Thanks to Paul Rajaguru and Shibu K. Mathew for useful discussions related to this work. Finally, we are very much thankful to the anonymous referee for the constructive and fruitful  comments and suggestions that improved the presentation of results and inclusion of relevant discussions in this paper.

\section*{Data availability}

The data underlying this article are available in Helioseismic and Magnetic  Imager (HMI) aboard the {\em Solar Dynamic Observatory} ({\em SDO}) data archive at  http://jsoc.stanford.edu/ajax/lookdata.html, Global Oscillation Network Group (GONG) data archive at https://gong2.nso.edu/archive/patch.pl?menutype=a and {\em Reuven Ramaty High Energy Solar Spectroscopic Imager} ({\em RHESSI}) data archive at http://hesperia.gsfc.nasa.gov/hessidata/. 









\end{document}


Online supplementary material related to the manuscript entitled {\bf ``On the seismic emission in sunspots associated with Lorentz force changes accompanying major solar flares"}.

\vspace{0.5cm}

\hspace{4cm} [A]   \bf{Active region NOAA 11261}
\vspace{1cm}

\begin{figure*}[h!]
\centering
\includegraphics[width=0.48\textwidth]{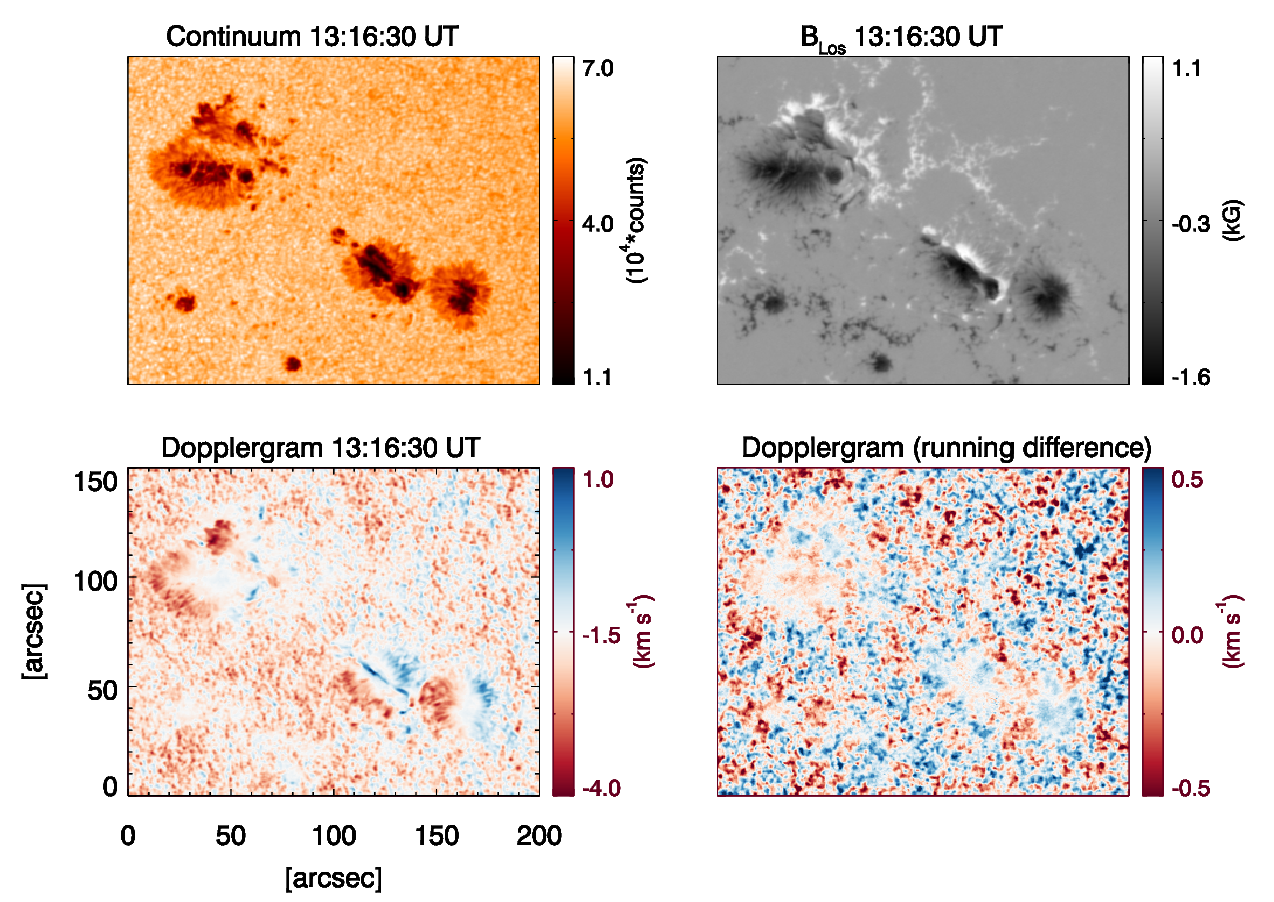}\hspace*{0.34cm}
\includegraphics[width=0.45\textwidth]{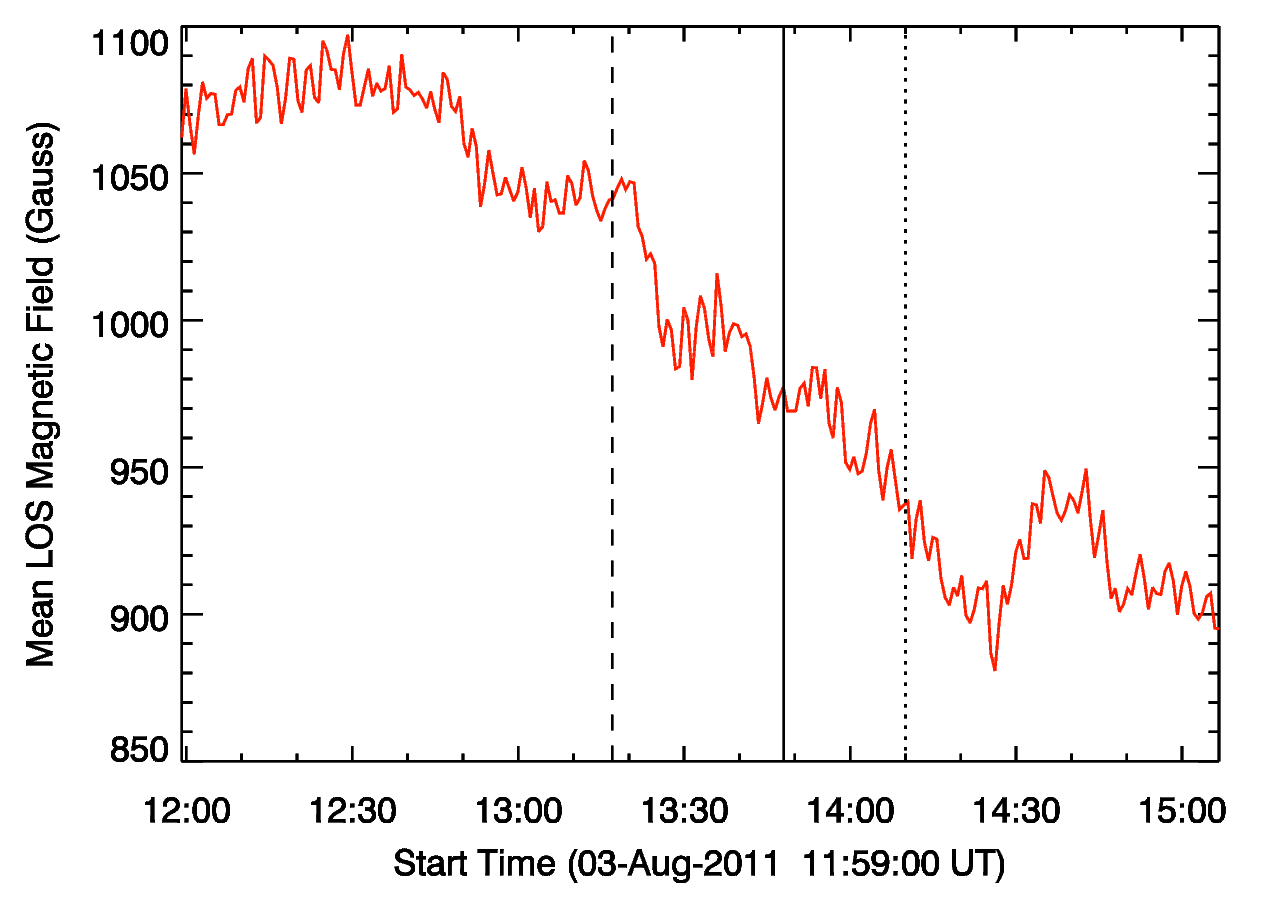}
\caption{\textit{Left panel}: Sample images of the active region NOAA 11261 showing continuum intensity (top left panel), photospheric line-of-sight  magnetic fields (top right panel), Dopplergram (bottom left panel) and running difference of Doppler images (bottom right panel) acquired from HMI instrument aboard the {\em SDO} spacecraft on 2011 August 03. \textit{Right panel}: Plot shows the temporal evolution of line-of-sight magnetic fields in the `D2' location of active region NOAA 11261. 
The dashed, solid and dotted vertical lines represents onset, peak and decay time of the flare.}
\label{fig: fourimage11261}
\end{figure*}

\begin{figure*}[h!]
\centering
\includegraphics[width=0.45\textwidth]{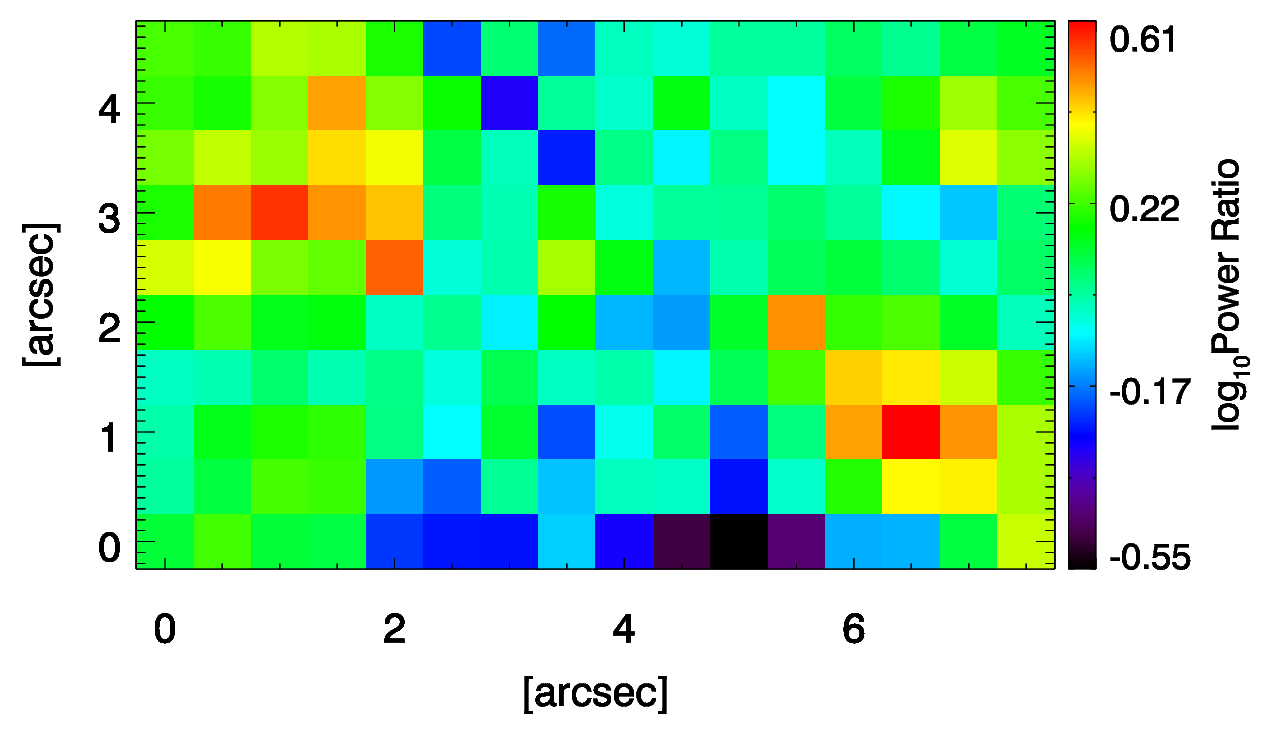}\hspace{0.34cm}
\includegraphics[width=0.45\textwidth]{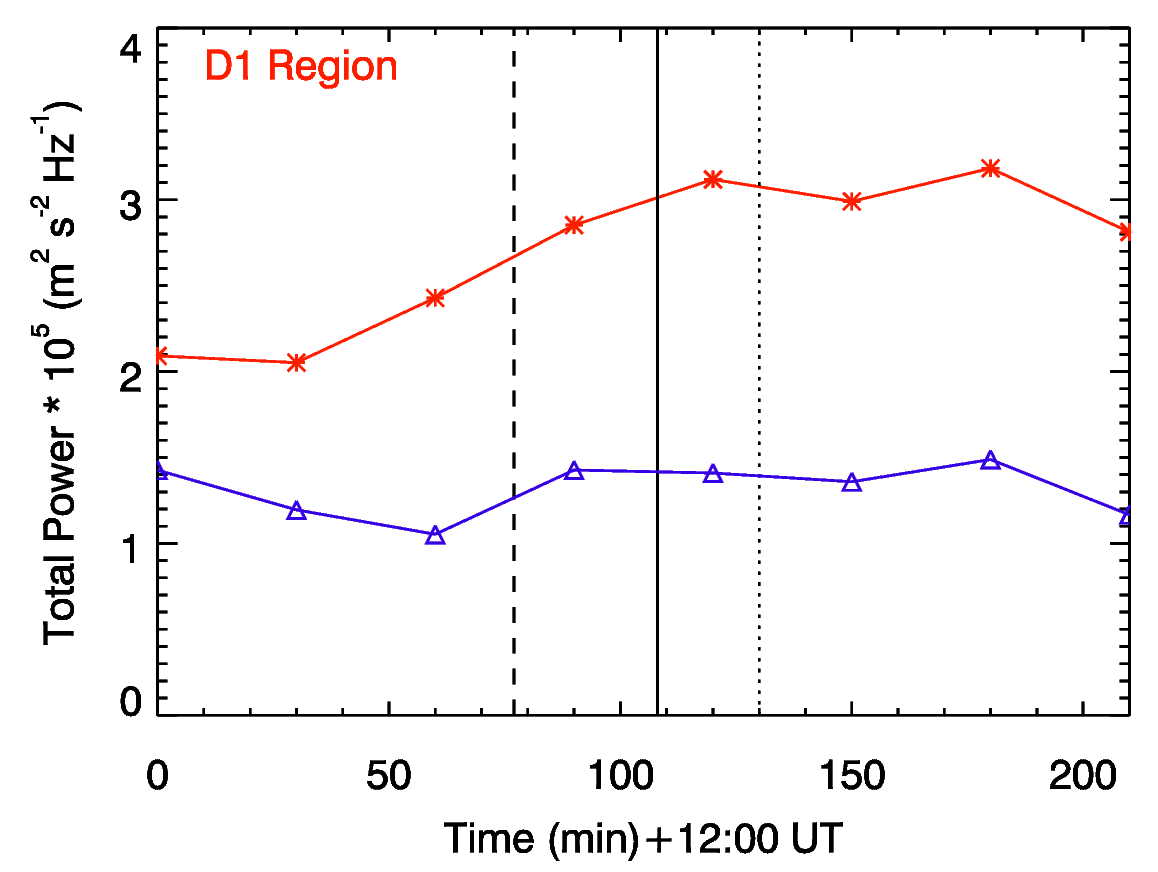}
\caption{\textit{Left panel}: Illustrates the blow-up region of `D1' enhanced location in the sunspot in the active region NOAA 11261, as indicated in the 
power map ratio in Figure 7 of the main manuscript. \textit{Right panel}: Plot shows the temporal evolution of integrated acoustic power over the ‘D1’ location (red colour with asterisks) whereas that shown in blue colour with triangles represents evolution of total acoustic power in an unaffected region in the same sunspot. It is to be noted that there is a time offset of about $\pm$ 30-minutes between the acoustic power variation and the GOES flare-time.}
\label{fig: D1_Power_11261}
\end{figure*}

\begin{figure*}
\centering
\includegraphics[width=0.44\textwidth]{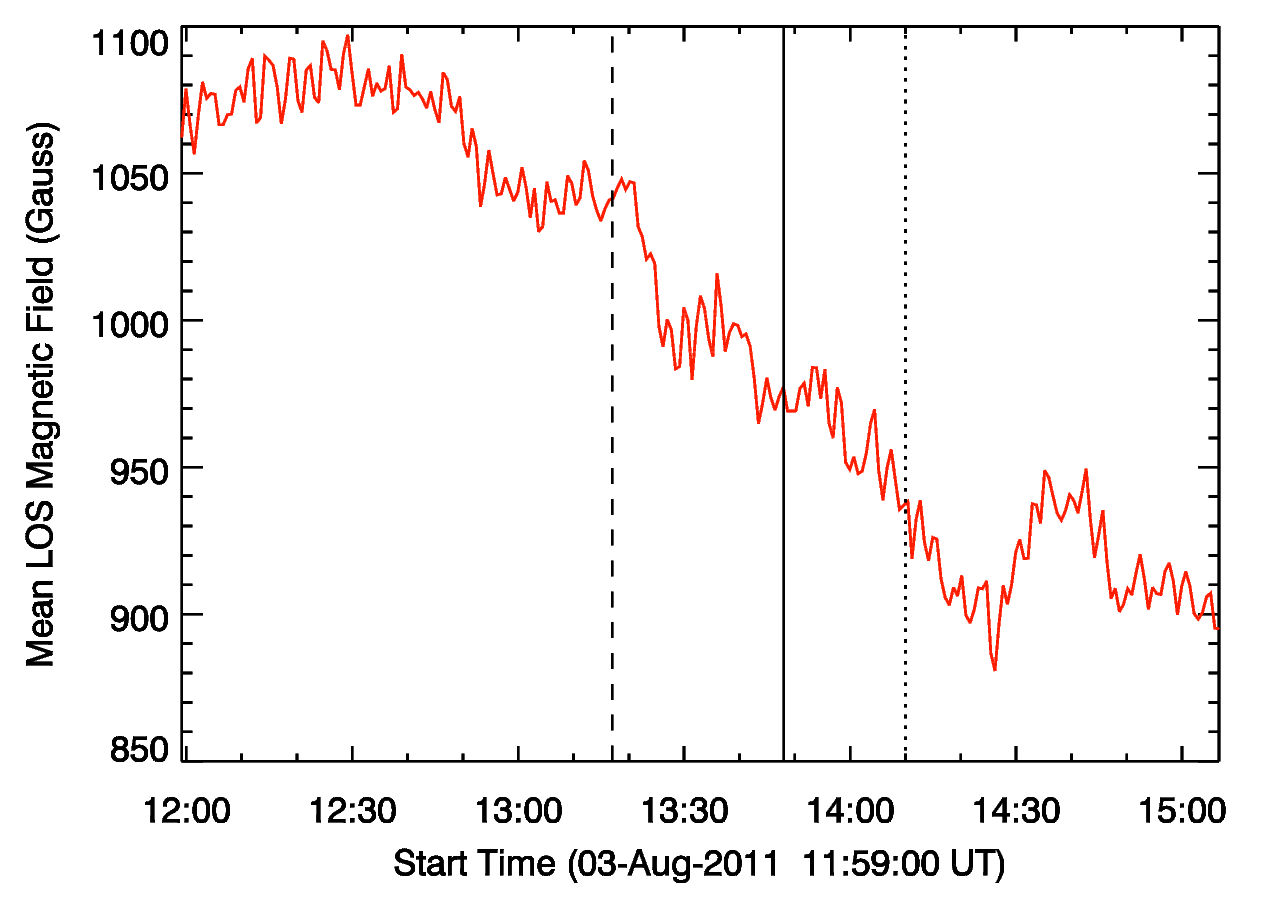}\hspace*{0.34cm}
\includegraphics[width=0.44\textwidth]{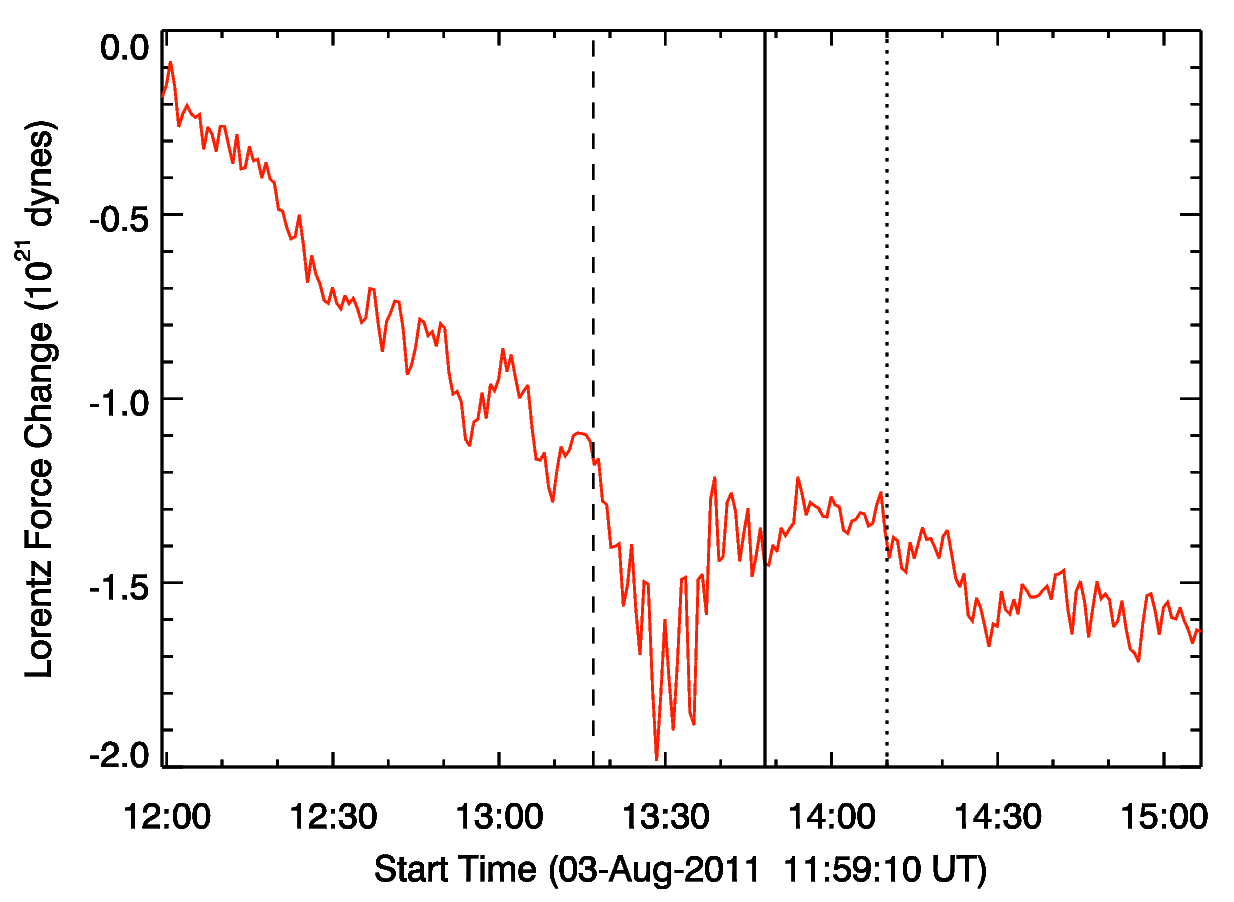}
\caption{\textit{Left panel}: Plot showing the temporal evolution of line-of-sight magnetic fields in the `D1' location of active region 11261.  \textit{Right panel}: Plot showing the temporal evolution of change in radial component of Lorentz force in the `D1' location of active region 11261. The dashed, solid and dotted vertical lines represents onset, peak and decay time of the flare.}
\label{Fig: D12_mag}
\end{figure*}
\break

\title{Active region NOAA 11882}
\hspace{4cm} [B] \bf{Active region NOAA 11882}
\vspace{1cm}

\begin{figure*}[h!]
\centering
\includegraphics[width=0.47\textwidth]{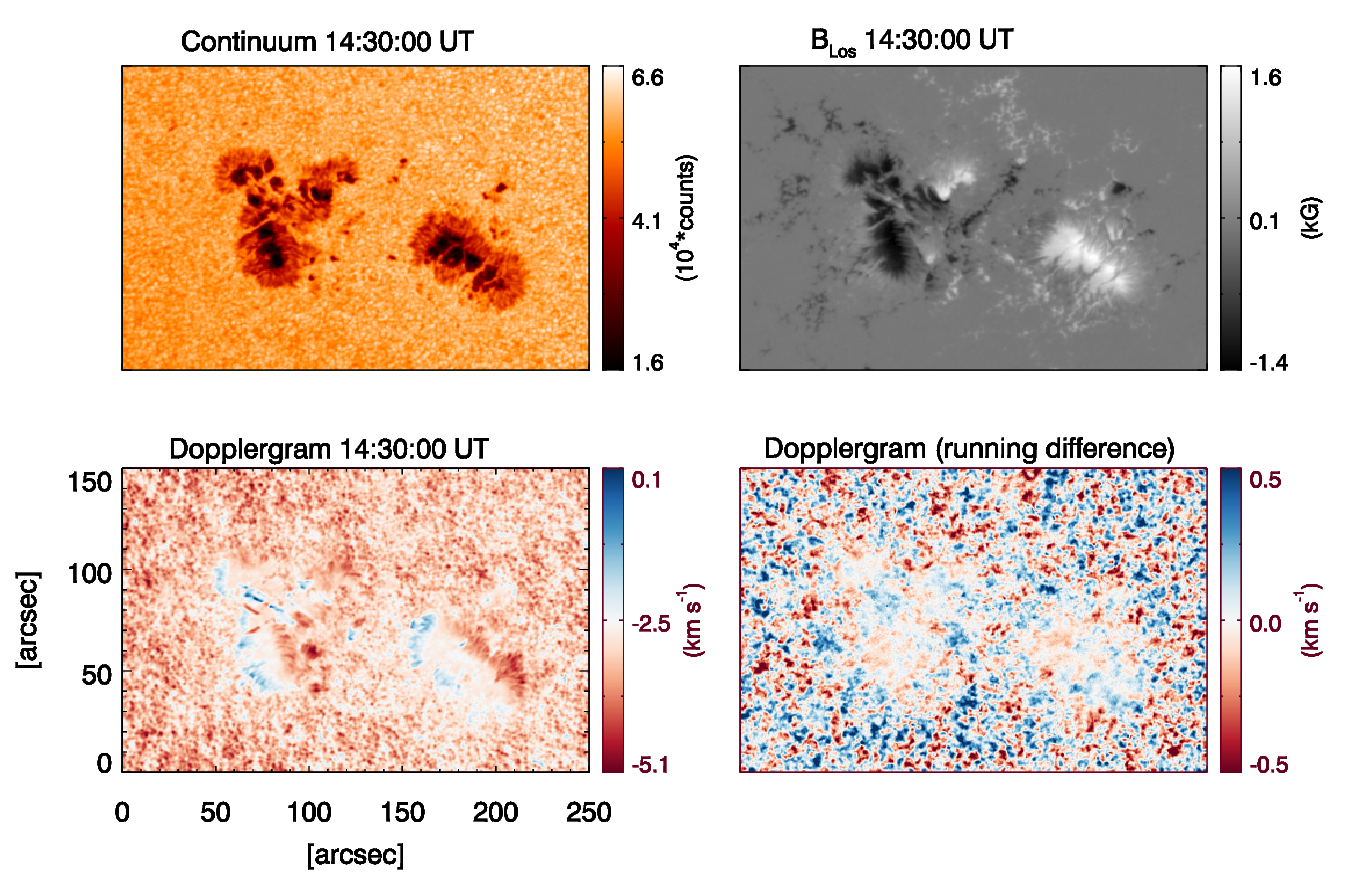}\hspace*{0.34cm}
\includegraphics[width=0.44\textwidth]{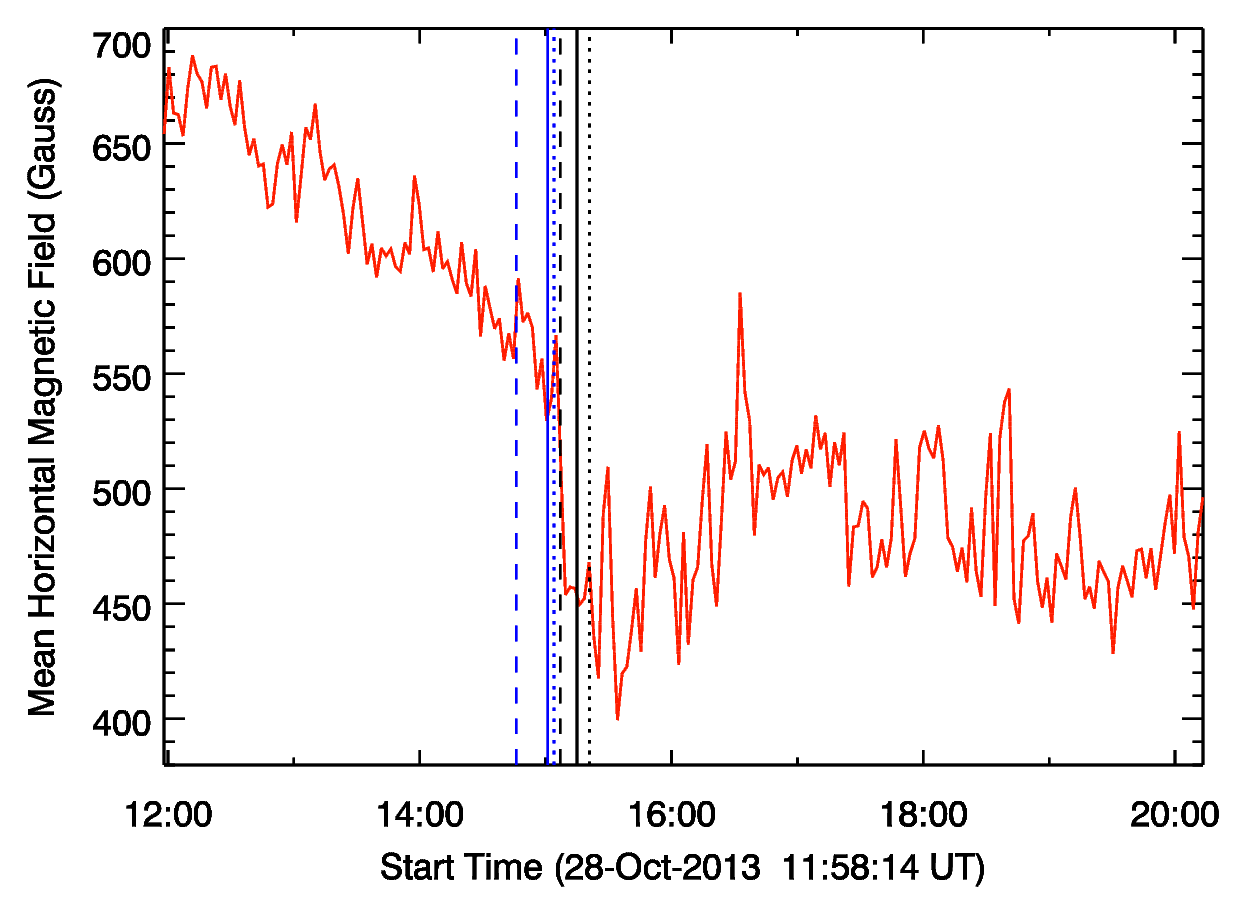}
\caption{\textit{Left panel}: Sample images of the active region NOAA 11882 showing continuum intensity (top left panel), photospheric line-of-sight magnetic fields (top right panel), Dopplergram (bottom left panel) and running difference of Doppler images (bottom right panel) acquired from HMI instrument aboard the {\em SDO} spacecraft on 2013 October 28. \textit{Right panel}: Plot shows the temporal evolution of horizontal magnetic fields in the `N1' location of active region NOAA 11882. The dashed, solid and dotted vertical blue colour and black colour lines represent the onset, peak and decay time of M2.7 $\&$ M4.4 class flares, respectively.}
\label{fig: fourimage11882}
\end{figure*}

\begin{figure*}[h!]
\centering
\includegraphics[width=0.4\textwidth]{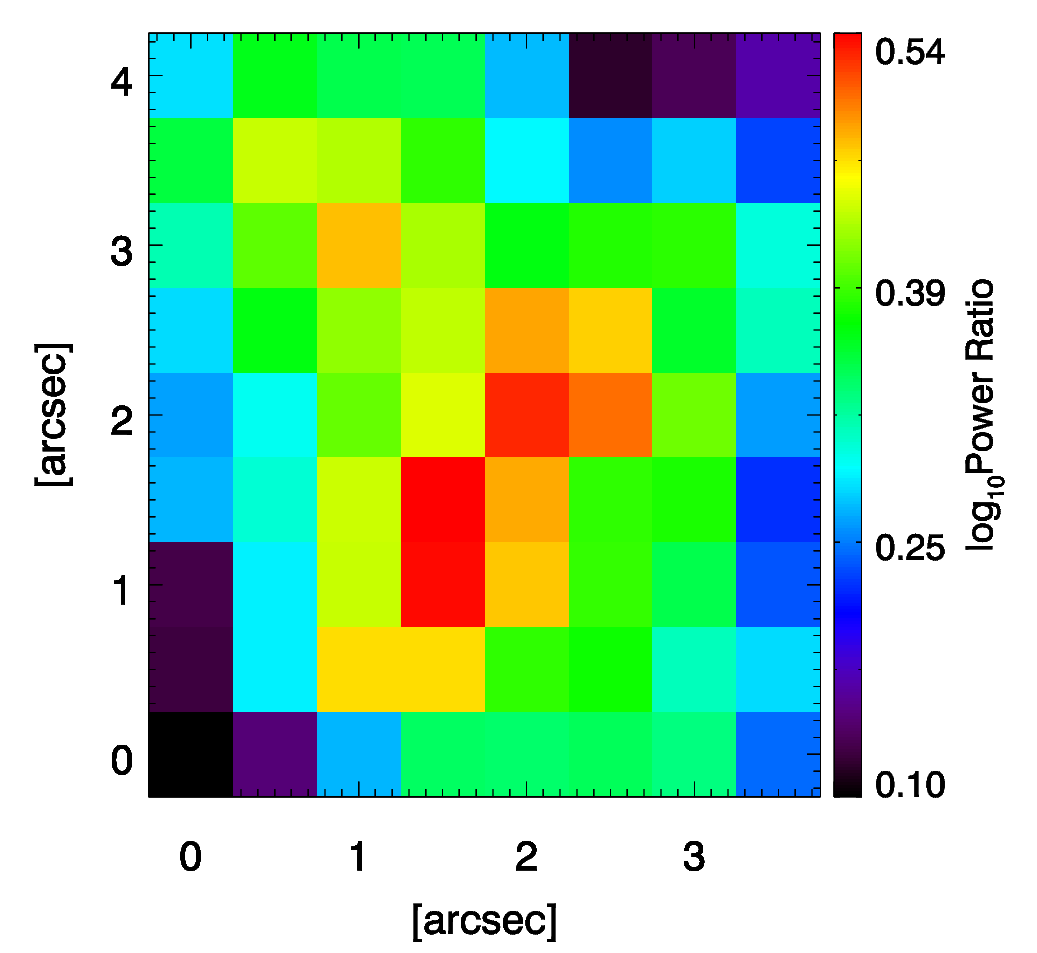}\hspace*{0.34cm}
\includegraphics[width=0.44\textwidth]{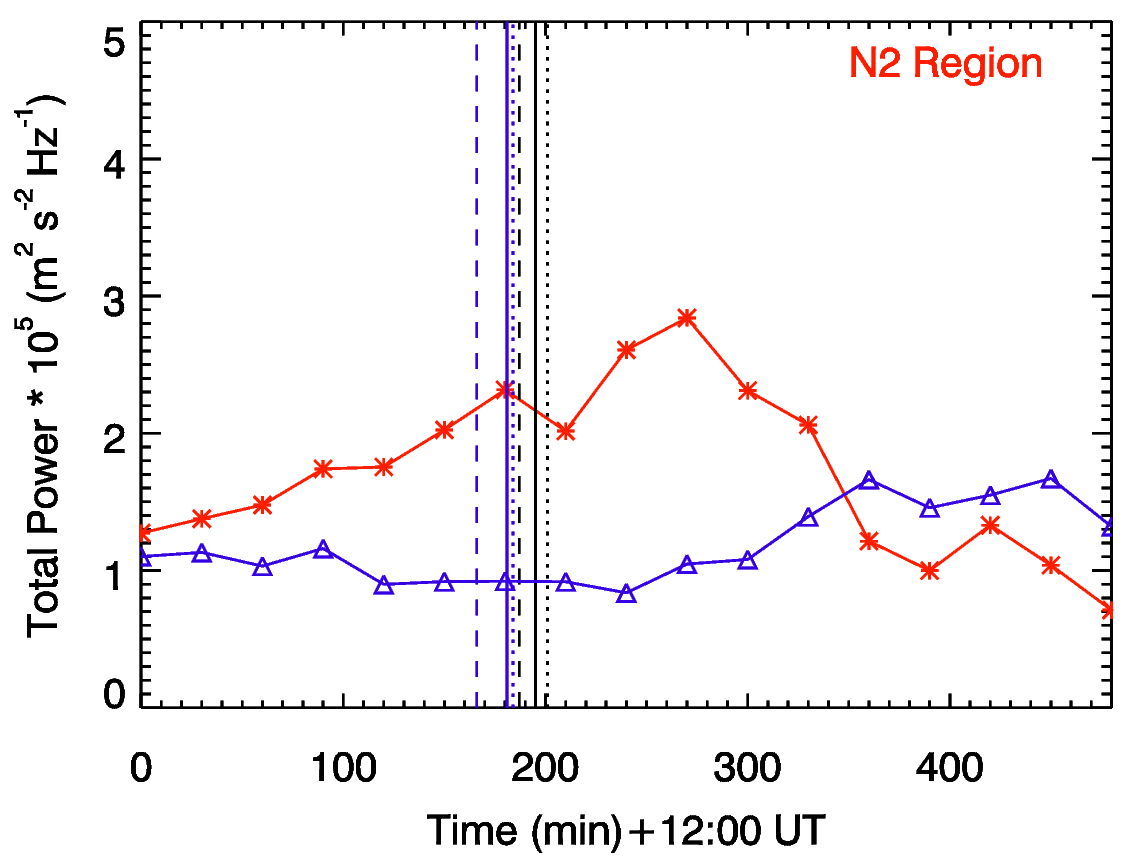}
\caption{\textit{Left panel}: Illustrates the blow-up region of `N2' enhanced location of active region 11882, as indicated in the power map ratio in Figure 8 of the main manuscript. \textit{Right panel}: Plot showing the temporal evolution of integrated acoustic power over the ‘N2’ location (red colour with asterisks) whereas that shown in blue colour with triangles represents evolution of total acoustic power in an unaffected region in the same sunspot. It is to be noted that there is a time offset of about $\pm$ 30-minutes between the acoustic power variation and the GOES flare-time. The dashed, solid and dotted vertical blue colour and black colour lines represents the onset, peak and decay time of M2.7 $\&$ M4.4 class flares, respectively.}
\label{fig: N2_11882}
\end{figure*}

\begin{figure*}
\centering
\includegraphics[width=0.44\textwidth]{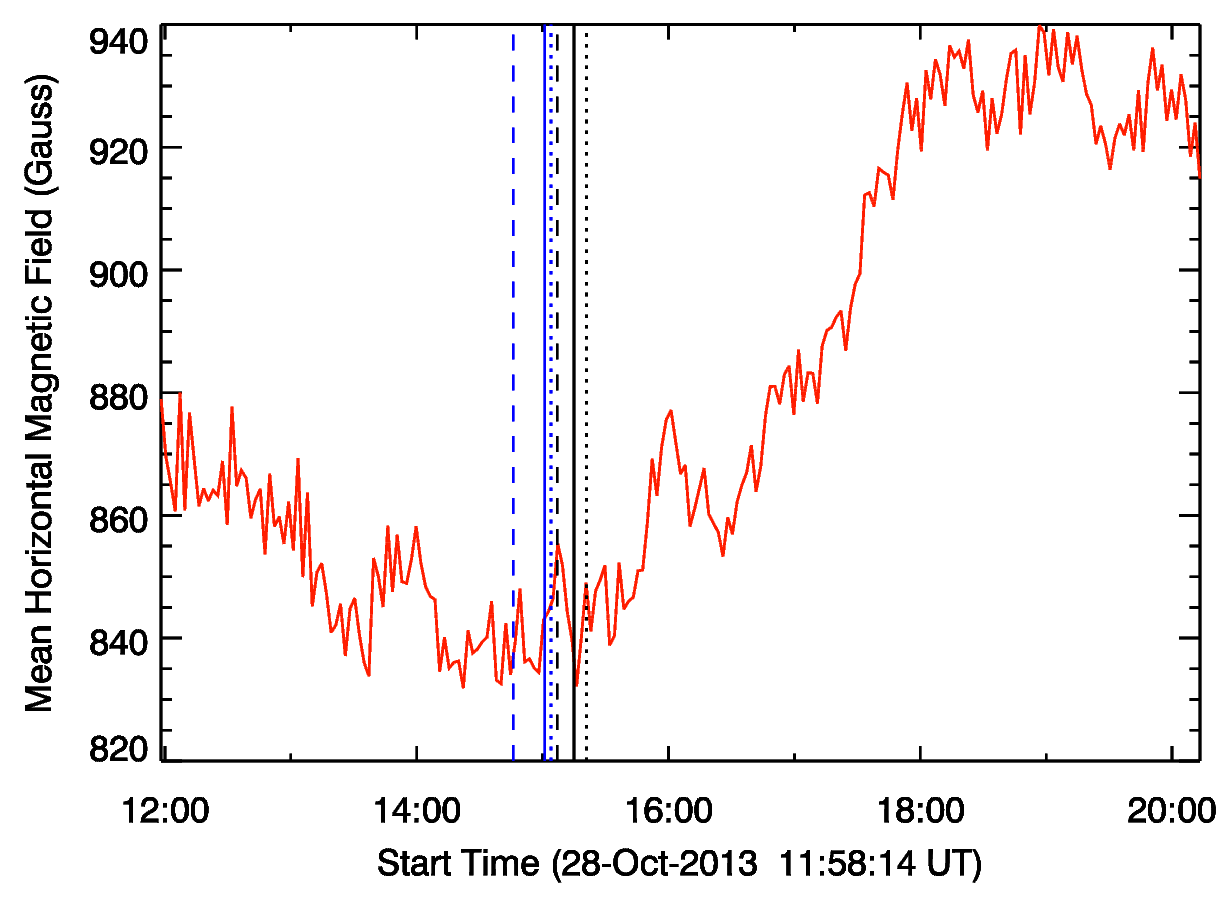}\hspace*{0.34cm}
\includegraphics[width=0.44\textwidth]{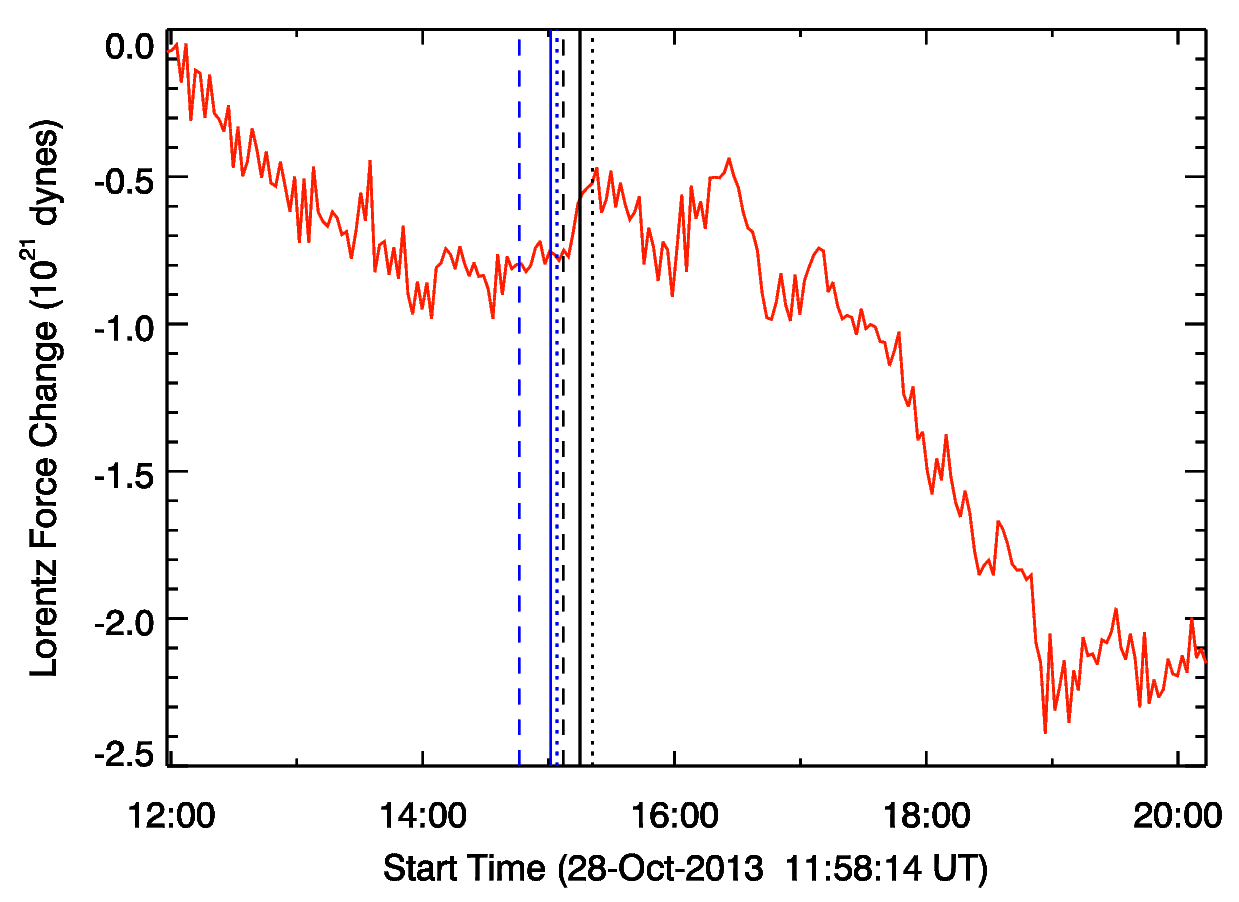}
\caption{\textit{Left panel}: Plot showing the temporal evolution of horizontal magnetic fields of `N2' location. \textit{Right Panel}: Plot showing the temporal evolution of change in radial component of Lorentz force in the `N2' location of active region 11882. The dashed, solid and dotted vertical blue colour and black colour lines represent the onset, peak and decay time of M2.7 $\&$ M4.4 class flares, respectively.}
\end{figure*}

\clearpage


\hspace{4cm} [C] \bf{Active region NOAA 12222}
\vspace{1cm}

\begin{figure*}[h!]
\centering
\includegraphics[width=0.47\textwidth]{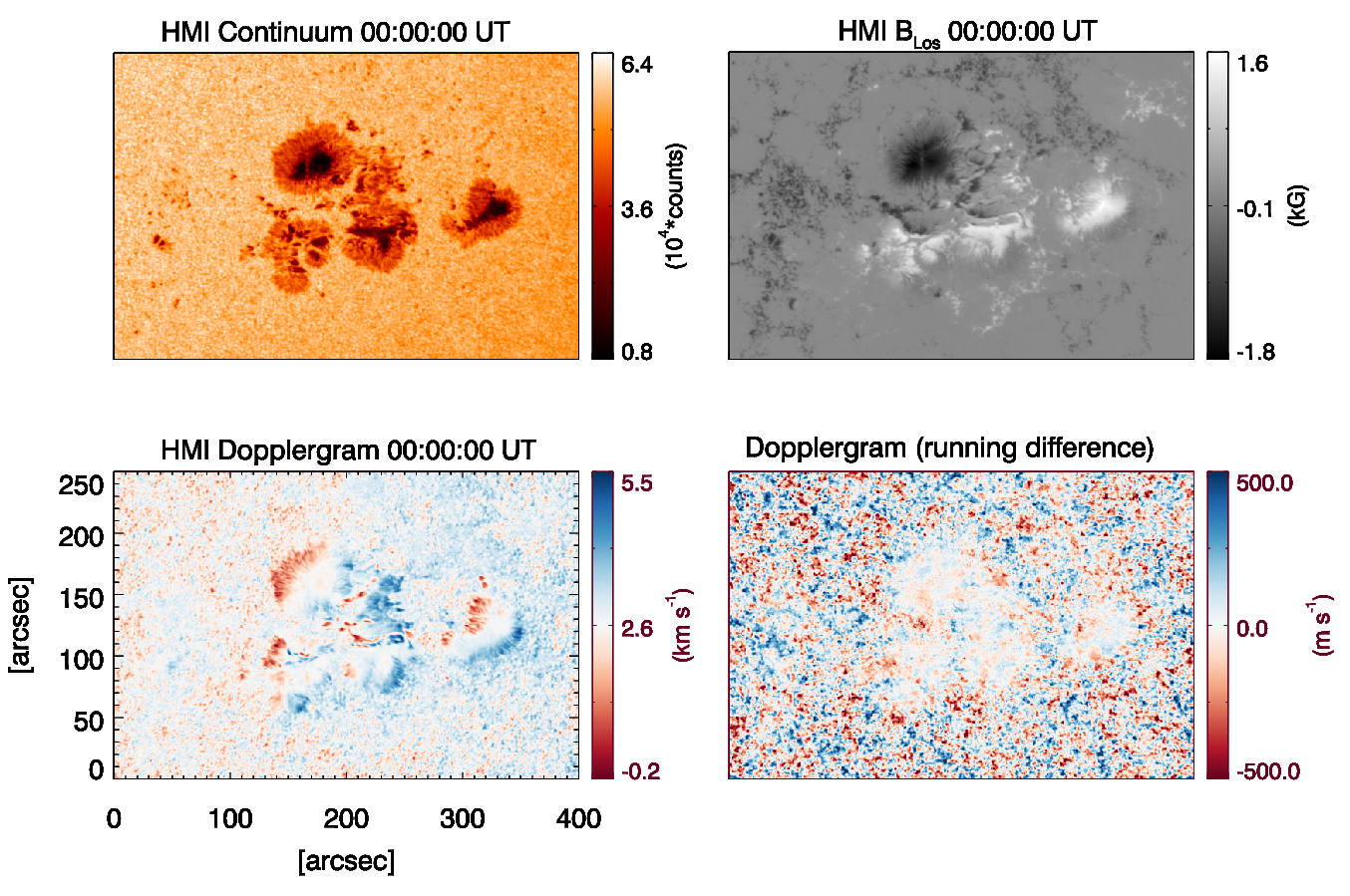}\hspace*{0.34cm}
\includegraphics[width=0.43\textwidth]{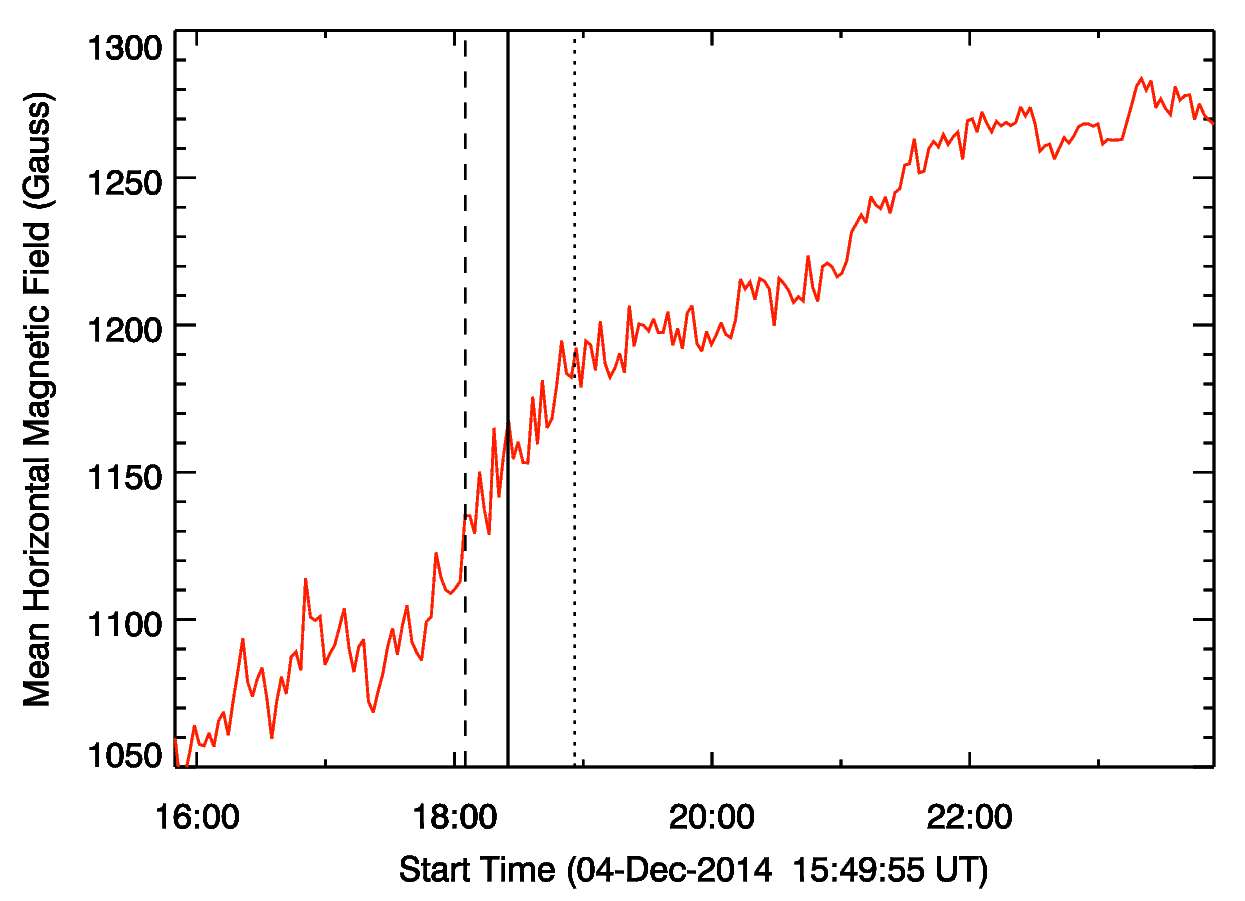}
\caption{\textit{Left panel}: Sample images of the active region NOAA 12222 showing continuum intensity (top left panel), photospheric line-of-sight magnetic fields (top right panel), Dopplergram (bottom left panel) and running difference of Doppler images (bottom right panel) acquired from HMI instrument aboard the {\em SDO} spacecraft on 2014 December 04. \textit{Right panel}: Plot shows the temporal evolution of horizontal magnetic fields in the `M1' location of active region NOAA 12222. The dashed, solid and dotted vertical lines represent the onset, peak and decay time of the flare.}
\label{fig: fourimage12222}
\end{figure*}

\begin{figure*}[h!]
\centering
\includegraphics[width=0.48\textwidth]{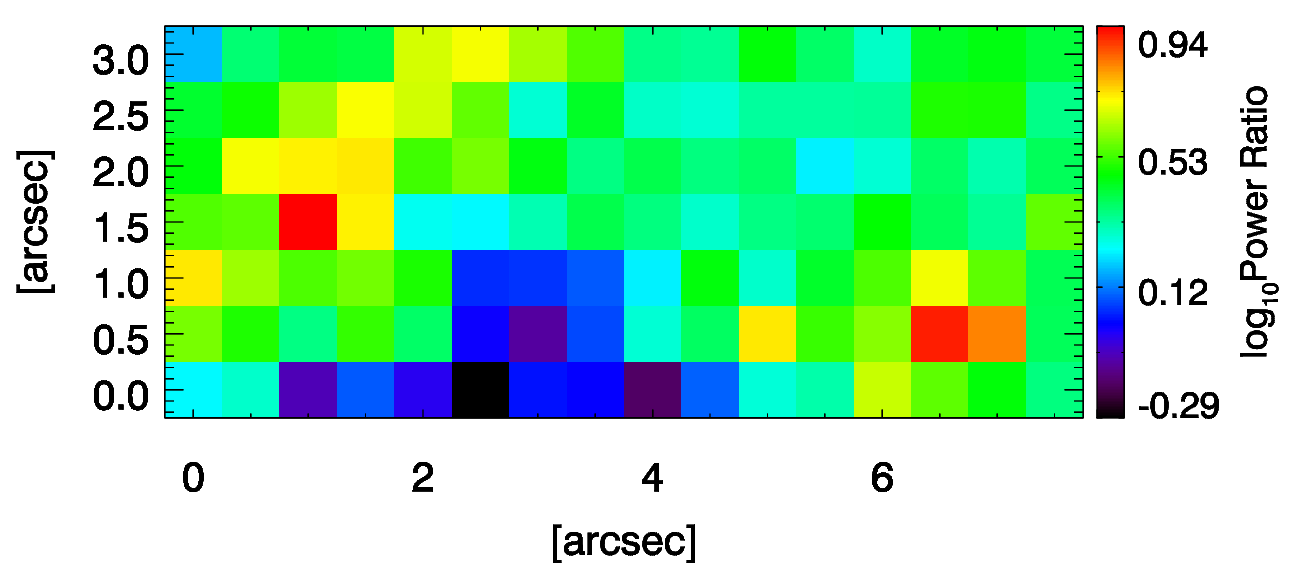}\hspace*{0.34cm}
\includegraphics[width=0.44\textwidth]{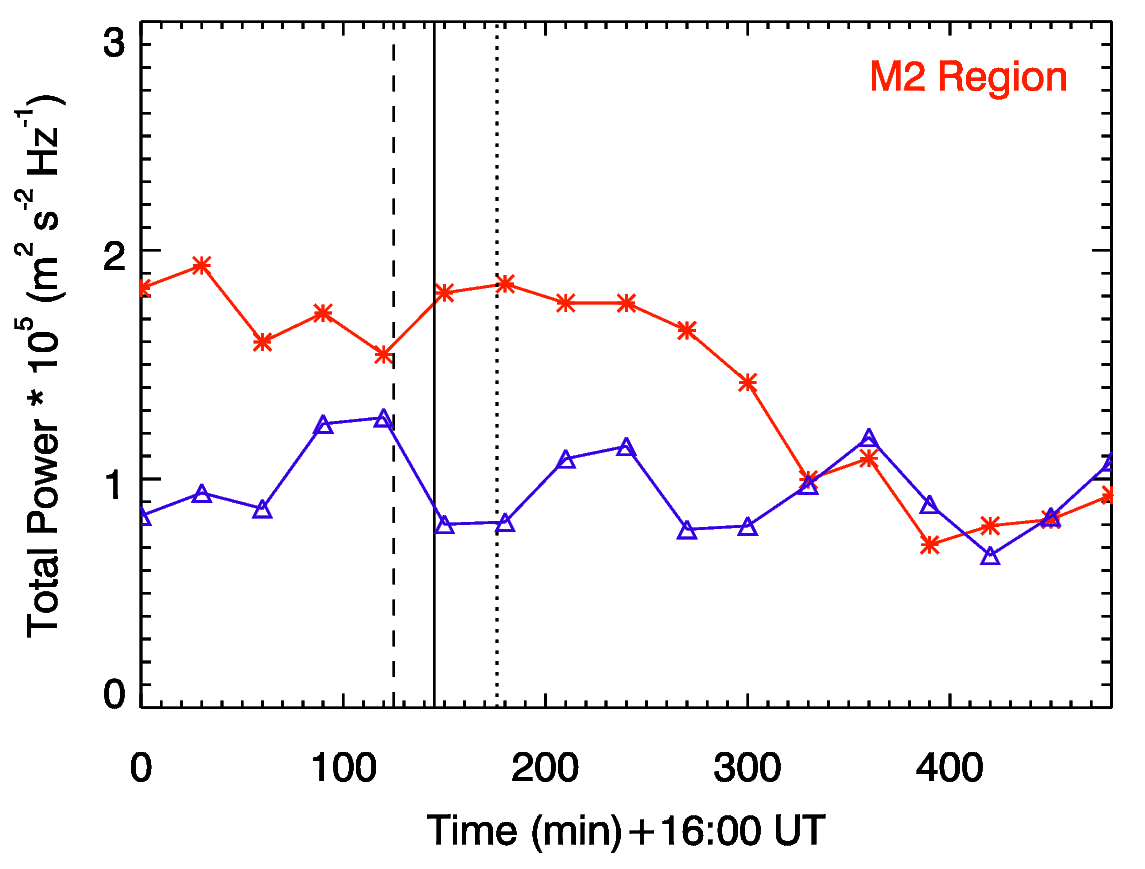}
\caption{\textit{Left panel}: Illustrates the blow-up region of `M2' enhanced location of active region 12222, as indicated in the power map ratio in Figure 9 of the main manuscript. \textit{Right panel}: Plot showing the temporal evolution of integrated power over the ‘M2’ location (red colour with asterisks) whereas that shown in blue colour with triangles represents evolution of total power in an unaffected region in the same sunspot. It is to be noted that there is a time offset of about $\pm$ 30-minutes between the acoustic power variation and the GOES flare-time.}
\label{fig: prespanning12222}
\end{figure*}

\begin{figure*}
\centering
\includegraphics[width=0.44\textwidth]{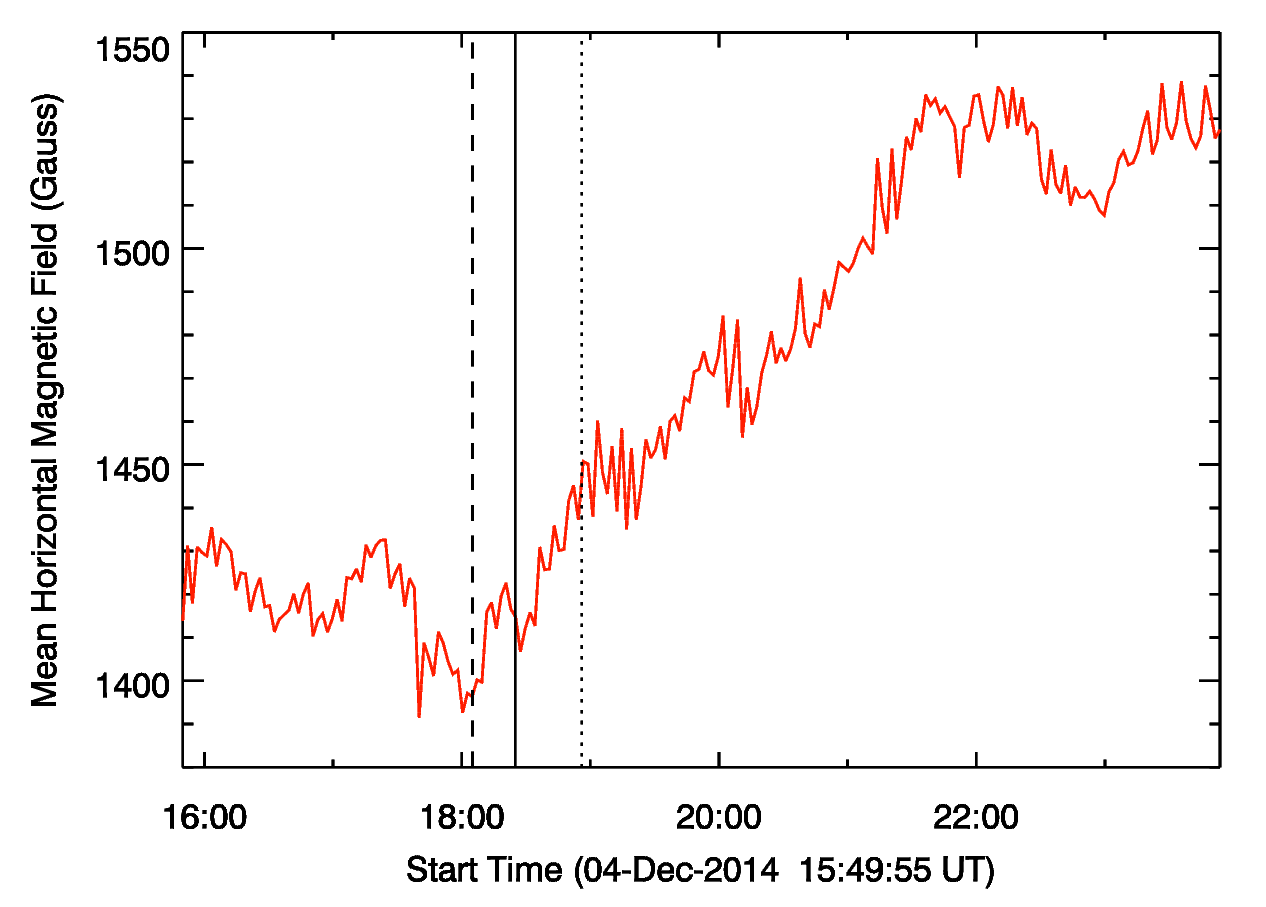}\hspace*{0.34cm}
\includegraphics[width=0.44\textwidth]{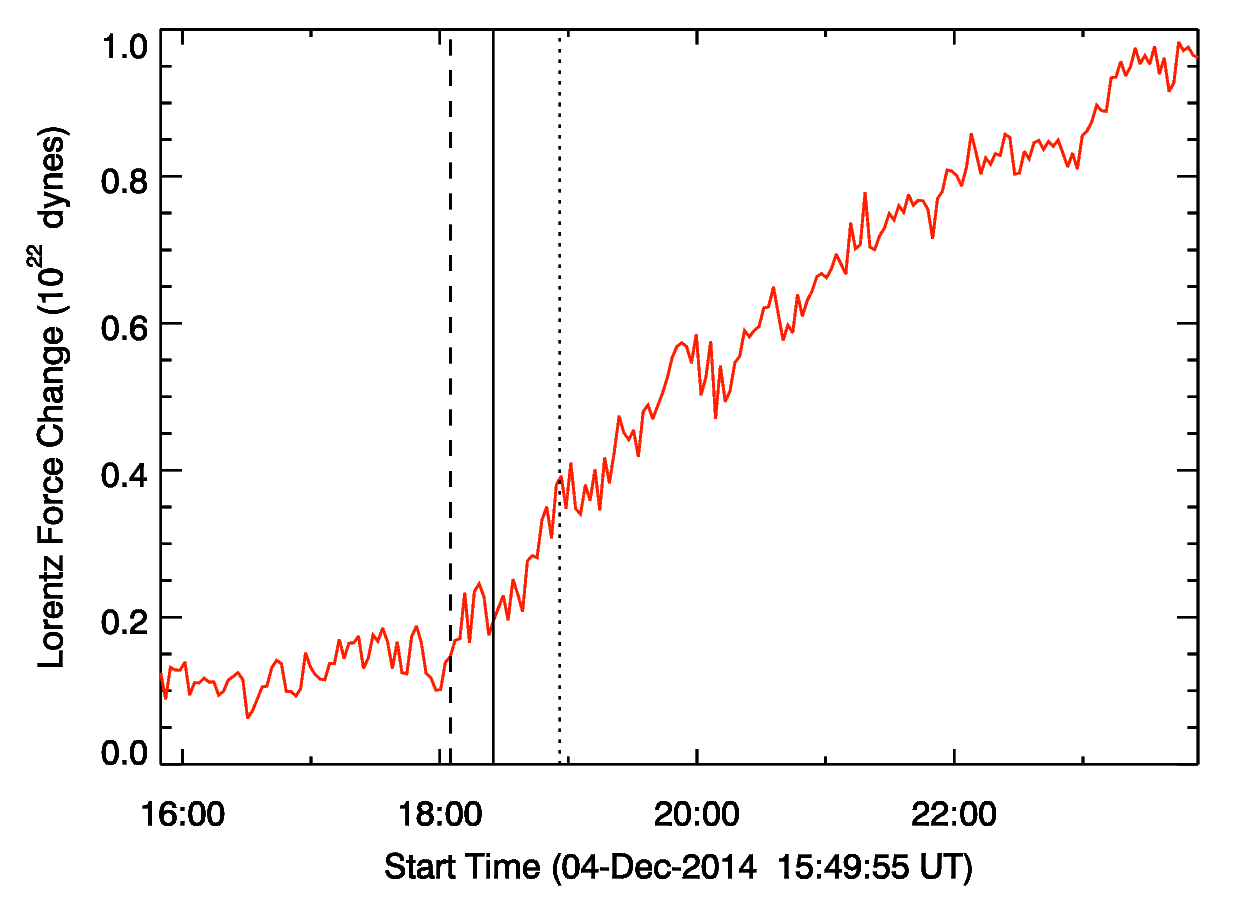}
\caption{\textit{Left panel}: Plot showing the temporal evolution of horizontal magnetic fields of `M2' location. \textit{Right panel}: Plot showing the temporal evolution of change in horizontal component of Lorentz force in the `M2' location of active region NOAA 12222. The dashed, solid and dotted vertical lines represent the onset, peak and decay time of the flare.}
\end{figure*}

\break


\hspace{4cm} [D] \bf{Active region NOAA 12241}
\vspace{1cm}

\begin{figure*}[h!]
\centering
\includegraphics[width=0.54\textwidth]{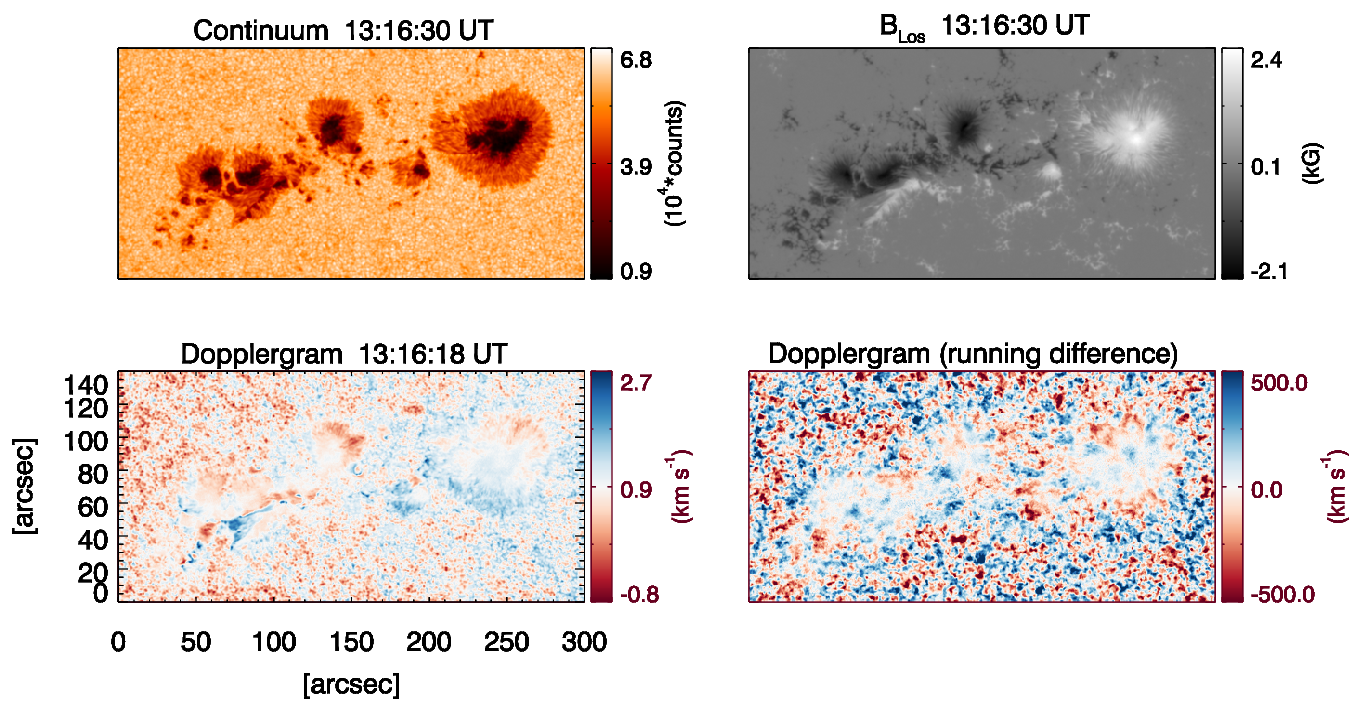}\hspace*{0.34cm}
\includegraphics[width = 0.43\textwidth]{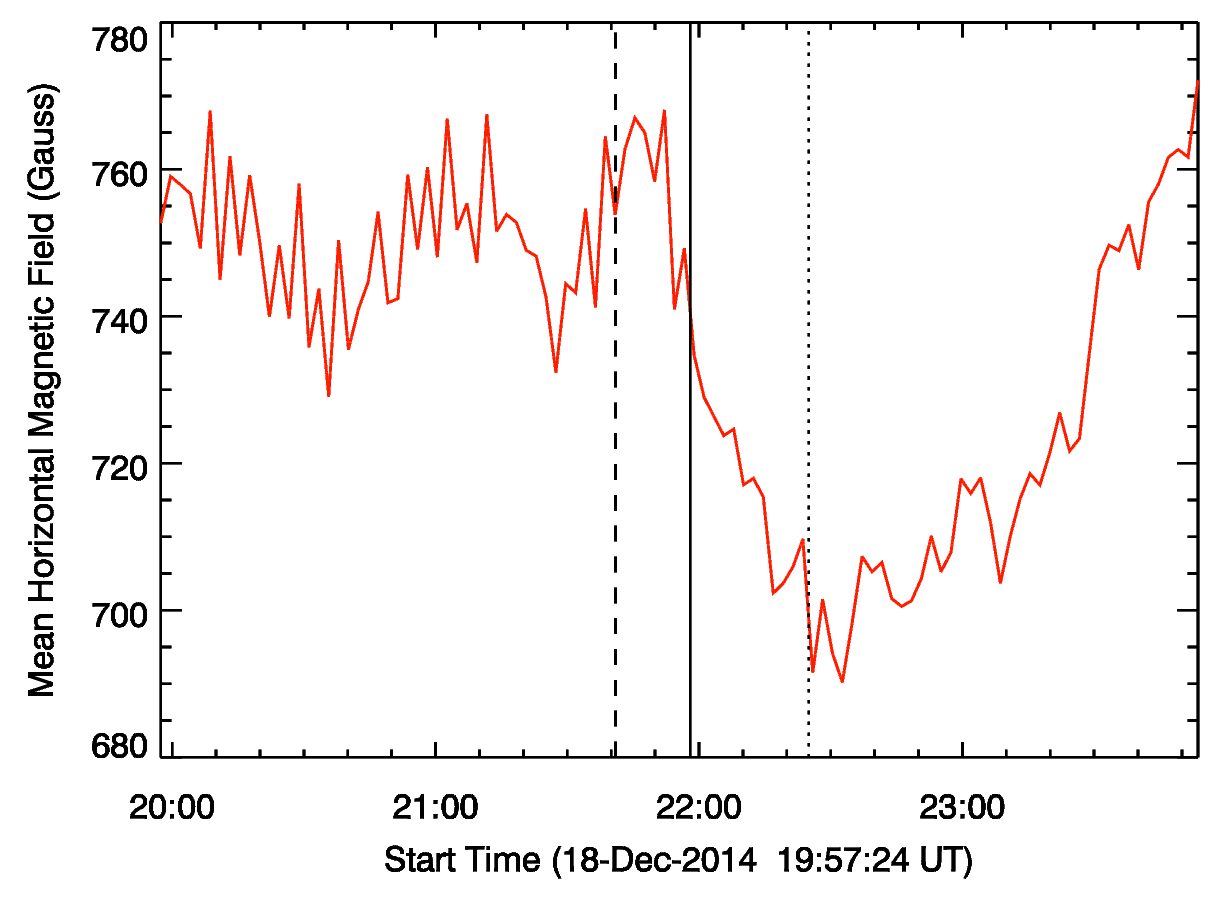}
\caption{\textit{Left panel}: Sample images of the active region NOAA 12241 showing continuum intensity (top left panel), photospheric line-of-sight magnetic fields (top right panel), Dopplergram (bottom left panel) and running difference of Doppler images (bottom right panel) acquired from HMI instrument aboard the {\em SDO} spacecraft on 2014 December 18. \textit{Right panel}: Plot shows the temporal evolution of horizontal magnetic fields in the `Q1' location of the active region NOAA 12241.}
\label{fig: fourimage12241}
\end{figure*}

\begin{figure*}
\centering
\includegraphics[width=0.44\textwidth]{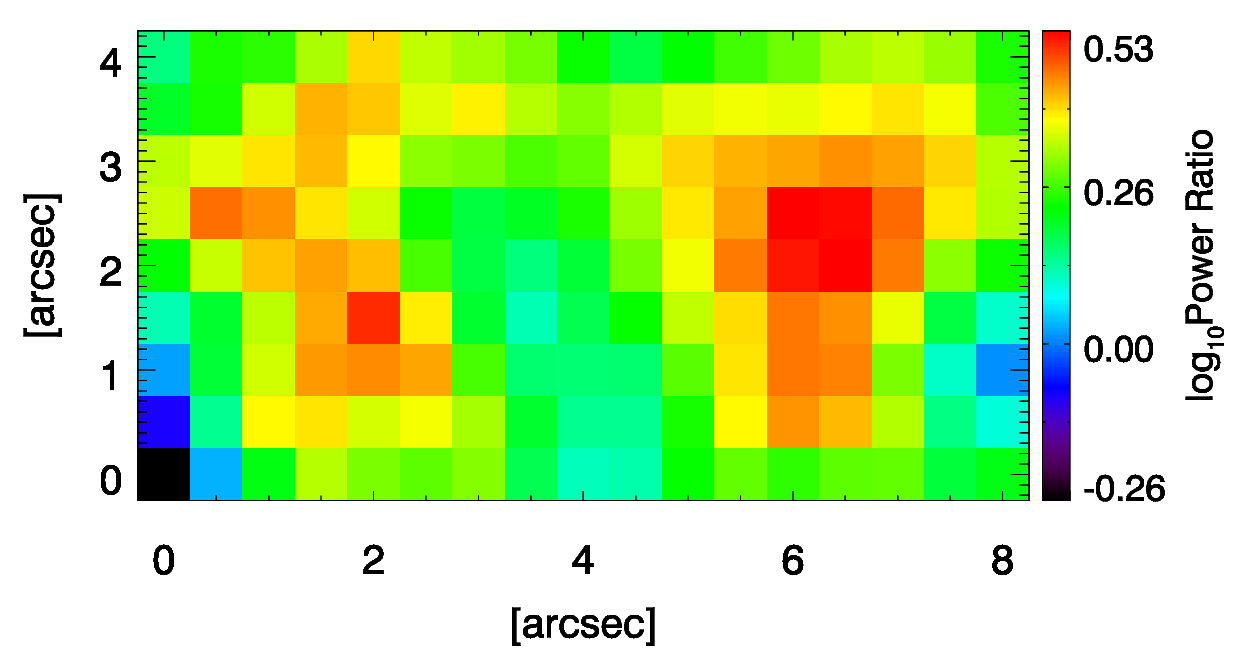}\hspace*{0.34cm}
\includegraphics[width=0.44\textwidth]{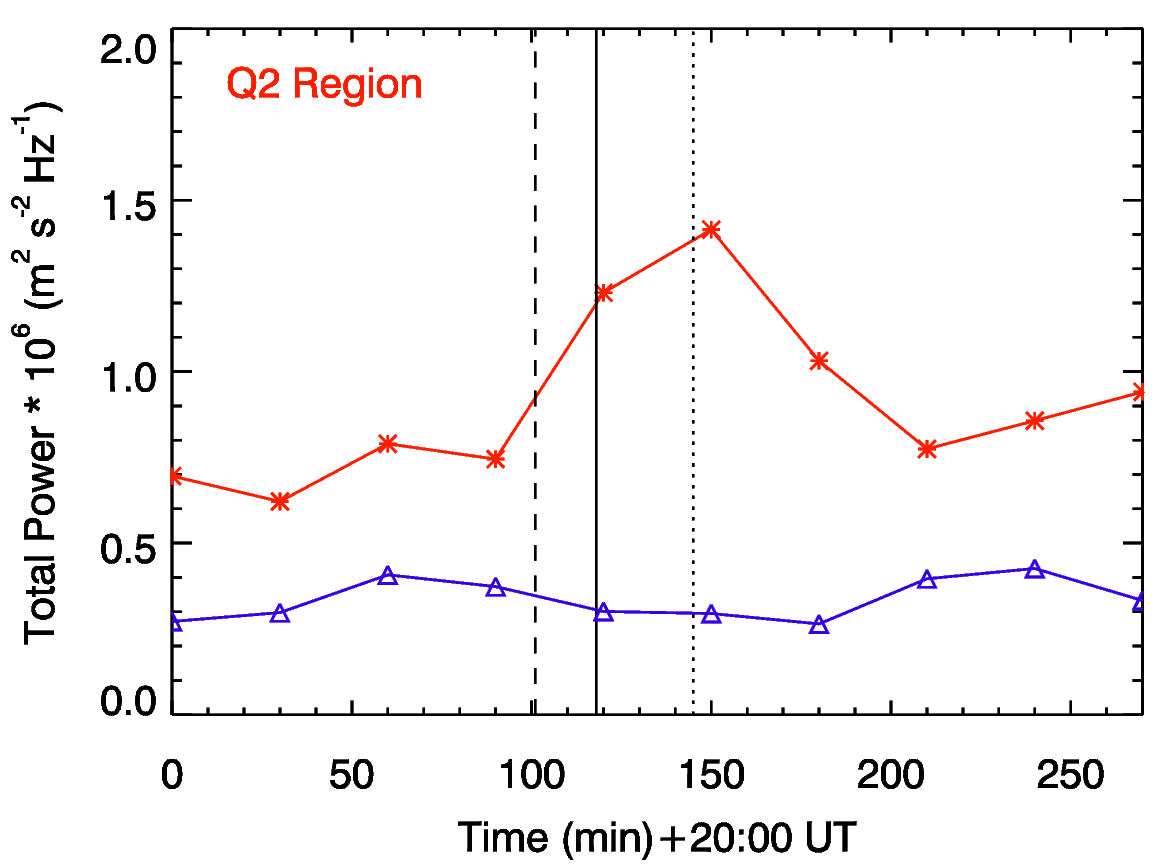}
\caption{\textit{Left panel:} Illustrates the blow-up region of `Q2' enhanced location of active region 12241, as indicated in the power map ratio in Figure 10 of the main manuscript. \textit{Right panel}: Plot showing the temporal evolution of integrated power over the ‘Q2’ location (red colour with asterisks) whereas that shown in blue colour with triangles represents evolution of total power in an unaffected region in the same sunspot. It is to be noted that there is a time offset of about $\pm$ 30-minutes between the acoustic power variation and the GOES flare-time.}
\end{figure*}

\begin{figure*}
\centering
\includegraphics[width=0.44\textwidth]{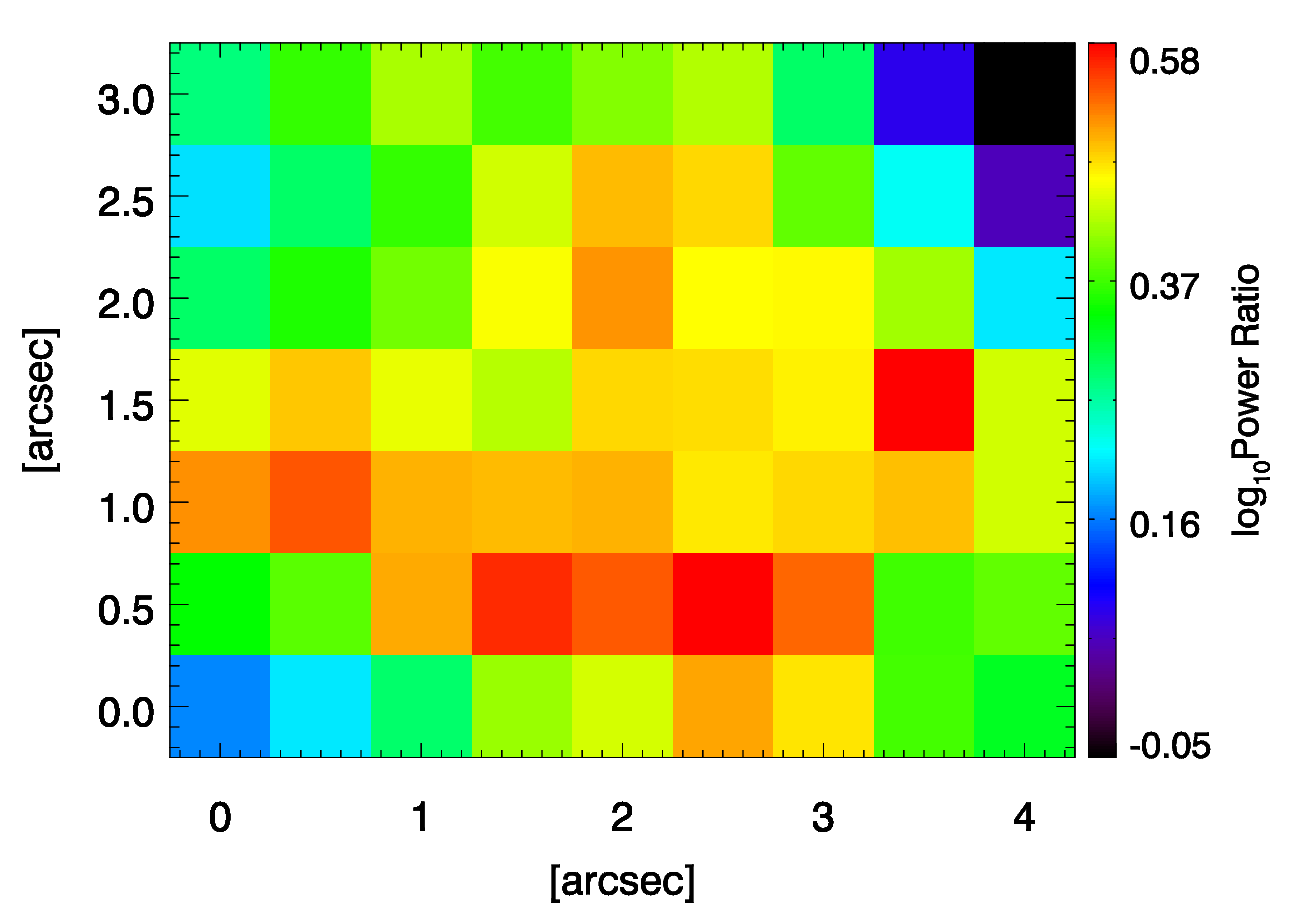}\hspace*{0.34cm}
\includegraphics[width=0.44\textwidth]{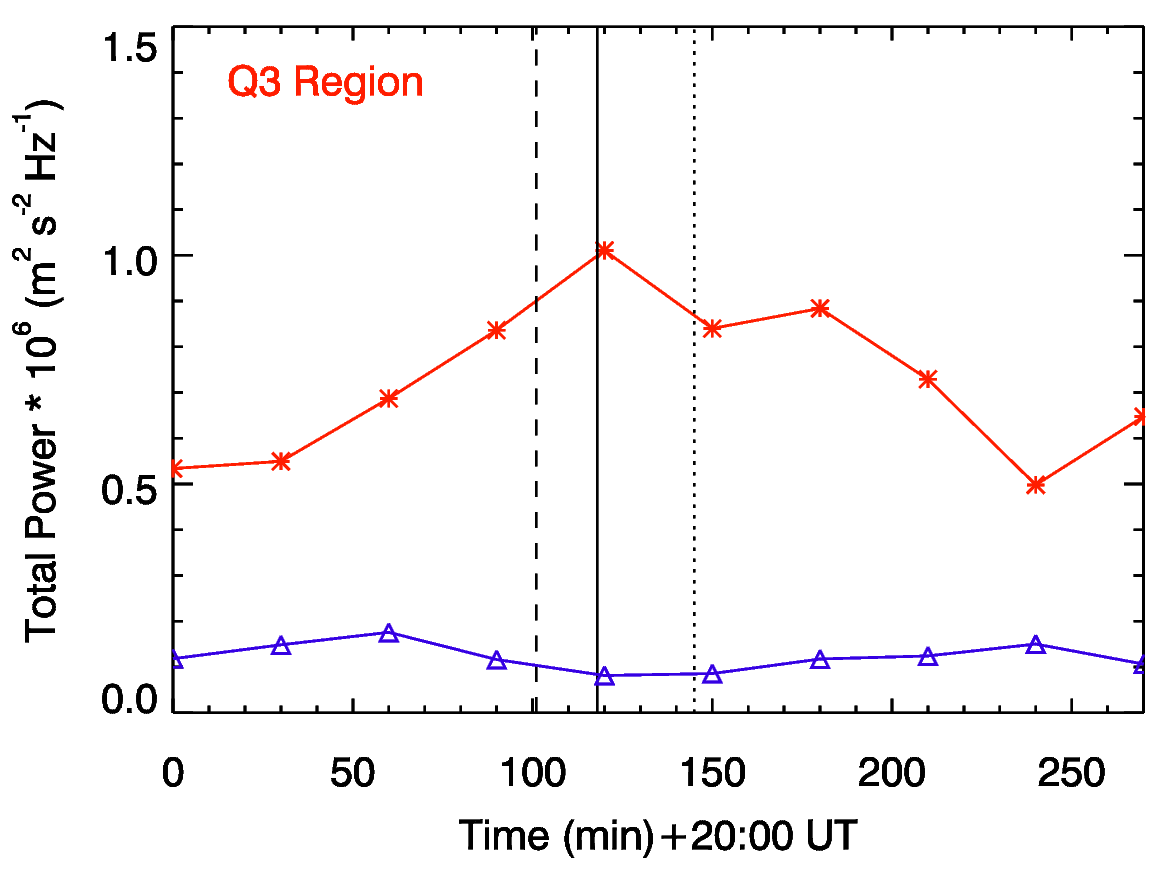}
\caption{Same as Figure 11, but for `Q3' location.}
\end{figure*}

\begin{figure*}
\centering
\includegraphics[width=0.44\textwidth]{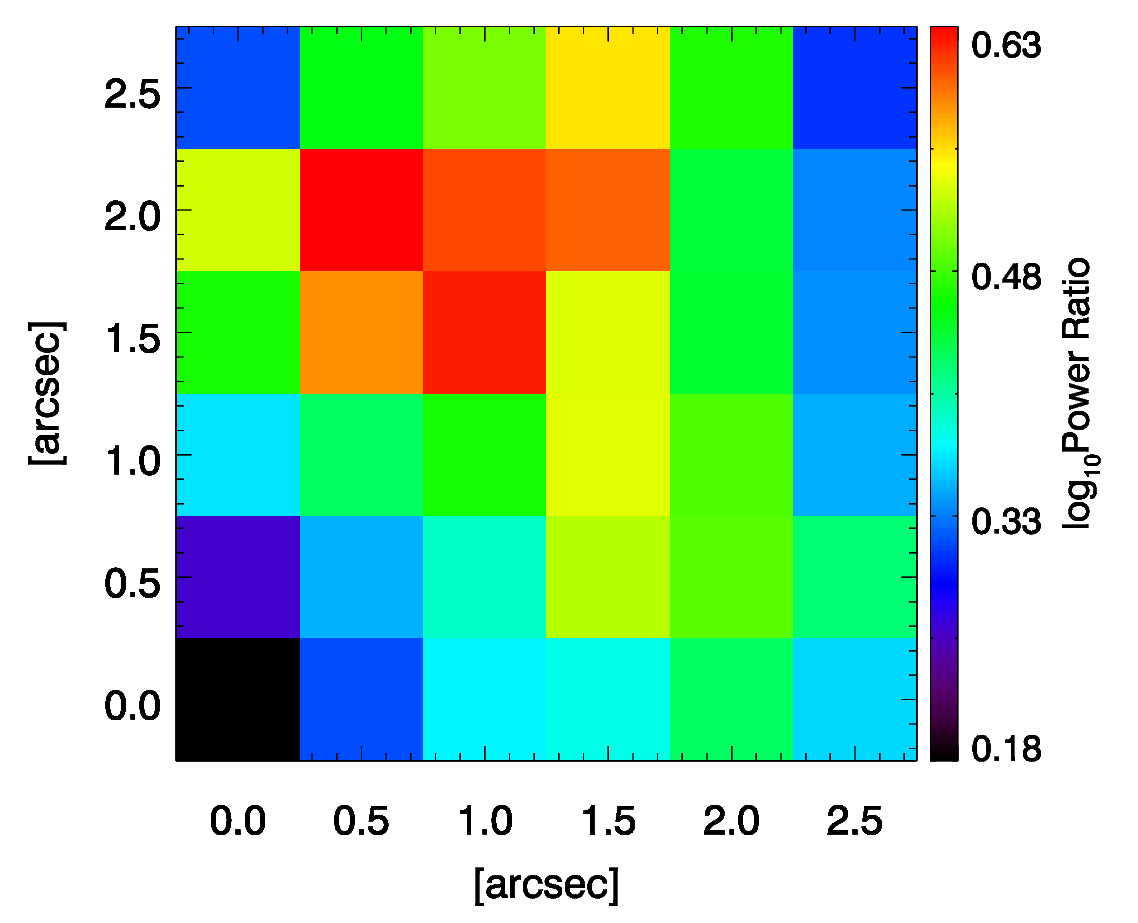}\hspace*{0.34cm}
\includegraphics[width=0.44\textwidth]{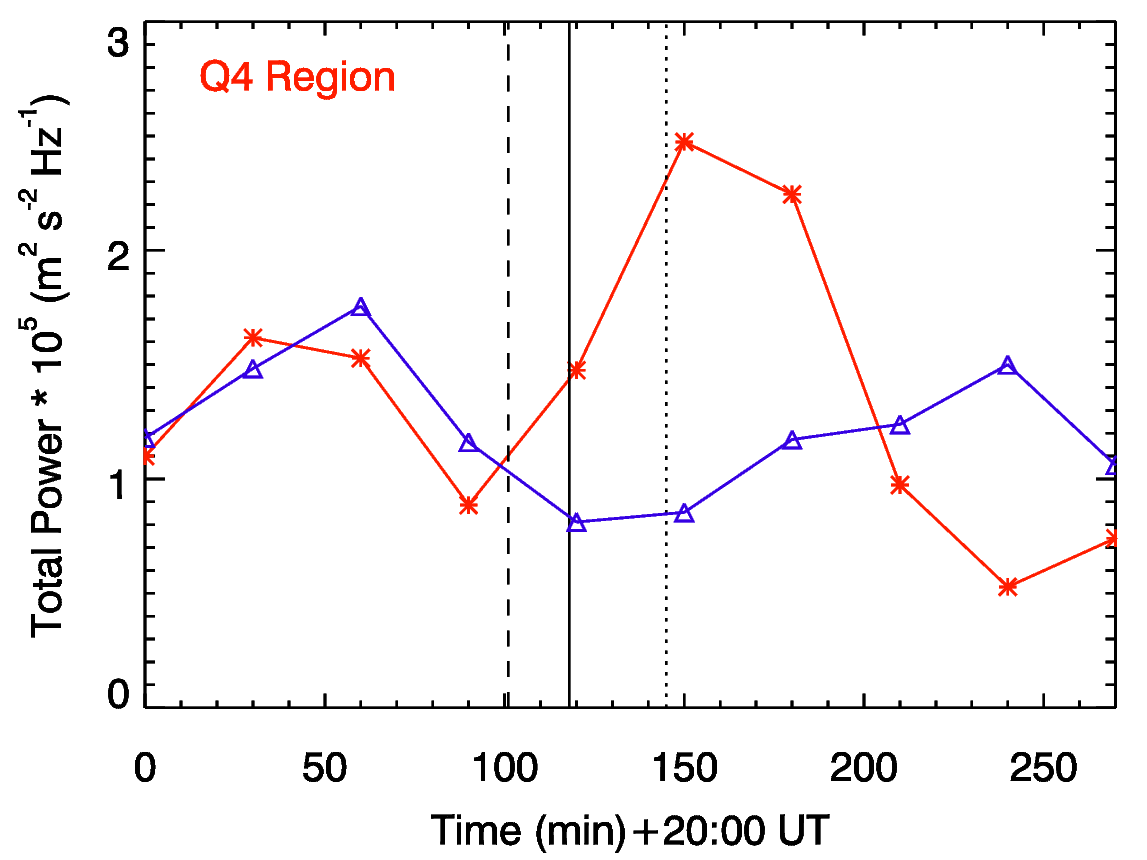}
\caption{Same as Figure 11, but for `Q4' location.}
\end{figure*}

\begin{figure*}
\centering
\includegraphics[width=0.44\textwidth]{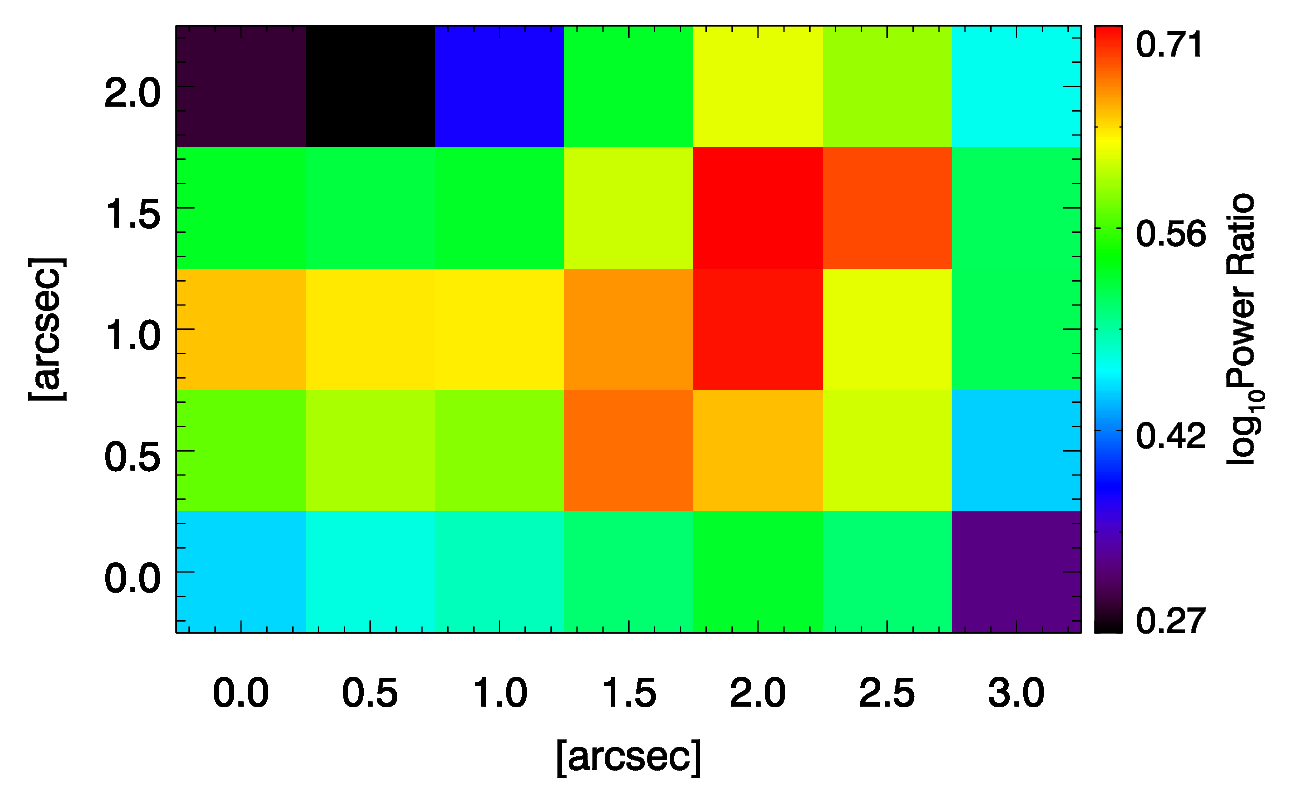}\hspace*{0.34cm}
\includegraphics[width=0.44\textwidth]{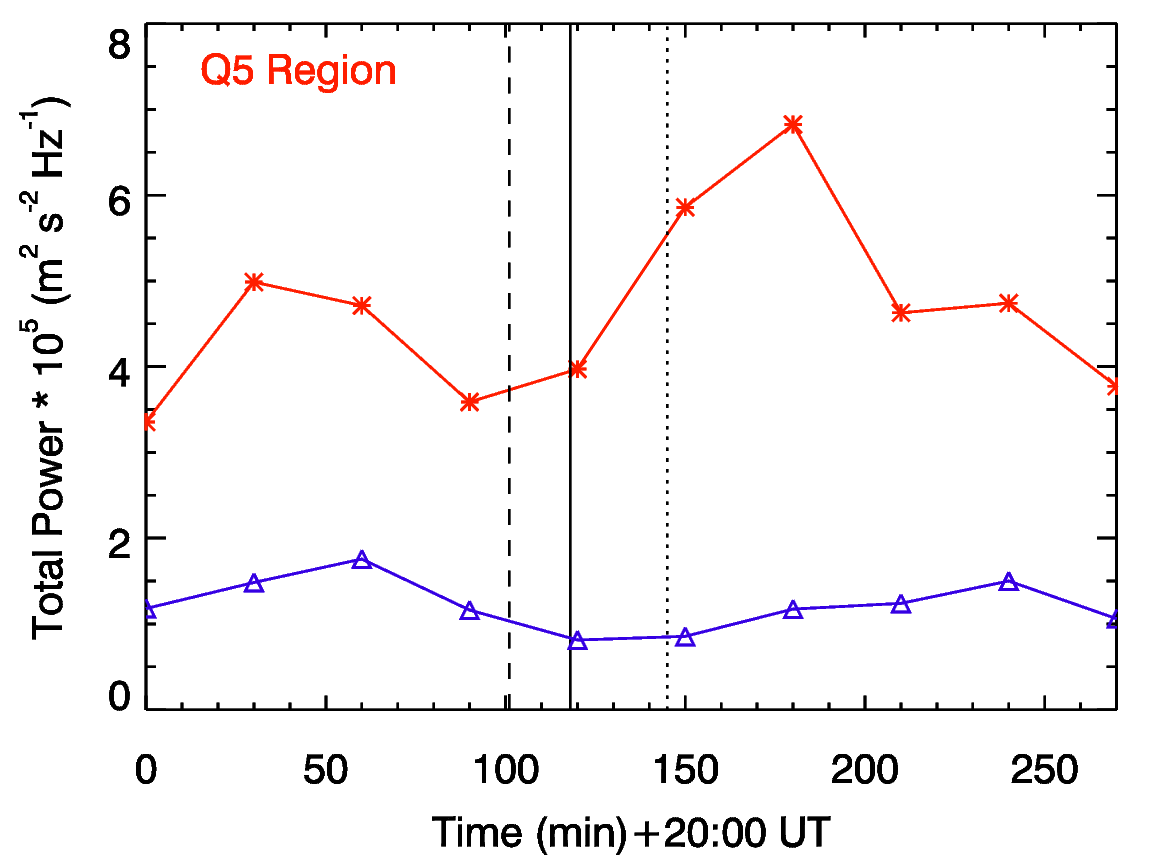}
\caption{Same as Figure 11, but for `Q5' location.}
\end{figure*}

\begin{figure*}
\centering
\includegraphics[width=0.44\textwidth]{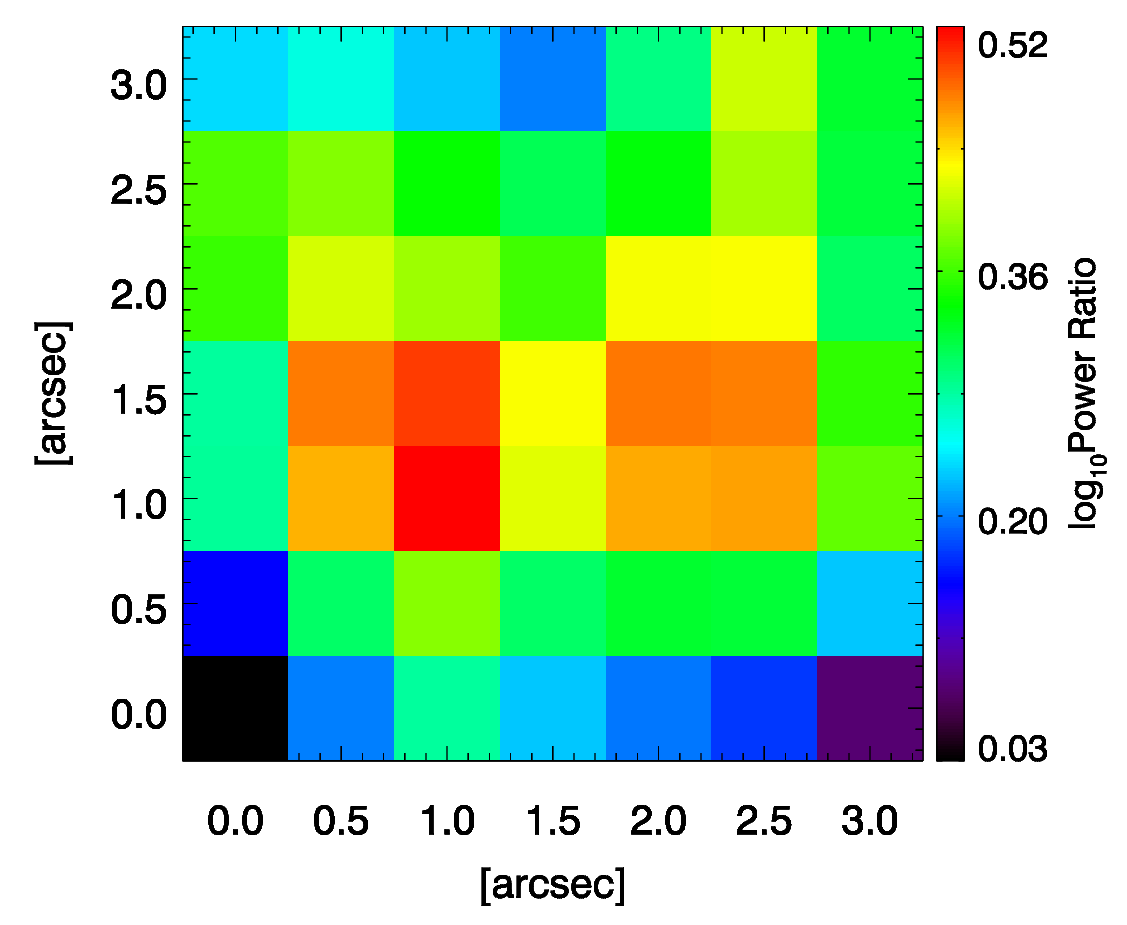}\hspace*{0.34cm}
\includegraphics[width=0.44\textwidth]{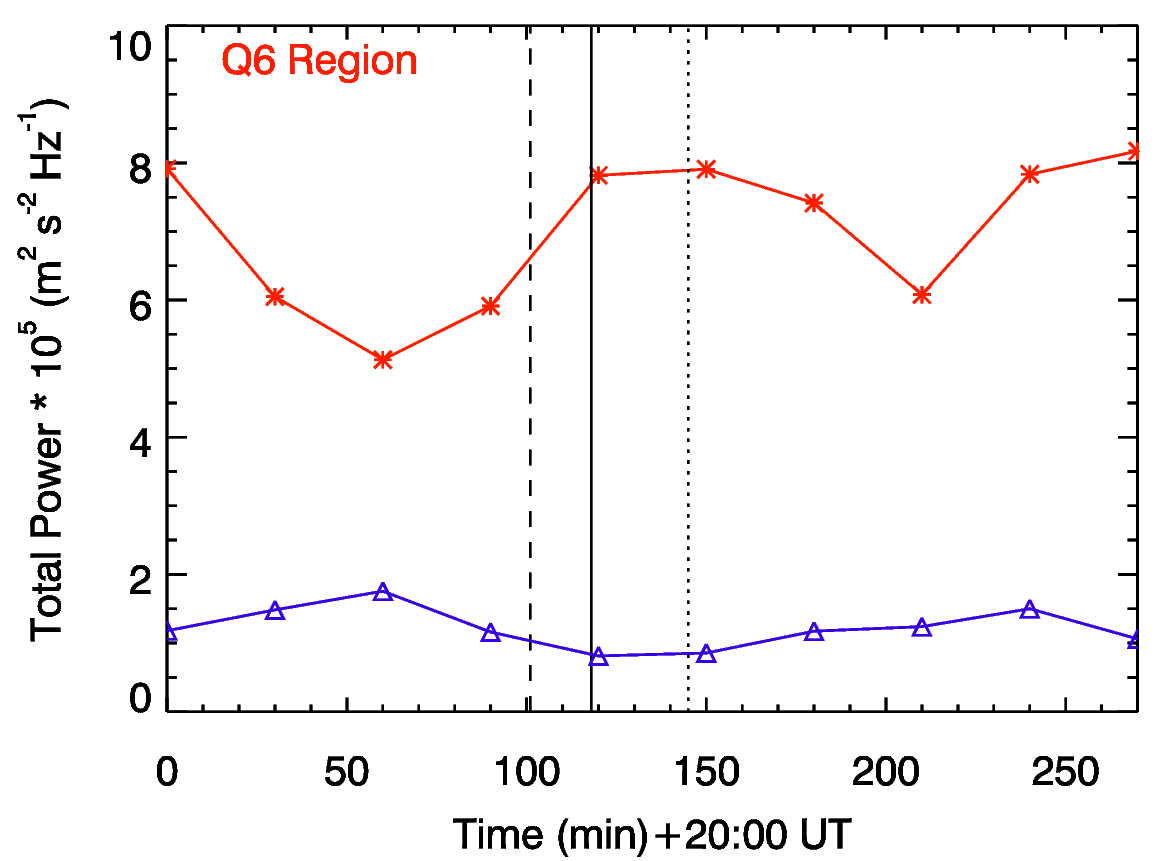}
\caption{Same as Figure 11, but for `Q6' location.}
\end{figure*}

\begin{figure*}
\centering
\includegraphics[width = 0.44\textwidth]{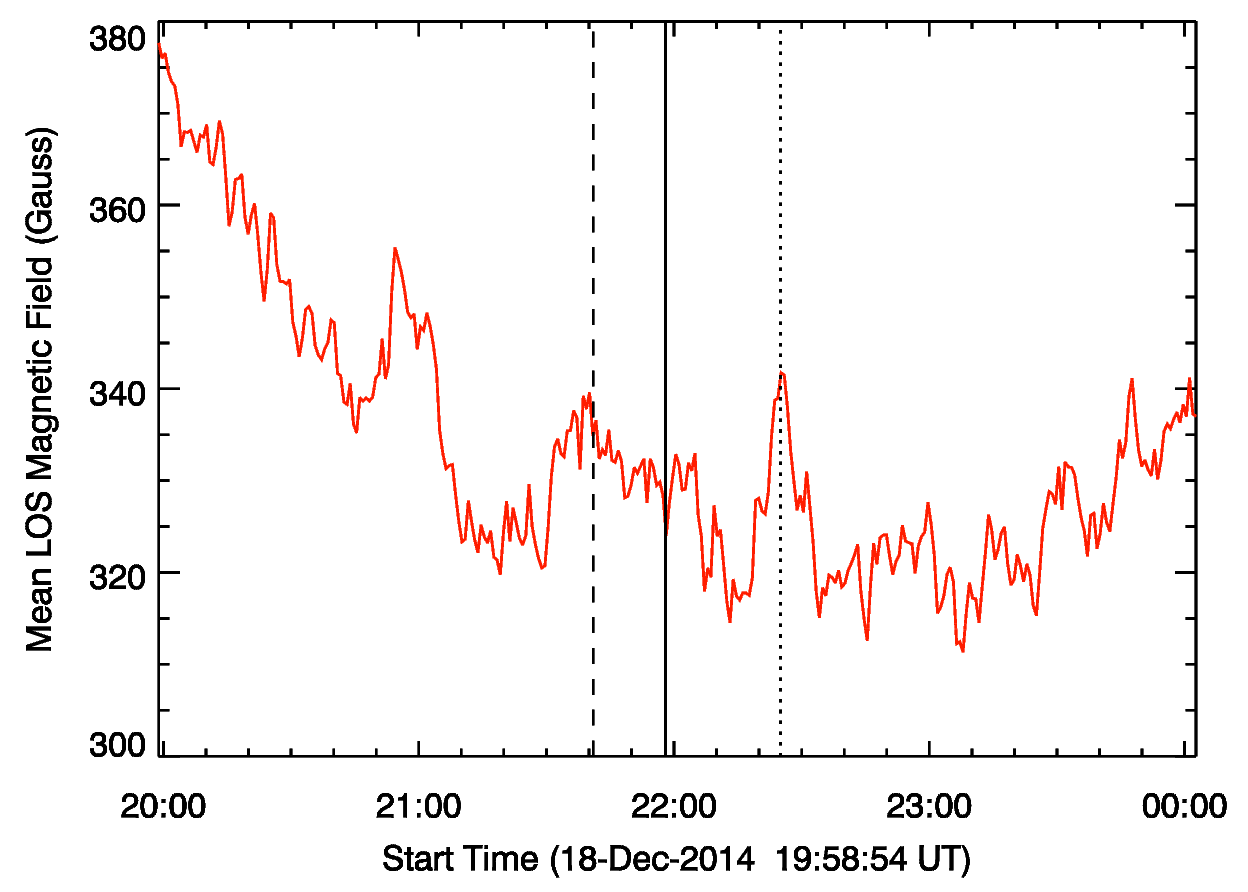}\hspace*{0.34cm}
\includegraphics[width = 0.44\textwidth]{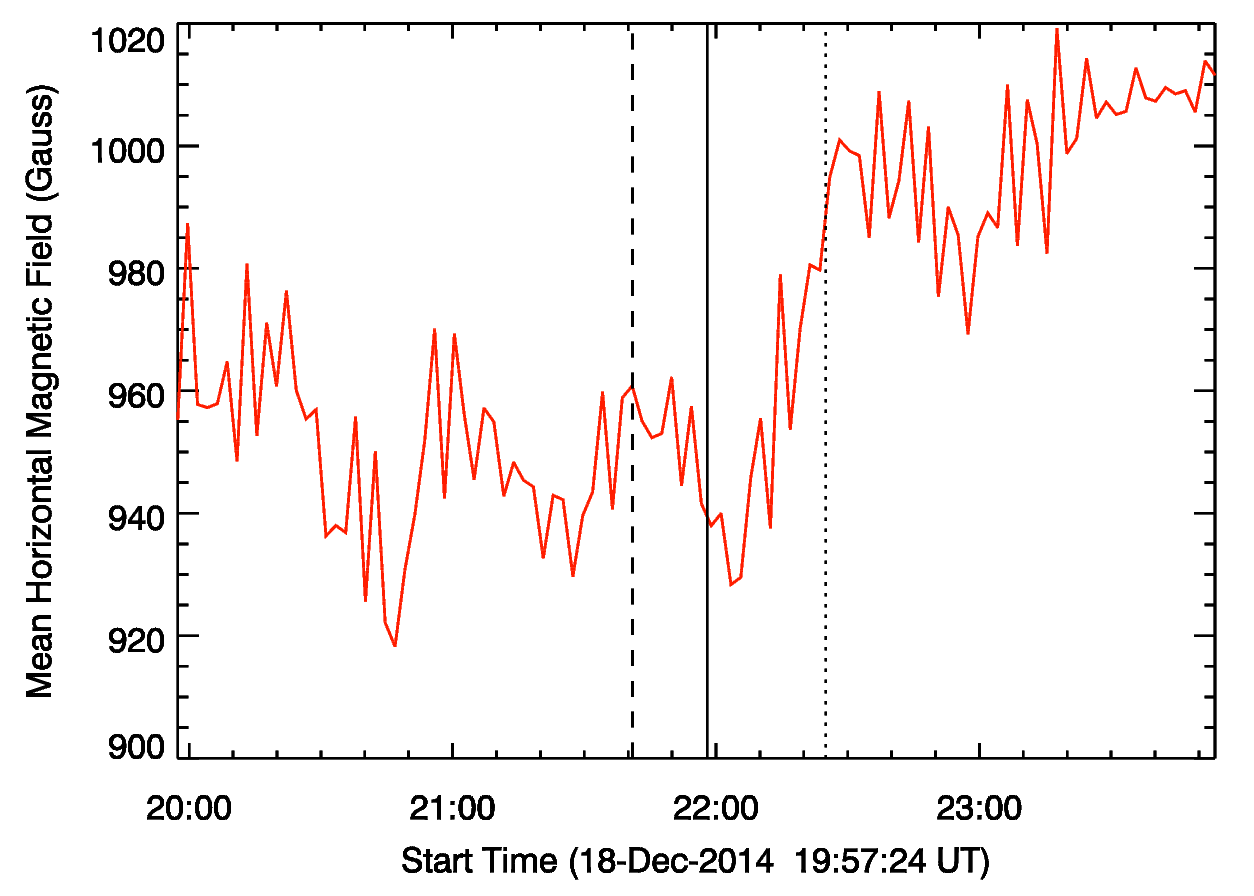}
\caption{Plot showing the temporal evolution of line-of-sight and horizontal magnetic fields of `Q3' and `Q4' locations, respectively.}
\end{figure*}

\begin{figure*}
\centering
\includegraphics[width = 0.44\textwidth]{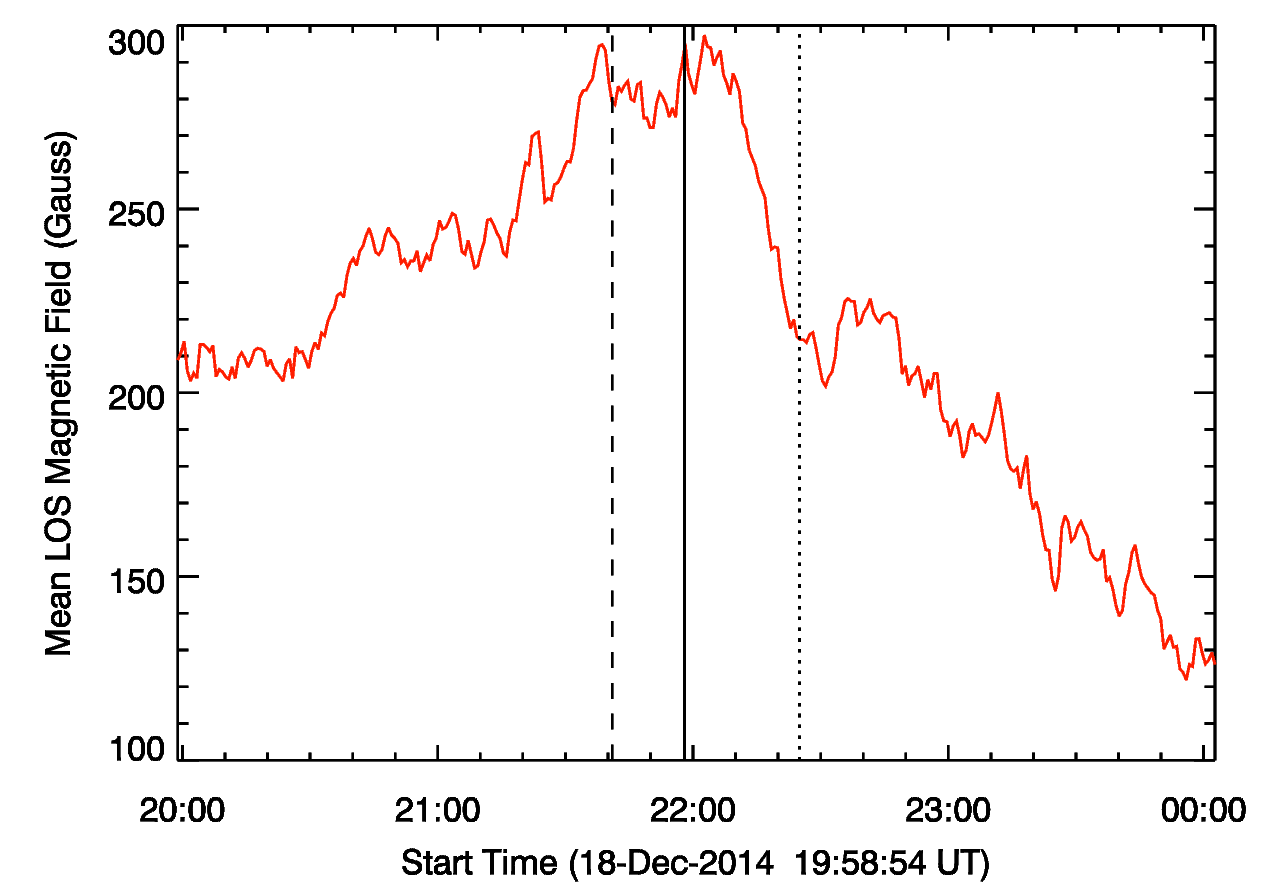}\hspace*{0.34cm}
\includegraphics[width = 0.44\textwidth]{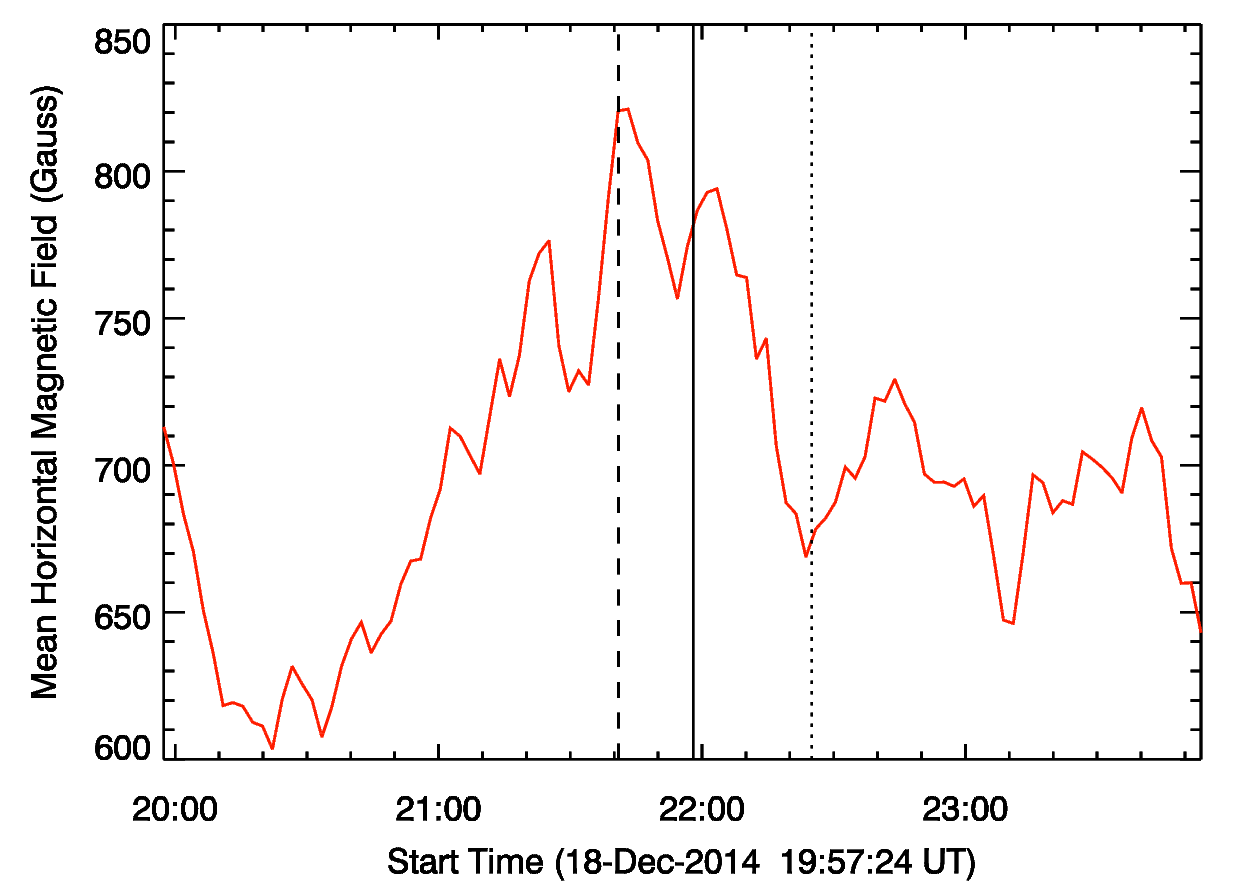}
\caption{Plot showing the temporal evolution of line-of-sight and horizontal magnetic fields of `Q5' and `Q6' locations, respectively.}
\end{figure*}

\begin{figure*}
\centering
\includegraphics[width=0.47\textwidth]{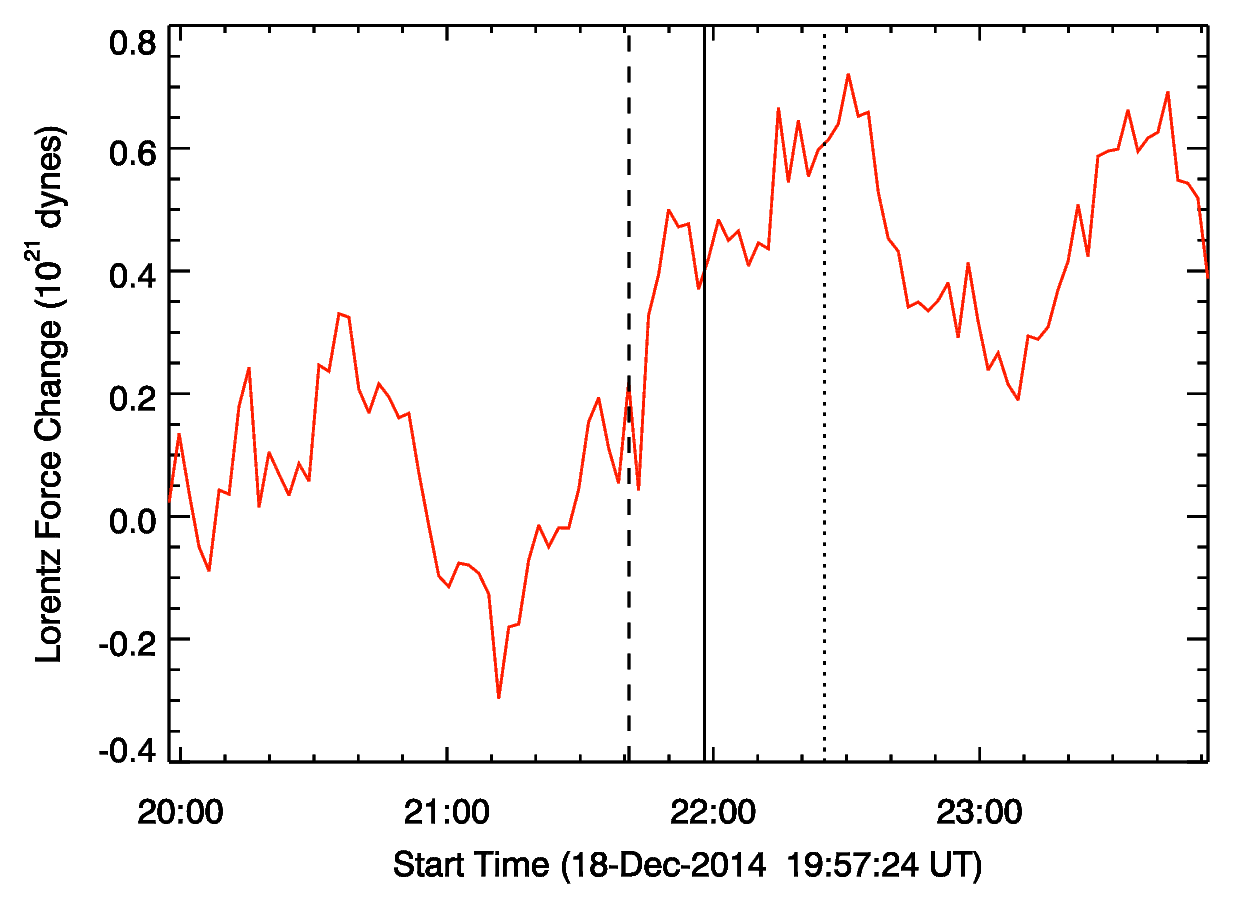}\hspace*{0.34cm}
\includegraphics[width=0.44\textwidth]{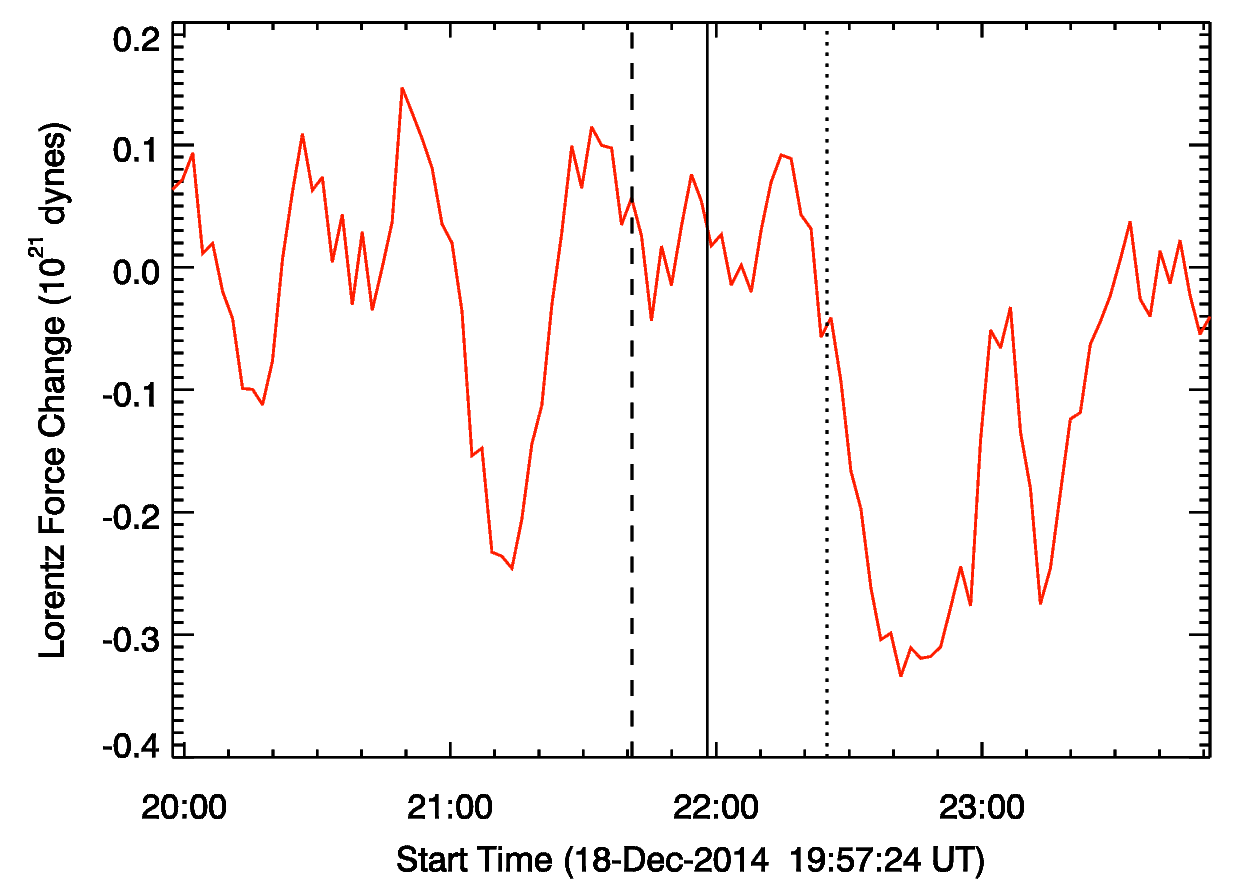}
\caption{\textit{Left panel}: Plot showing the temporal evolution of change in radial component of Lorentz force in the `Q2' location of active region NOAA 12241. 
\textit{Right panel}: Same as shown in the left panel but for `Q3' location. The dashed, solid and dotted vertical lines represent the onset, peak and decay time of the flare.}
\end{figure*}

\begin{figure*}
\centering
\includegraphics[width=0.44\textwidth]{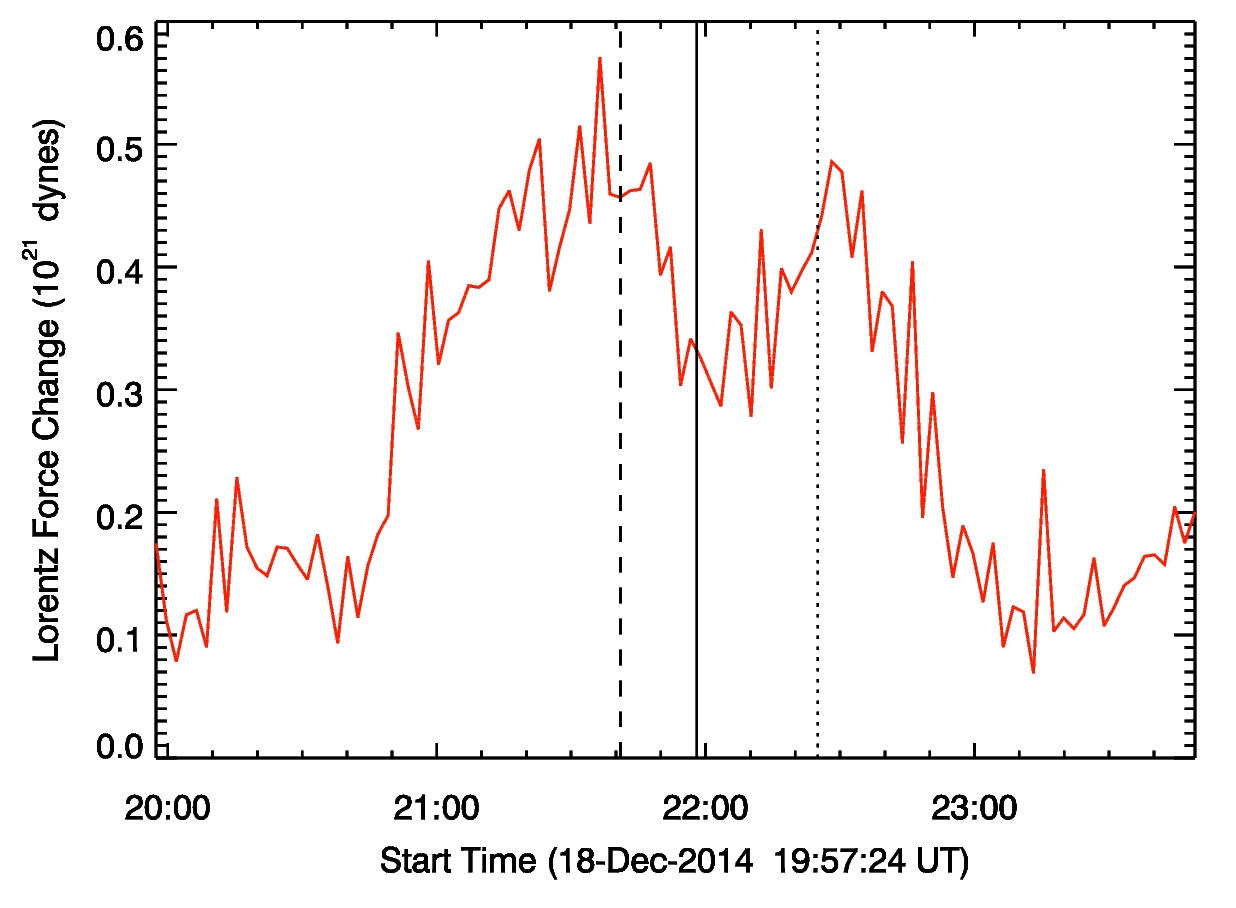}\hspace*{0.34cm}
\includegraphics[width=0.44\textwidth]{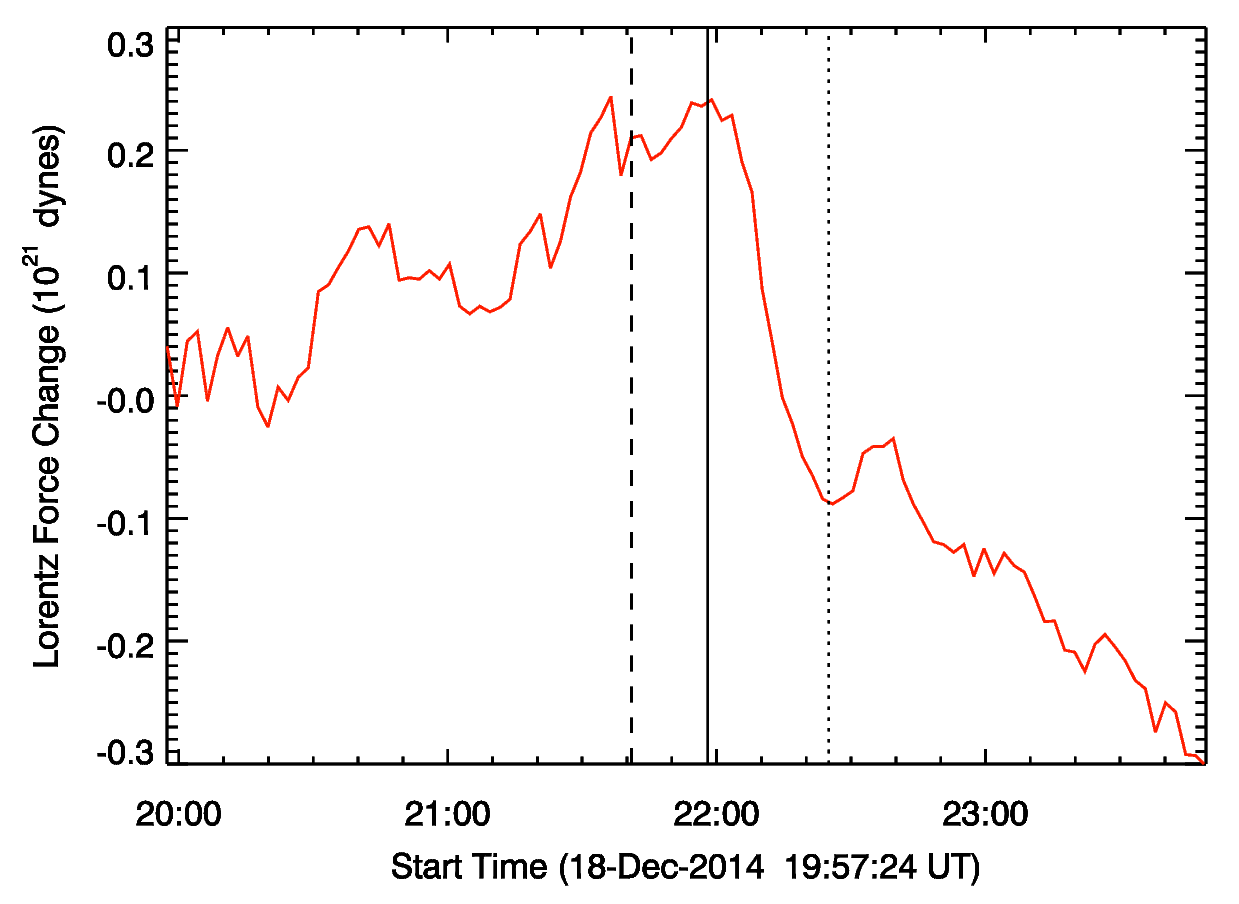}
\caption{\textit{Left panel}: Plot showing the temporal evolution of change in horizontal component of Lorentz force in the `Q4' location of active region NOAA 12241. \textit{Right panel}: Same as shown in the left panel but for `Q5' location. The dashed, solid and dotted vertical lines represent onset, peak and decay time of the flare.}
\end{figure*}

\begin{figure*}
\centering
\includegraphics[width=0.44\textwidth]{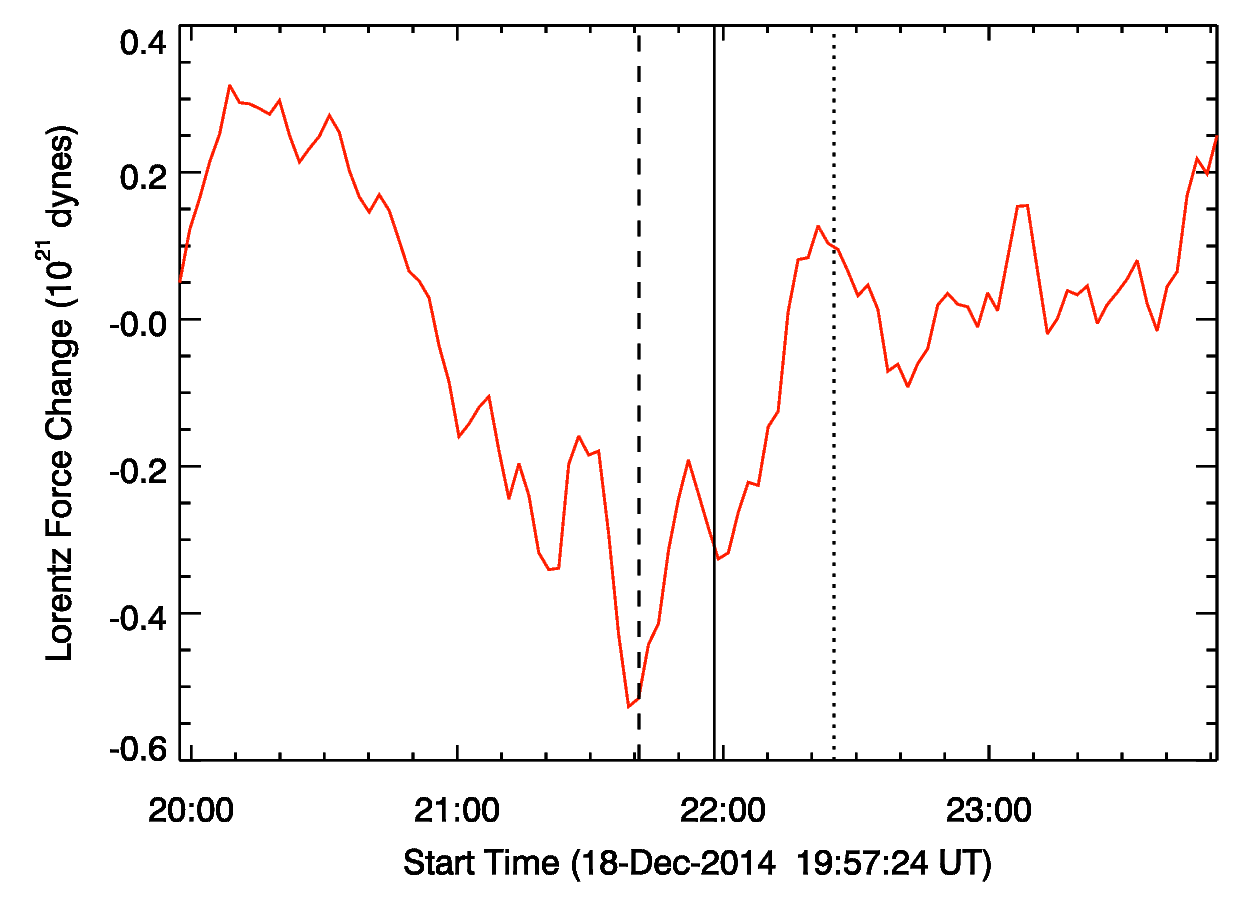}
\caption{Plot showing the temporal evolution of change in radial component of Lorentz force in the `Q6' location of active region NOAA 12241. The dashed, solid and dotted vertical lines represent onset, peak and decay time of the flare.}
\end{figure*}
\break


\hspace{4cm} [E] \bf{Active region NOAA 12242}

\begin{figure*}[h!]
\centering
\includegraphics[width=0.48\textwidth]{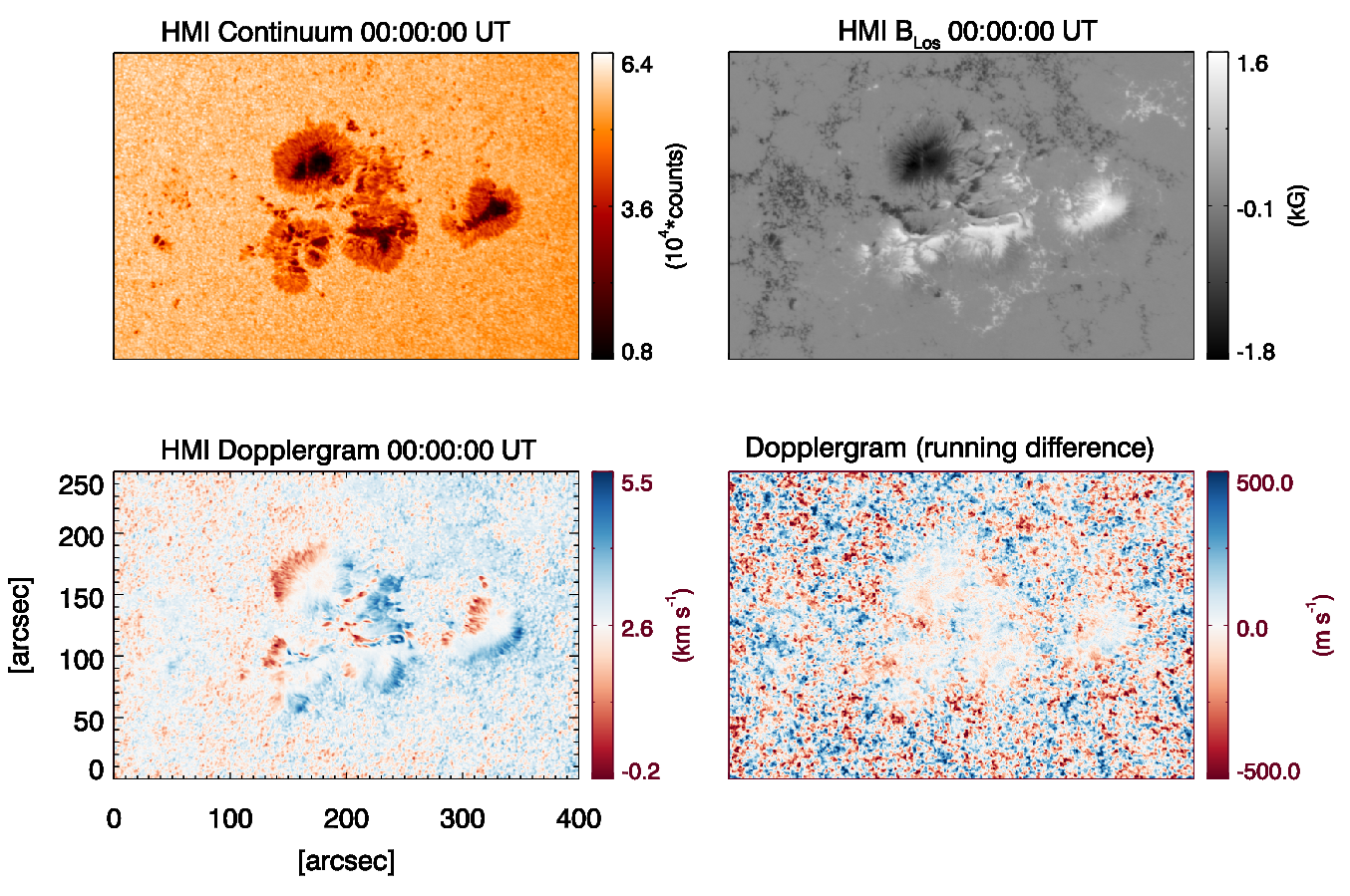}\hspace*{0.34cm}
\includegraphics[width=0.44\textwidth]{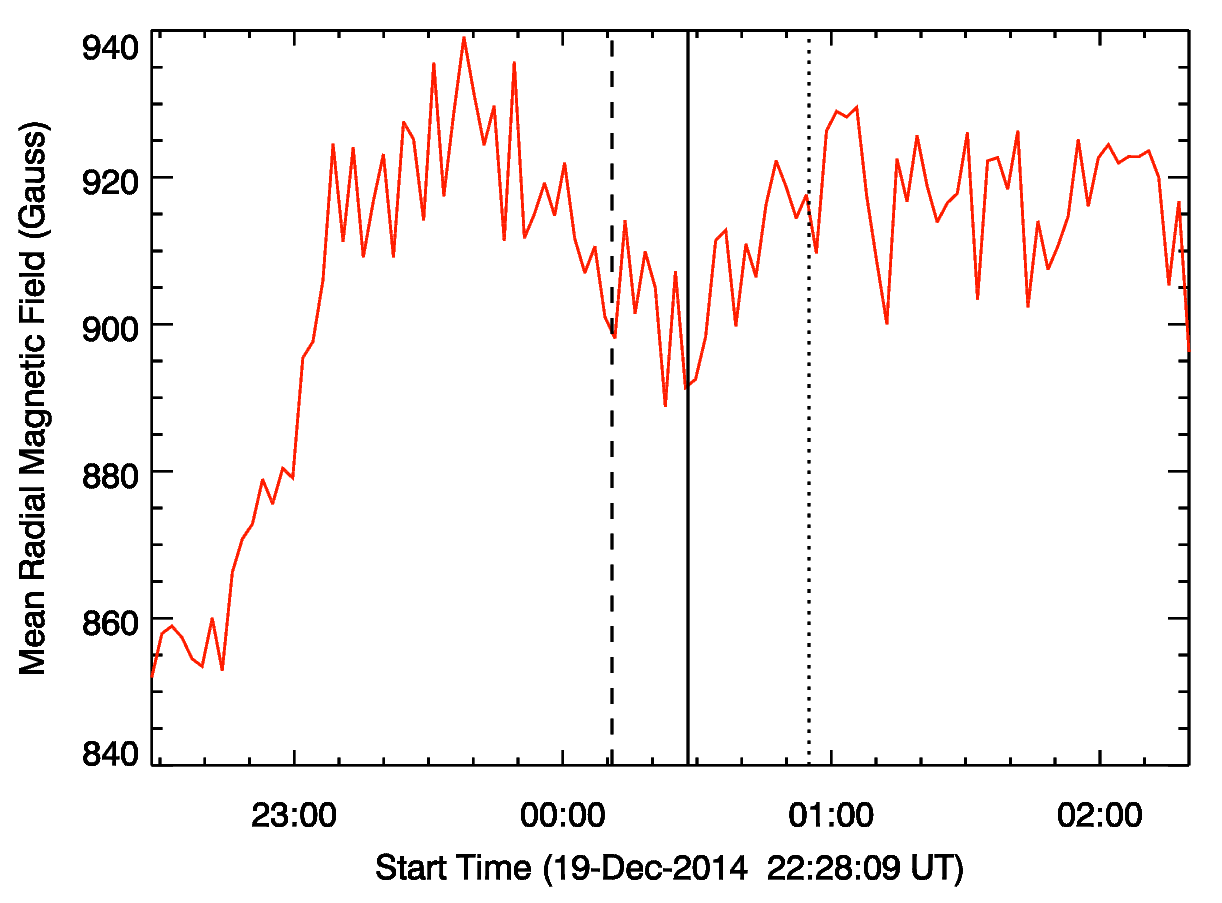}
\caption{\textit{Left panel}: Sample images of the active region NOAA 12242 showing continuum intensity (top left panel), photospheric line-of-sight magnetic fields (top right panel), Dopplergram (bottom left panel) and running difference of Doppler images (bottom right panel) acquired from HMI instrument aboard the {\em SDO} spacecraft on 2014 December 20. \textit{Right panel}: Plot shows the temporal evolution of radial magnetic fields in the `K1' location of active region NOAA 12242. The dashed, solid and dotted vertical lines represent the onset, peak and decay time of the flare.}
\label{fig: fourimage12242}
\end{figure*}

\begin{figure*}[h!]
\centering
\includegraphics[width=0.44\textwidth]{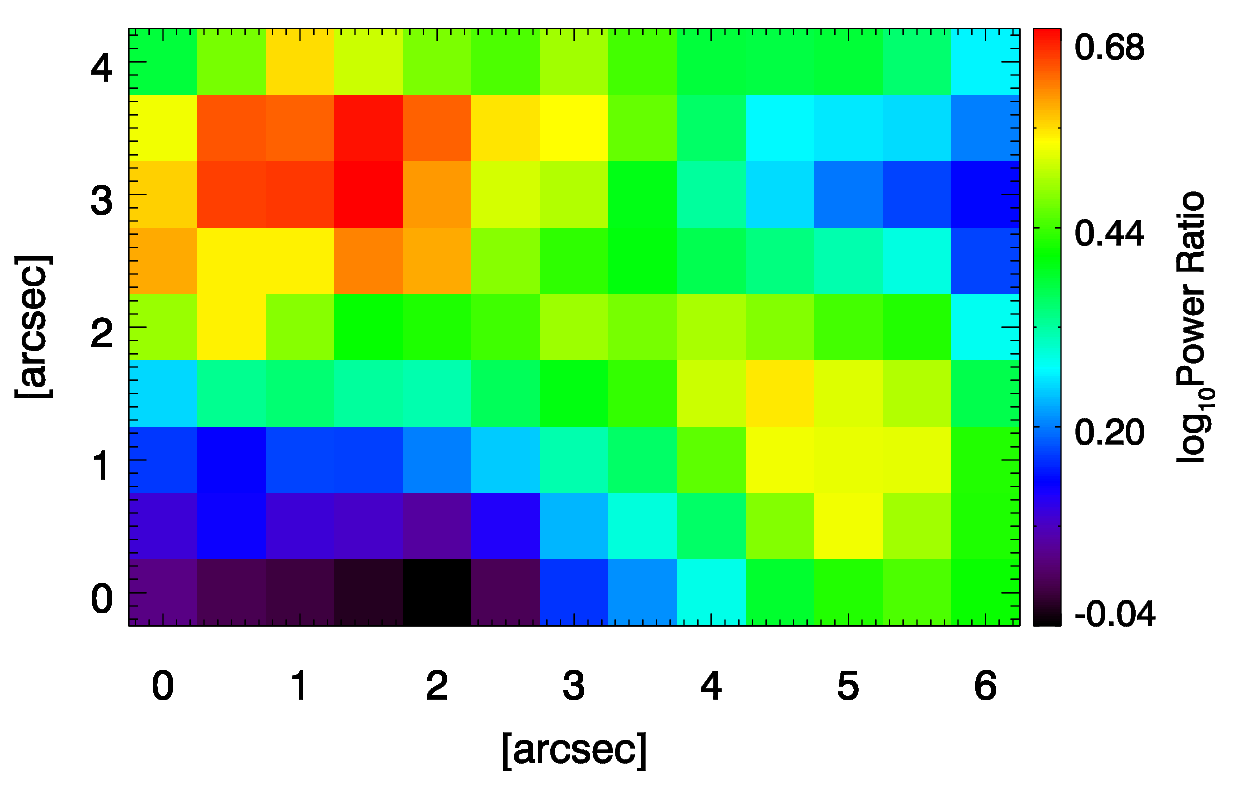}\hspace*{0.34cm}
\includegraphics[width=0.44\textwidth]{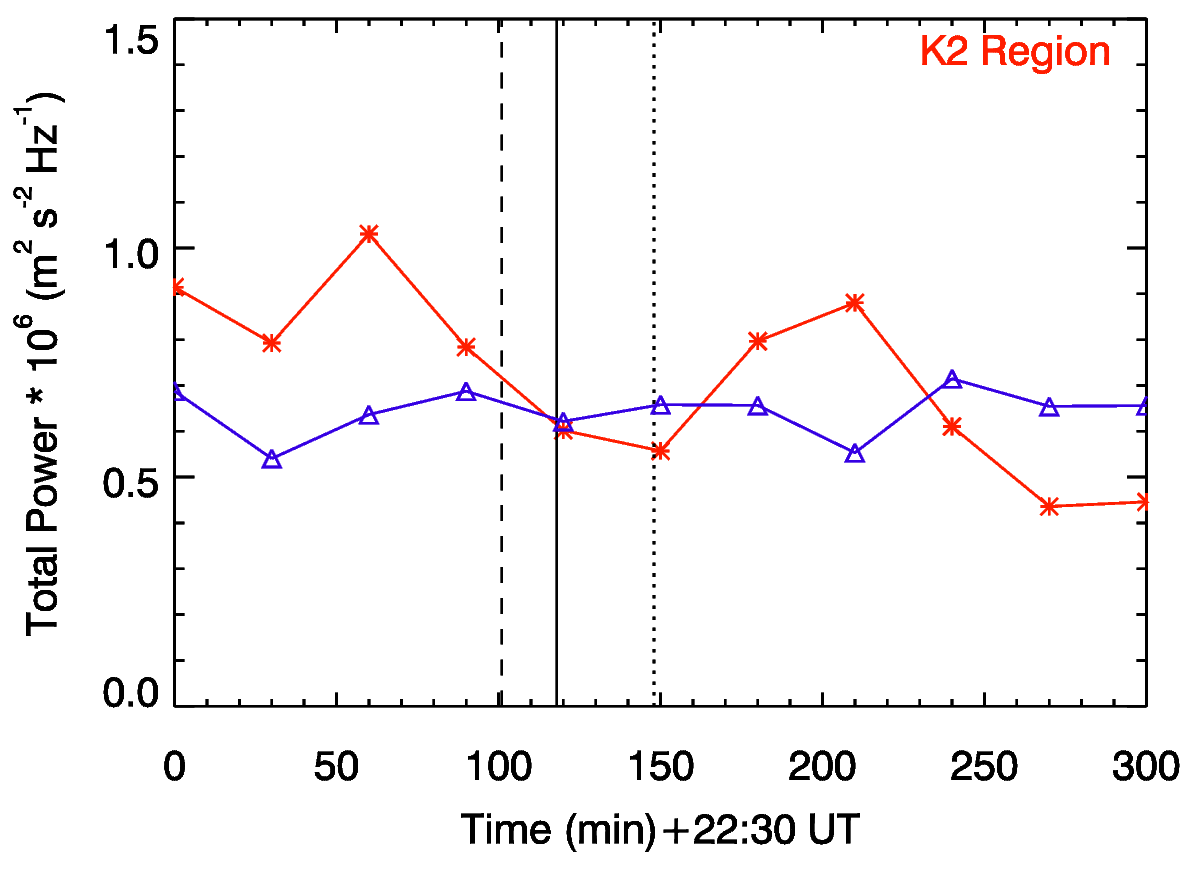}
\caption{\textit{Left panel}: Illustrates the blow-up region of `K2' enhanced location as indicated in the power map ratio in Figure 11 of the main manuscript. \textit{Right panel}: Plots showing the temporal evolution of integrated acoustic power over the ‘K2’ location (red colour with asterisks) whereas that shown in blue colour with triangles represents evolution of total acoustic power in an unaffected region in the same sunspot. It is to be noted that there is a time offset of about $\pm$ 30-minutes between the acoustic power variation and the GOES flare-time.}
\label{fig: prespanning12242}
\end{figure*}

\begin{figure*}[h!]
\centering
\includegraphics[width=0.44\textwidth]{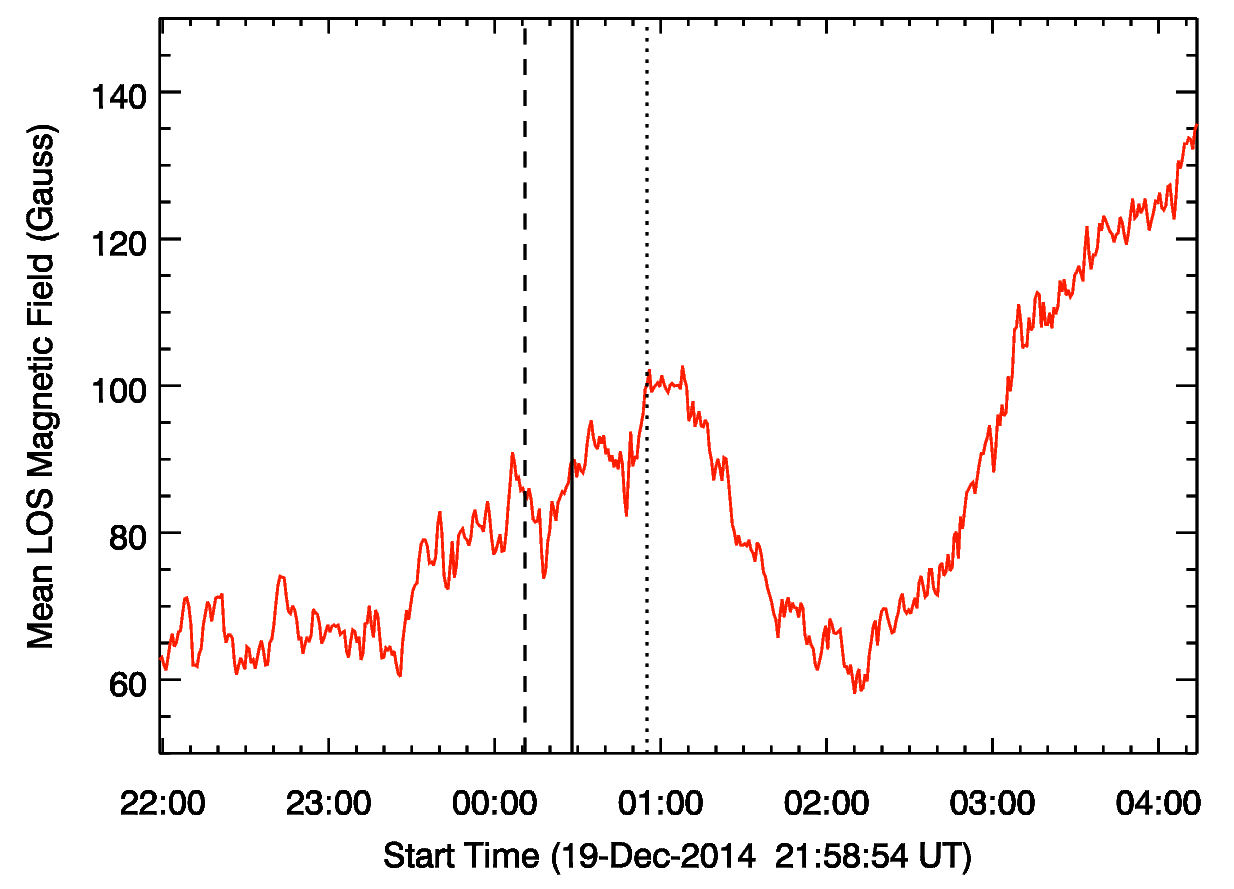}\hspace*{0.34cm}
\includegraphics[width=0.44\textwidth]{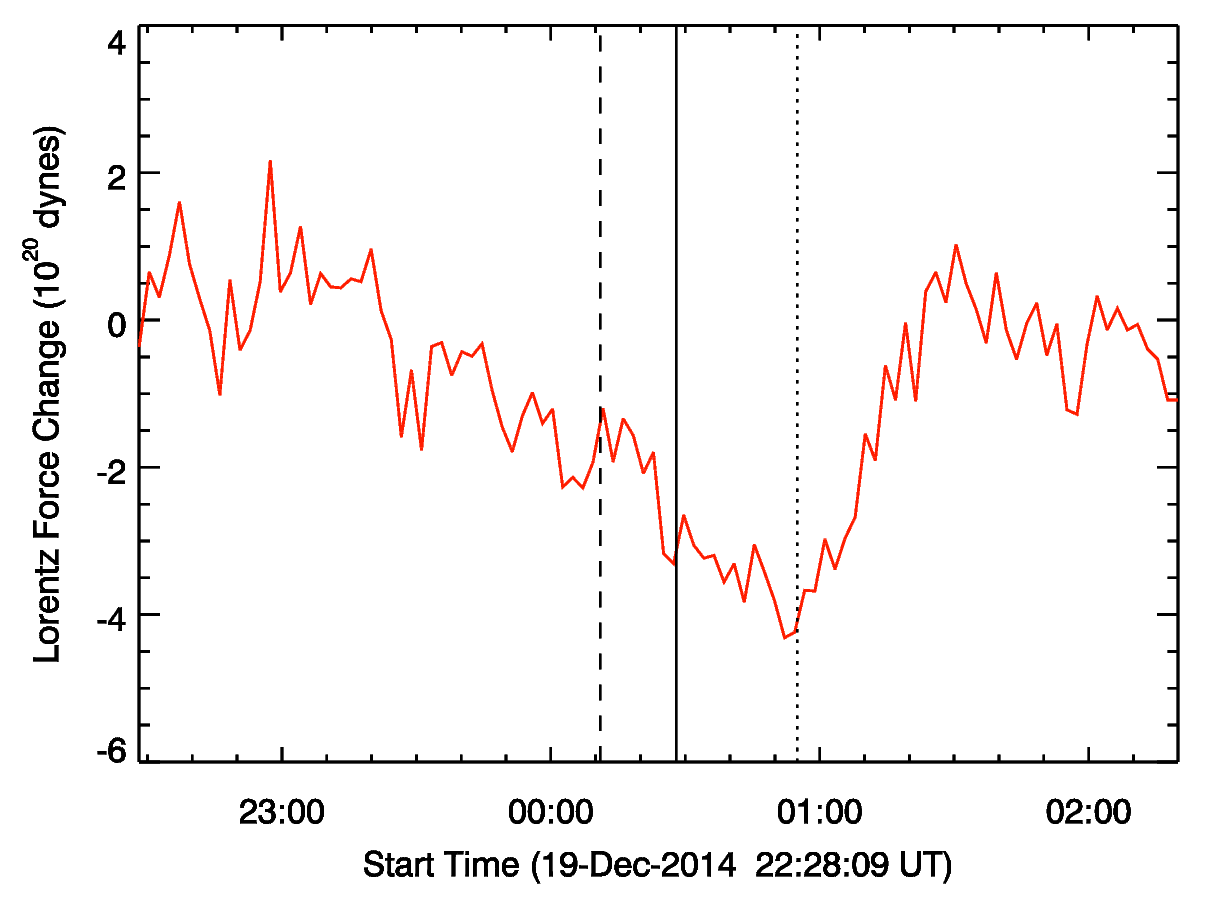}
\caption{\textit{Left panel}: Plot shows the temporal evolution of line-of-sight magnetic fields of `K2' location. \textit{Right panel}: Plot shows the temporal evolution of change in radial component of Lorentz force in the `K2' location. The dashed, solid and dotted vertical lines represent the onset, peak and decay time of the flare.}
\end{figure*}

\clearpage

\hspace{4cm} [F] \bf{Active region NOAA 12297}

\begin{figure*}[h!]
\centering
\includegraphics[width=0.44\textwidth]{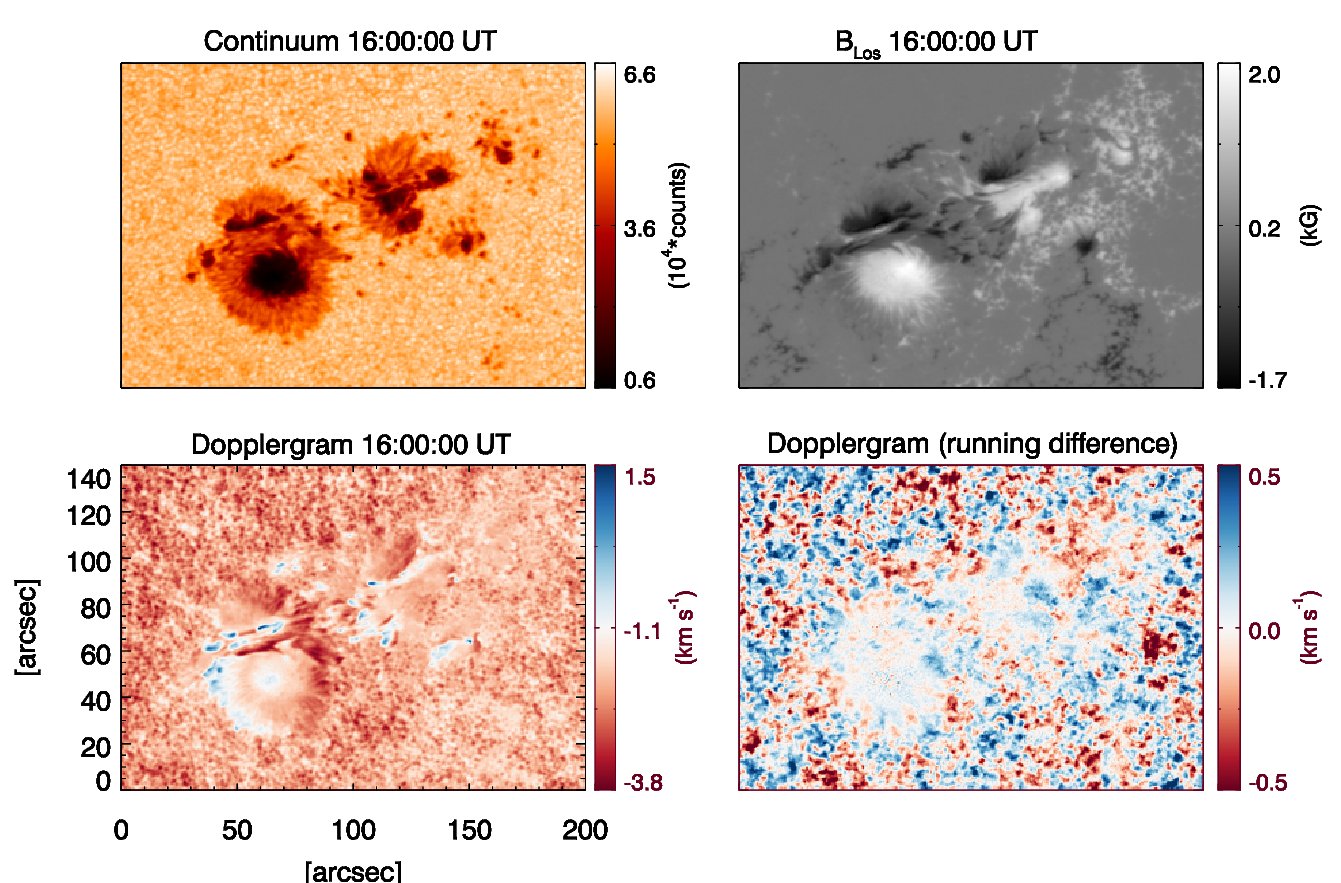}\hspace*{0.34cm}
\includegraphics[width=0.44\textwidth]{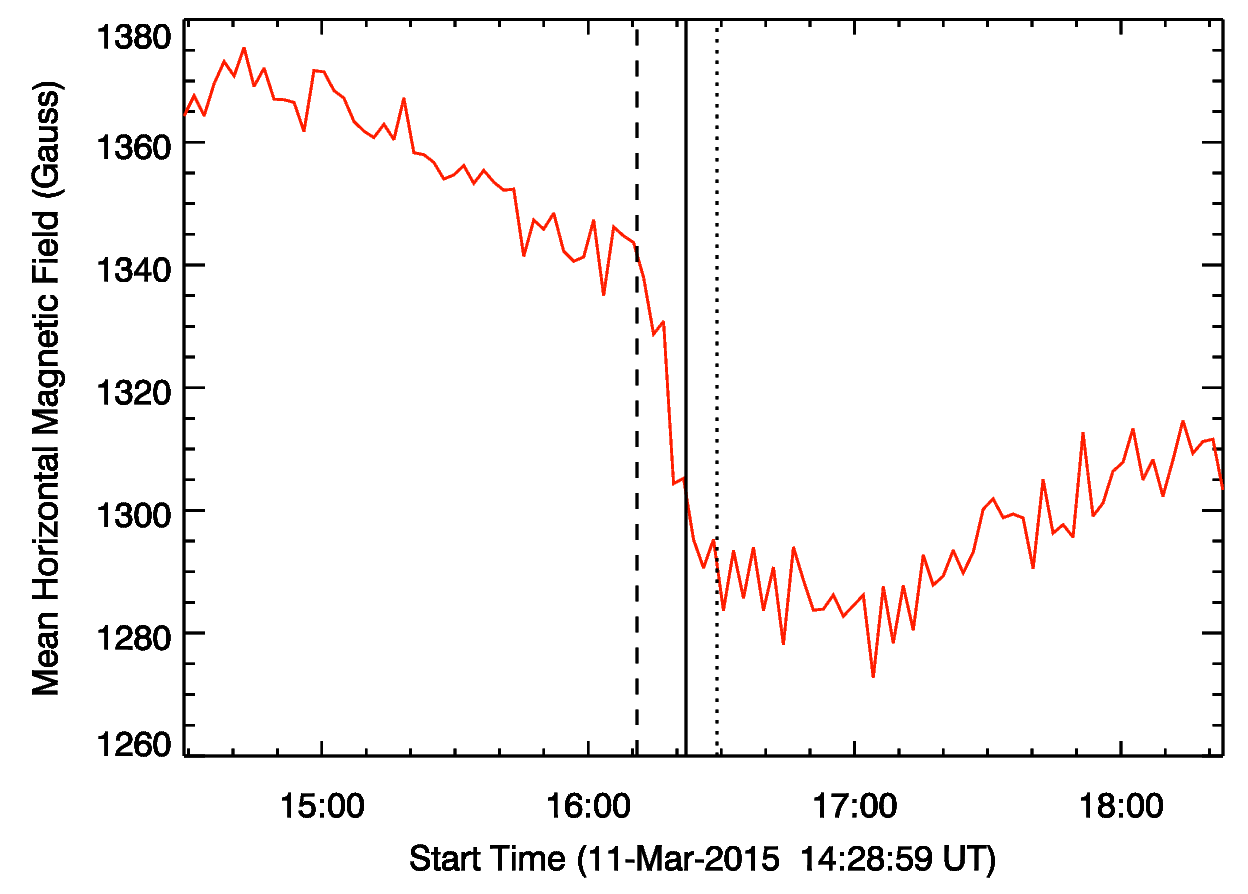}
\caption{\textit{Left panel}: Sample images of the active region NOAA 12297 showing continuum intensity (top left panel), photospheric line-of-sight magnetic fields (top right panel), Dopplergram (bottom left panel) and running difference of Doppler images (bottom right panel) acquired from HMI instrument aboard the {\em SDO} spacecraft on 2015 March 11. \textit{Right panel}: Plot shows the temporal evolution of horizontal magnetic fields in the `F2' location of active region NOAA 12297. The dashed, solid and dotted vertical lines represent the onset, peak and decay time of the flare.}
\label{fig: fourimage12297}
\end{figure*}

\begin{figure*}[h!]
\centering
\includegraphics[width=0.44\textwidth]{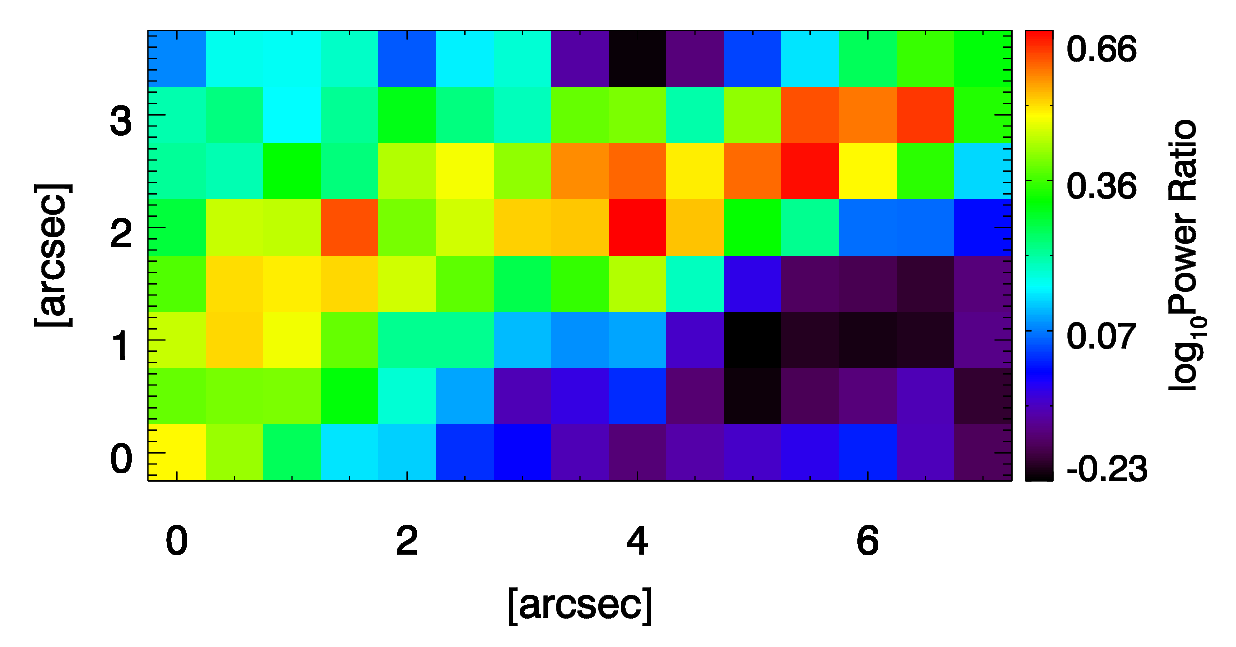}\hspace*{0.34cm}
\includegraphics[width=0.44\textwidth]{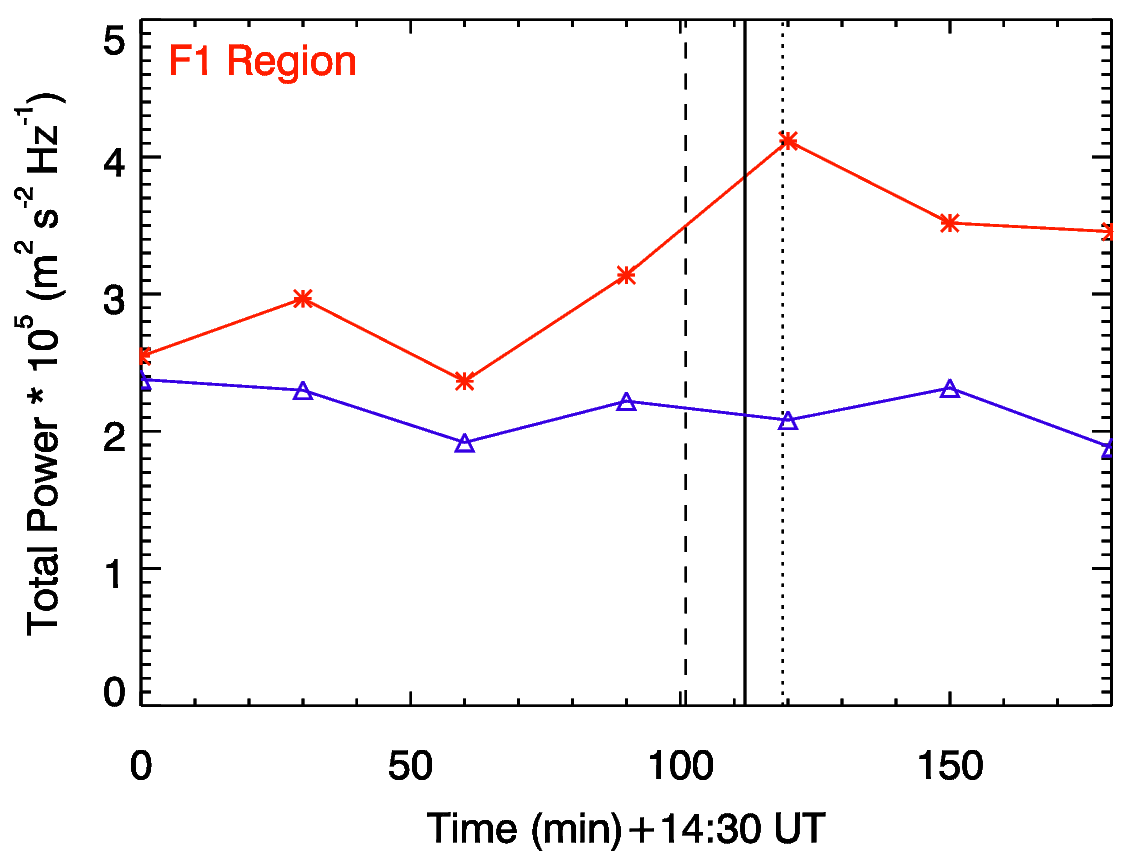}
\caption{\textit{Left panel}: Illustrates the blow-up region of `F1'  location as indicated in the power map ratio in Figure 12 of the main manuscript. \textit{Right panel}: Plot showing the temporal evolution of integrated acoustic power over the ‘F1’ location (red colour with asterisks) whereas that shown in blue colour with triangles represents evolution of total power in an unaffected region in the same sunspot. It is to be noted that there is a time offset of about $\pm$ 30-minutes between the acoustic power variation and the GOES flare-time.}
\label{fig: prespanning12297}
\end{figure*}

\begin{figure*}[h!]
\centering
\includegraphics[width=0.44\textwidth]{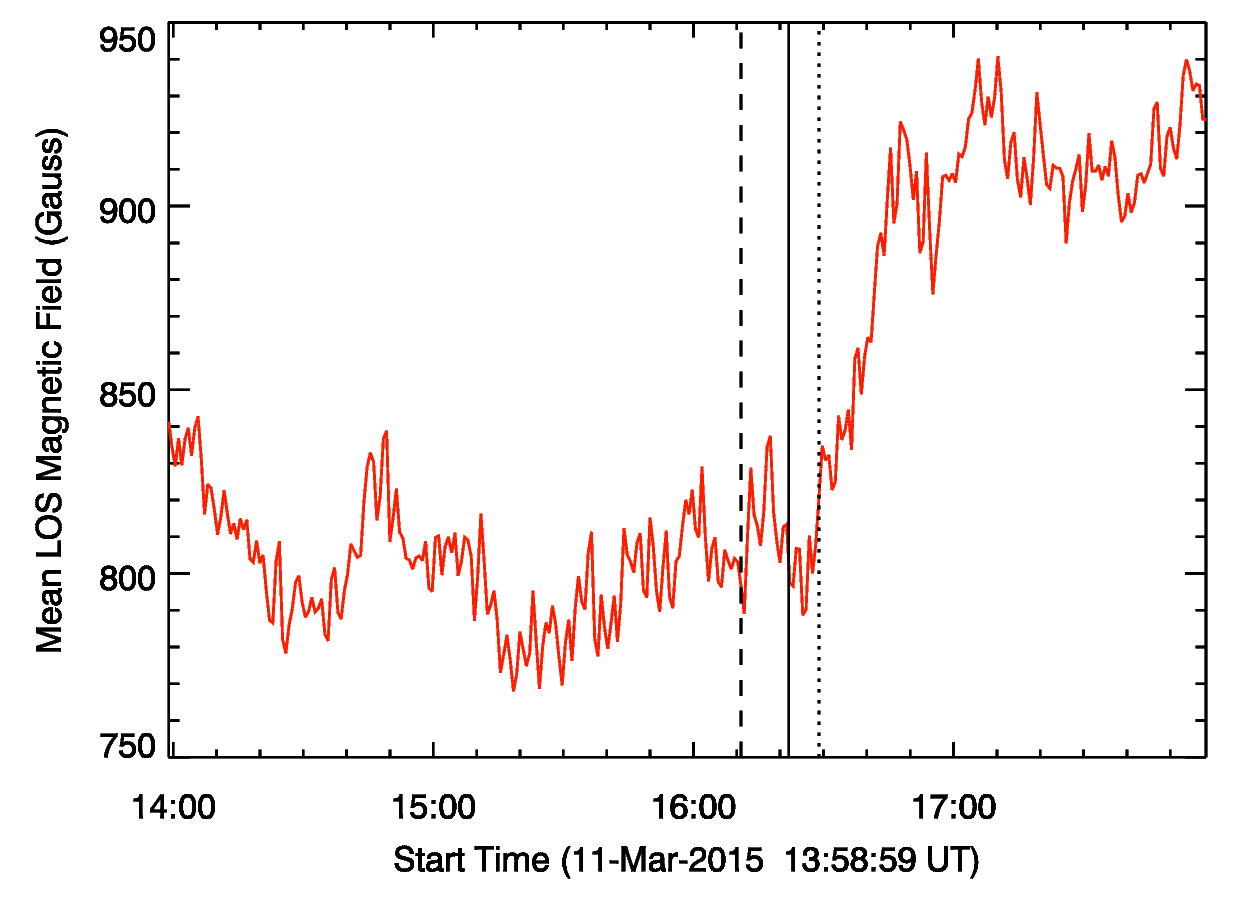}\hspace*{0.34cm}
\includegraphics[width=0.44\textwidth]{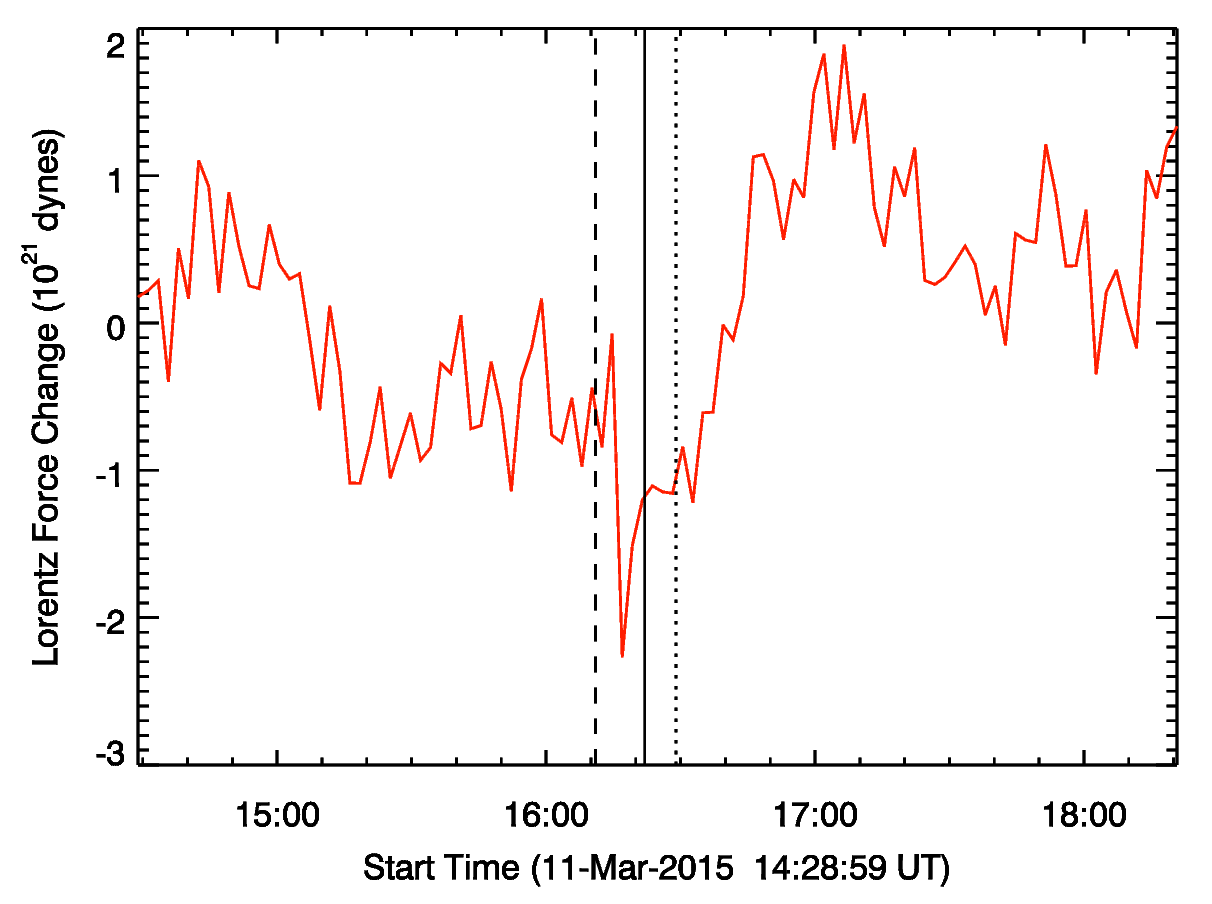}
\caption{\textit{Left panel}: Plot shows the temporal evolution of line-of-sight magnetic field in the `F1' location of active region NOAA 12297. \textit{Right panel}: Plot shows the temporal evolution of change in radial component of Lorentz force in the `F1' location. The dashed, solid and  dotted vertical lines represent onset, peak and decay time of the flare.}
\end{figure*}

\break


\hspace{4cm} [G] \bf{Active region NOAA 12371}

\begin{figure*}[h!]
\centering
\includegraphics[width=0.47\textwidth]{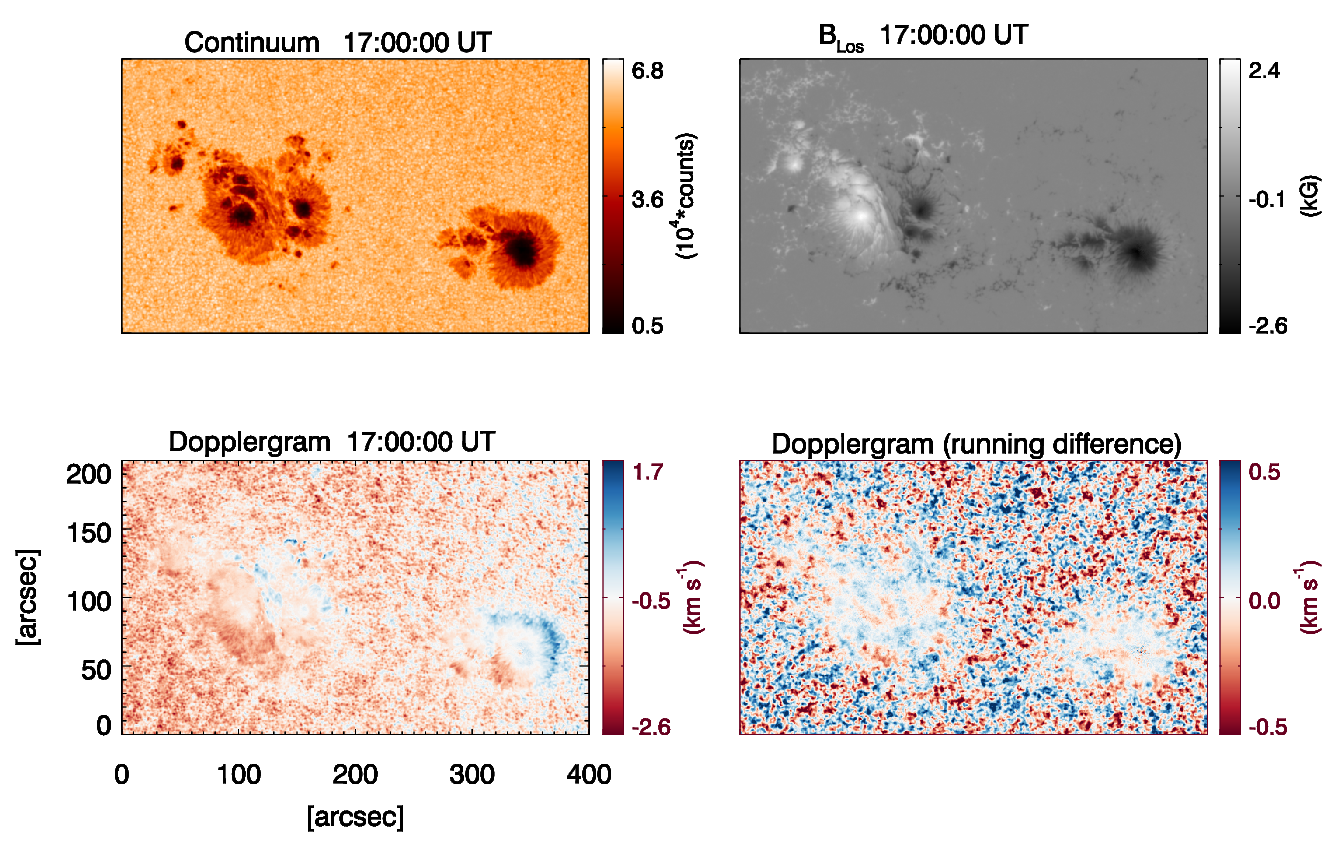}\hspace*{0.34cm}
\includegraphics[width=0.44\textwidth]{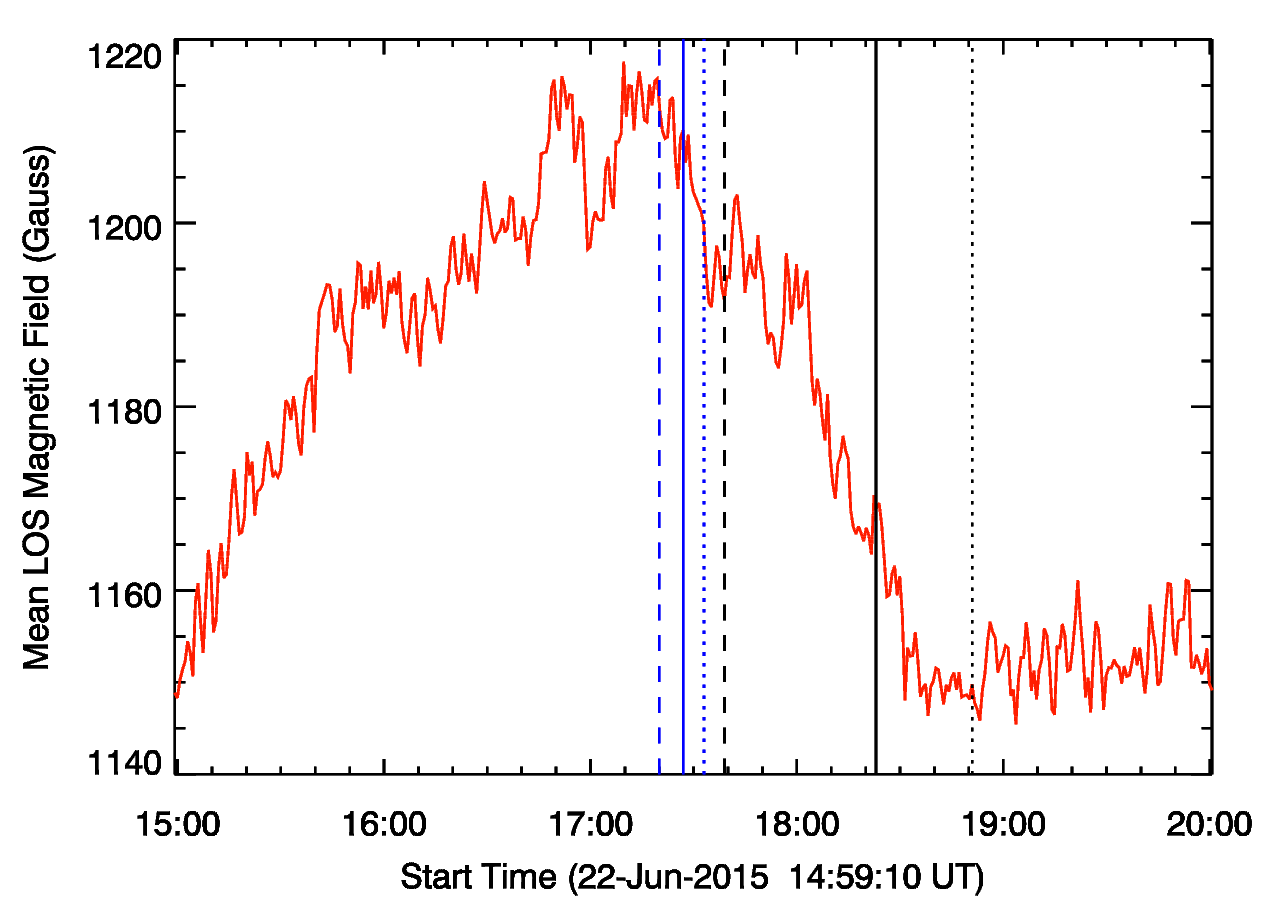}
\caption{\textit{Left panel}: Sample images of the active region NOAA 12371 showing continuum intensity (top left panel), photospheric line-of-sight magnetic fields (top right panel), Dopplergram (bottom left panel) and running difference of Doppler images (bottom right panel) acquired from HMI instrument aboard the {\em SDO} spacecraft on 2015 June 22. \textit{Right panel}: Plot shows the temporal evolution of line-of-sight magnetic fields in the `P1' location. The dashed, solid and dotted vertical lines represent the onset, peak and decay time of the flare.}
\label{fig: fourimage12371}
\end{figure*}

\begin{figure*}[h!]
\centering
\includegraphics[width=0.4\textwidth]{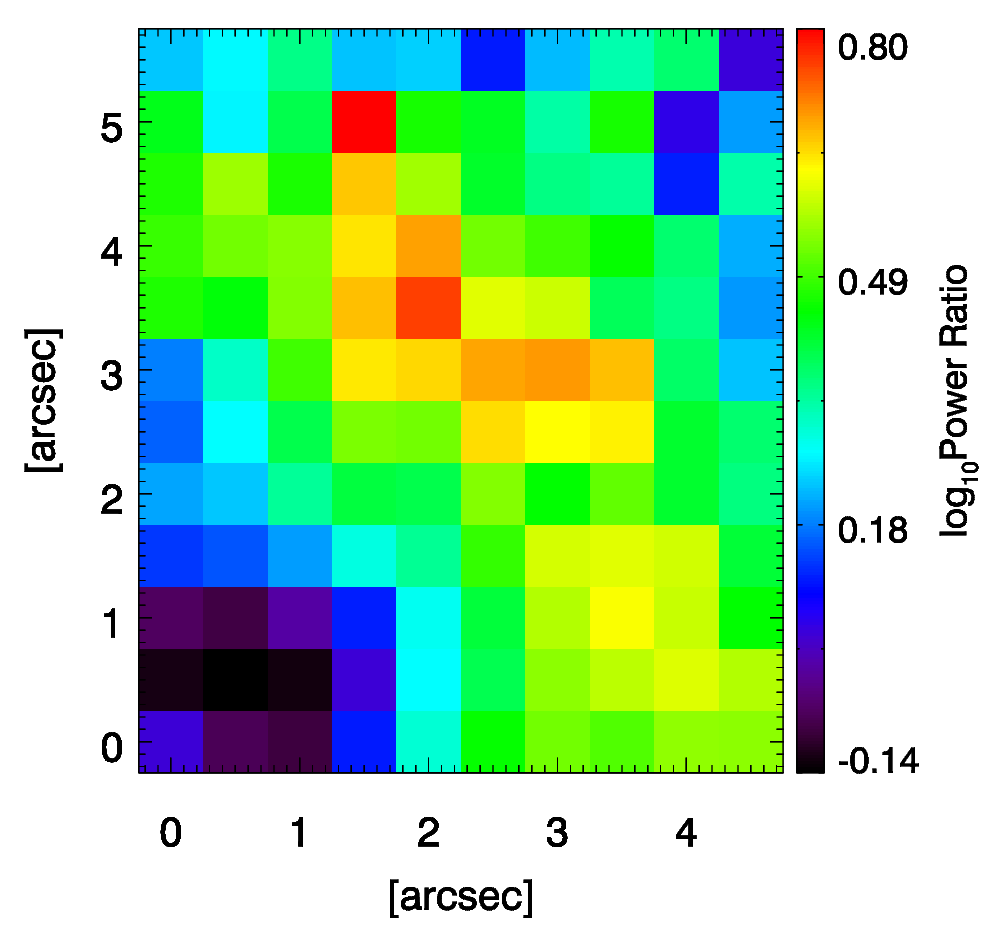}\hspace*{0.34cm}
\includegraphics[width=0.44\textwidth]{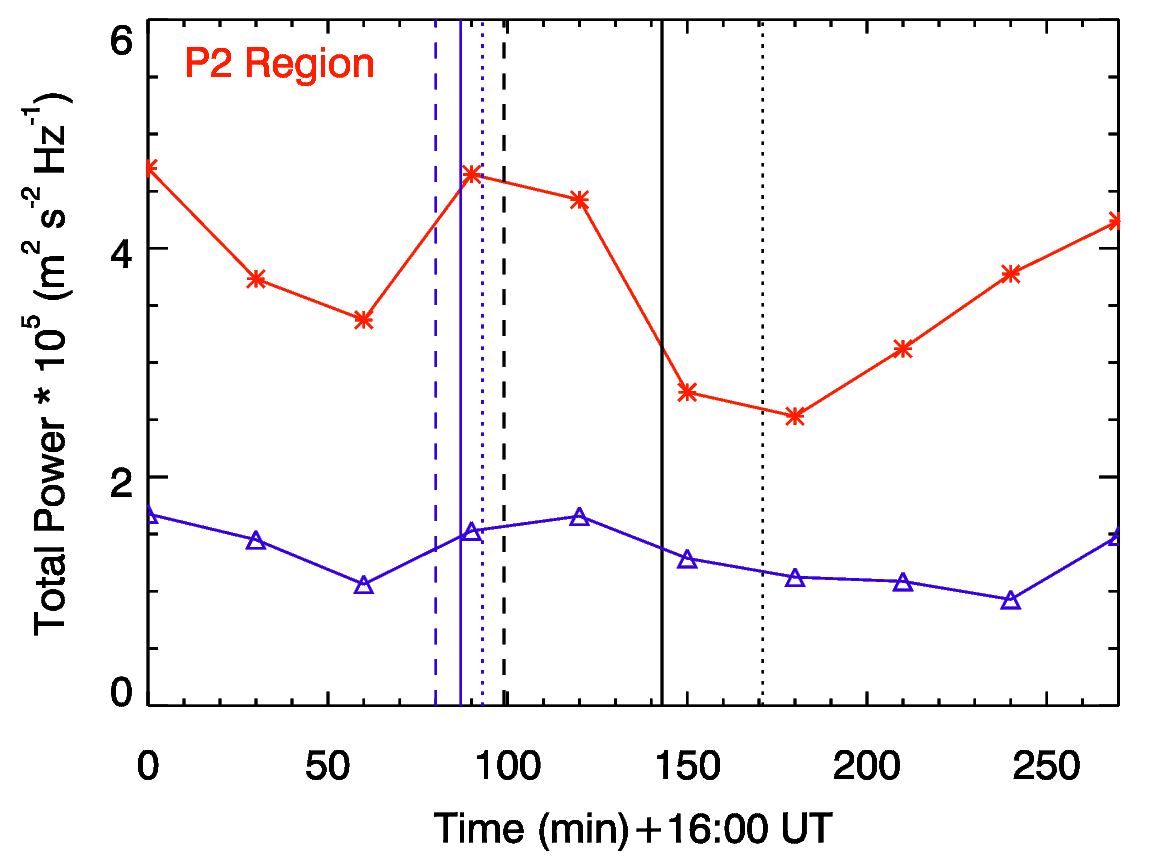}
\caption{\textit{Left panel}: Illustrates the blow-up region of `P1' enhanced location as indicated in the power map ratio in Figure 13 of the main manuscript. 
\textit{Right panel}: Plot showing the temporal evolution of integrated acoustic power over the ‘P2’ location (red colour with asterisks) whereas that shown in blue colour with triangles represents evolution of total power in an unaffected region in the same sunspot. It is to be noted that there is a time offset of about $\pm$ 30-minutes between the acoustic power variation and the GOES flare-time.}
\label{fig: prespanning12371}
\end{figure*}

\begin{figure*}
\centering
\includegraphics[width=0.44\textwidth]{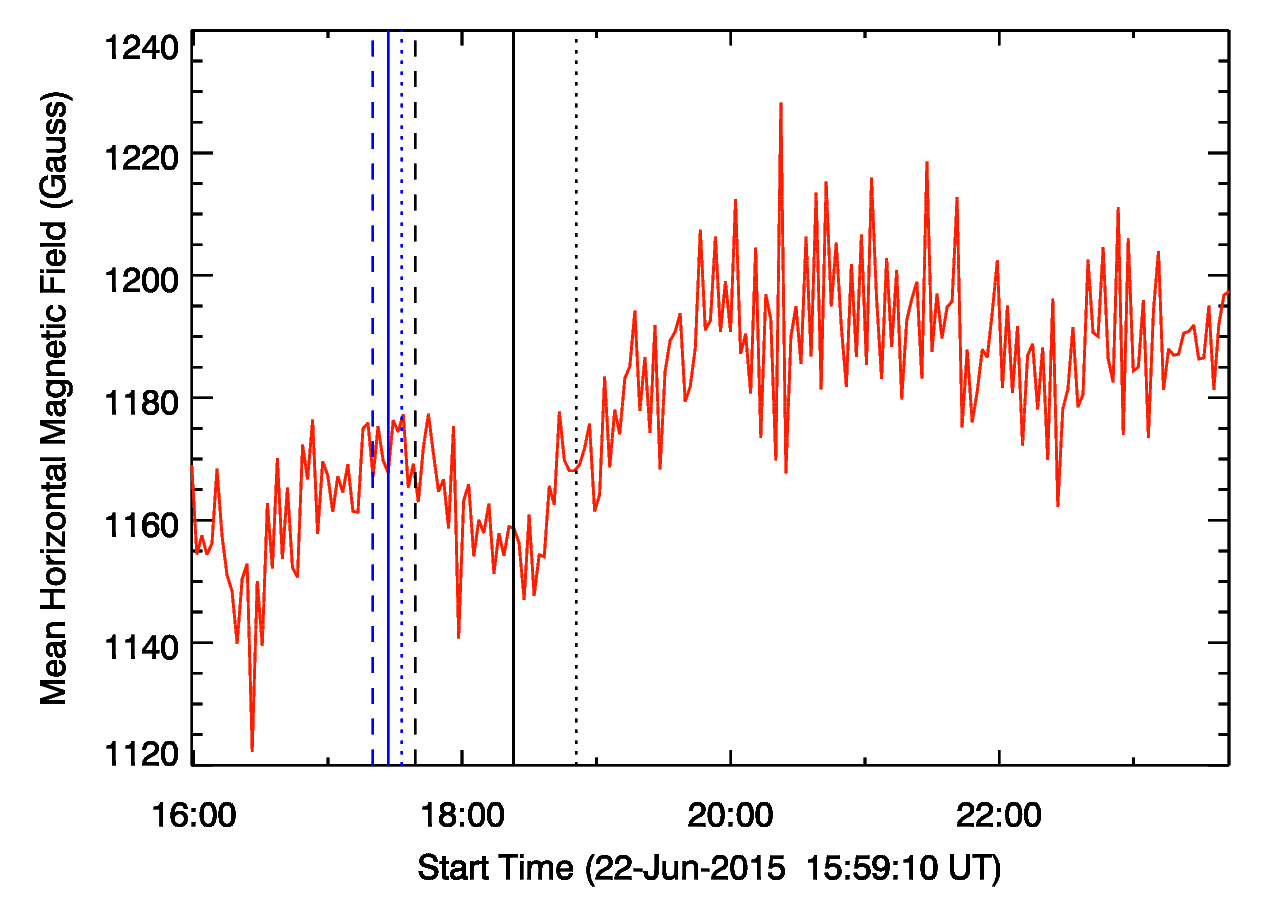}\hspace*{0.34cm}
\includegraphics[width=0.44\textwidth]{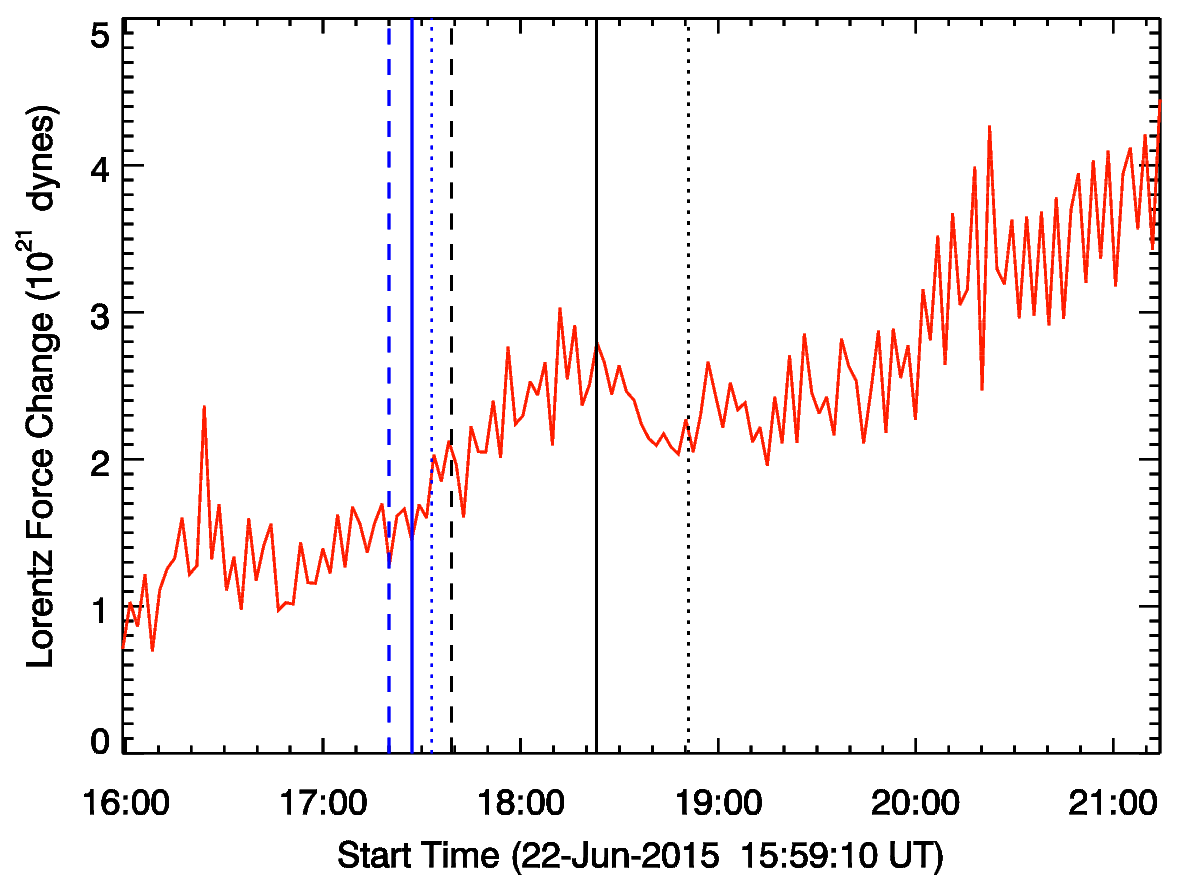}
\caption{\textit{Left panel}: Plot showing the temporal evolution of horizontal magnetic fields in the `P2' location. \textit{Right panel}: Plot showing the temporal evolution of change in horizontal component of Lorentz force in the `P2' location of active region 12371. The dashed, solid and dotted vertical lines represent the onset, peak and decay time of the flare.}
\end{figure*}

\clearpage